\documentclass[12pt
]{iopart}

\usepackage{mathrsfs,bm}
\usepackage[pdftex]{graphicx}
\usepackage{amssymb}%
\usepackage{natbib,aas_macros}

\usepackage[unicode=true,pdfusetitle,bookmarks=true,bookmarksnumbered=false,bookmarksopen=false,breaklinks=true,pdfborder={0 0 0},backref=false,colorlinks=true]{hyperref}
\hypersetup{citecolor=blue,filecolor=blue,linkcolor=blue,urlcolor=blue}

\usepackage{cleveref}
\usepackage{etoolbox}
\crefname{equation}{}{}
\patchcmd{\numparts}{\addtocounter{equation}{1}}{\refstepcounter{equation}}{}{}

\usepackage[perpage]{footmisc}

\usepackage{subcaption}

\def\Eu{ \mathfrak{H} }

\def\bea{\begin{eqnarray}}
\def\eea{\end{eqnarray}}
\newcommand{\lcdm}{$\Lambda {\rm CDM}$}
\newcommand{\flrwsolver}{\texttt{FLRWSolver}}
\newcommand{\mesc}{\texttt{mescaline}}
\newcommand{\alp}{\alpha}
\newcommand{\gam}{\gamma}
\newcommand{\Gam}{\Gamma}
\newcommand{\pd}{\partial}
\newcommand{\ph}{\phantom}

\newcommand{\mH}{\mathcal{H}}
\newcommand{\Si}{\Sigma}
\newcommand{\lam}{\lambda}

\newcommand{\sn}{SN~Ia}
\newcommand{\sne}{SNe~Ia}

\begin{document}

\title[Cosmological distances with ray tracing in GR]{Cosmological distances with general-relativistic ray tracing: framework and comparison to cosmographic predictions}

\author{Hayley J Macpherson}
\address{Department of Applied Mathematics and Theoretical Physics, University of Cambridge, Cambridge CB3 0WA, UK}
\ead{hayleyjmacpherson@gmail.com}
\vspace{10pt}
\begin{indented}
\item[]\today
\end{indented}

\begin{abstract}
In this work we present the first results from a new ray-tracing tool to calculate cosmological distances in the context of fully nonlinear general relativity. 
We use this tool to study the ability of the general 
cosmographic representation of luminosity distance, as truncated at third order in redshift, 
to accurately capture anisotropies in the ``true'' luminosity distance. 
We use numerical relativity simulations of cosmological large-scale structure formation which are free from common simplifying assumptions in cosmology. 
We find the general, third-order cosmography is accurate to within 1\% for redshifts to $z\approx 0.034$ when sampling scales strictly above 100 $h^{-1}$ Mpc,
which is in agreement with an earlier prediction. 
We find the inclusion of small-scale structure generally spoils the ability of the third-order cosmography to accurately reproduce the full luminosity distance for wide redshift intervals, as might be expected. 
For a simulation sampling small-scale structures, we find a $\sim \pm 5\%$ variance in the monopole of the ray-traced luminosity distance at $z\approx 0.02$. %
Further, all 25 observers we study here see a 9--20\% variance in the luminosity distance across their sky at $z\approx 0.03$, which reduces to 2--5\% by $z\approx 0.1$. 
These calculations are based on simulations and ray tracing which adopt fully nonlinear general relativity, and highlight the potential importance of %
fair sky-sampling in low-redshift isotropic cosmological analysis. 

\end{abstract}

\maketitle

\section{Introduction}

Standard cosmology commonly adopts the Copernican principle: the assumption that we are not privileged observers in the Universe. 
Combined with primary cosmic microwave background (CMB) radiation measurements, which strongly disfavour large deviations from isotropy in the early Universe \citep{PlanckCMB2020}, 
this leads us to conclude that the Universe is statistically homogeneous and isotropic on large scales.
These conditions comprise the \textit{cosmological principle} which forms the backbone of
the current standard $\Lambda$ cold dark matter (\lcdm) cosmological model.
This very simple standard model has enjoyed decades of success, providing a consistent fit to most of our cosmological observations to date.
However, as our cosmological data becomes increasingly more precise we are beginning to uncover some disagreements %
between theoretical predictions from \lcdm\ and our observations \citep[see][for a recent review]{Perivolaropoulos:2022}.

The Universe is inhomogeneous and anisotropic on small scales. 
The transition to statistical homogeneity in the galaxy distribution has been measured to occur at $\sim 70$--80$h^{-1}$ Mpc \citep[e.g.][]{Hogg:2005,Scrimgeour:2012}, %
however, some studies claim to have found cosmological structures above these scales \citep[e.g.][]{Clowes:2013,Horvath2015} --- potentially violating the cosmological principle. 
Tests of isotropy using late Universe large-scale structure probes do not offer a clear consensus --- with some works yielding consistency with the cosmological principle \citep[e.g.][]{Sarkar:2019,Bengaly:2017,Alonso:2015,Marinoni:2012,Gibelyou:2012,Hirata:2009} and some finding disagreement \citep[e.g.][]{Luongo:2022,Secrest2021,Migkas:2021zdo,Tiwari:2015,Appleby:2014,Rubart:2013,Singal:2011}.
See \citet{CPReview:2022} for an overview of observational tests and consistencies with the cosmological principle.

Maintaining the cosmological principle, we might further assume that the transition to statistical homogeneity and isotropy coincides with a transition to a Friedmann--Lema\^itre--Robertson--Walker (FLRW) geometrical description. 
This coincidence is not necessarily obvious either in the context of spatial averages \citep{Buchert:2000,Rasanen:2011,Buchert:2020a} or light cone averages \citep{Buchert:2022}. 

With the exception of several anomalies measured in CMB data 
(see \citet{Schwarz:2016}, Section~6 of \citet{PlanckCMB2020}, and Section~III of \citet{CPReview:2022}), 
we observe the CMB radiation to be statistically isotropic. However, this condition alone is not sufficient to guarantee a near-FLRW geometry in our vicinity. The latter also requires that the low-$\ell$ multipoles in the CMB --- as well as their time and space derivatives --- are small \citep{Clarkson:2010}. 

The FLRW metric assumption forms the basis of \lcdm\ and is prevalent throughout modern cosmological analysis. Particularly relevant to this work is the use of FLRW cosmography: a theoretical approximation of the luminosity distance using a Taylor series expansion in redshift, usually truncated at third order \citep[see][and Section~\ref{sec:flrwcg}]{Visser:2004}. This method allows constraints of FLRW parameters without assuming a particular cosmological expansion history, and is most commonly used in low-redshift supernova Type 1a (SN~Ia) analysis \citep[e.g.][]{Riess:2021,Freedman:2019}. 

Most state-of-the-art cosmological simulations implement \lcdm\ by adopting an \textit{a priori} assumed FLRW background space-time which is unaffected by the nonlinear collapse of structures. While these simulations have proven invaluable in developing our knowledge of structure formation beyond perturbation theory, this assumption limits our ability to study regimes beyond FLRW with these simulations. The weak field general-relativistic N-body code \textit{gevolution} offers an improvement by including the effects of perturbations on the background expansion \citep{Adamek:2013,Adamek:2016}. 
A more recent development is the use of numerical relativity (NR) in cosmological simulations \citep{giblin:2016a,Bentivegna:2016a,Macpherson:2017,east:2018a,Daverio:2019}, allowing the removal of the concept of a background space-time altogether. Such simulations provide the ideal testbed for the FLRW assumption in the late-time nonlinear Universe. 

Recently, \citet{Heinesen:2020b} derived a generalised version of the luminosity-distance redshift cosmography --- incorporating all sources of inhomogeneity and anisotropy due to local structures (see Section~\ref{sec:cg}). In \citet[][hereafter MH21]{Macpherson:2021}, the authors studied the impact of local anisotropy on the generalised cosmographic parameters using NR simulations. They found that %
such effects could bias an effective measurement of the parameters if the observer did not fairly sample their sky. 
In Appendix~A of MH21, the authors find that the general cosmography (as truncated to third order in redshift) may not converge for large redshift intervals. %
In this work, we extend the analysis of MH21 using ray-tracing in fully nonlinear GR. 
We will study the regime within which the general cosmography provides an accurate description of the ray-traced luminosity distance, comparing to the estimate given in MH21. 
We use the same set of observers in the very smooth simulations presented in MH21, as well as studying the anisotropies in more realistic simulations with more small-scale structure. 

Anisotropies in the luminosity distance have been studied in the context of perturbed FLRW models \citep[e.g.]{Sugiura:1999,Bonvin:2006,Bonvin:2006b,Ben-Dayan:2012}. Scatter in the Hubble diagram of luminosity distances has been studied using the \textit{gevolution} code \citep{Adamek:2019} as well as with NR simulations \citep{giblin2016b}. 
However, to the best of our knowledge, an analysis of the anisotropic signatures in the luminosity distance in simulations within the fully nonlinear context of GR has not yet been done \citep[though see][for an analysis of weak-lensing convergence in NR simulations]{Giblin:2017}. %

In Section~\ref{sec:flrwcg} and \ref{sec:cg} we provide a brief review of the FLRW and general cosmographies, respectively, in Section~\ref{sec:sims} we describe the NR simulations, and in Section~\ref{sec:analysis} we introduce our post-processing analysis software. In Section~\ref{sec:raytracing} we provide a detailed overview of the theory behind the ray-tracing code, including the propagation of geodesics, distance calculations, initial data, and some technical code details. In Section~\ref{sec:cosmographic_calc} we briefly review the calculation of the cosmographic parameters in MH21, and in Section~\ref{sec:results} we compare these calculations to our ray-tracing analysis. %
We discuss some important caveats in Section~\ref{sec:caveats} and conclude in Section~\ref{sec:conclude}.
Throughout this work, Greek indices represent space-time indices and take values $0\ldots3$, and Latin indices represent spatial indices and take values $1\ldots3$, and repeated indices imply summation. We use geometric units with $c=G=1$ throughout, where $c$ is the speed of light and $G$ is the gravitational constant.

\section{FLRW cosmography} \label{sec:flrwcg}

For the low-redshift cosmological analysis of standarisable candles, we are able to use the cosmographic relation between luminosity distance and redshift. Specifically, it is common to adopt the FLRW cosmography of \citet{Visser:2004}, %
which first involves performing a Taylor series expansion of the luminosity distance in redshift about the point of observation, namely, 
\bea\label{eq:dL_expand}
    d_L(z) = d_L^{(1)} z + d_L^{(2)} z^2 + d_L^{(3)} z^3 + \mathcal{O}(z^4),
\eea
as truncated at third order. For a strictly FLRW metric tensor, the coefficients $d_L^{(i)}$ are
\numparts\label{eqs:dLi_FLRW}
\bea
    d_{L,{\rm FLRW}}^{(1)} &= \frac{1}{H_o}, \quad\quad d_{L,{\rm FLRW}}^{(2)} = \frac{1-q_o}{2H_o}, \\
    d_{L,{\rm FLRW}}^{(3)} &= \frac{-1 + 3 q_o^2 + q_o - j_o + \Omega_{ko}}{6 H_o},
\eea\endnumparts
which are expressed in terms of the Hubble ($H_o$), deceleration ($q_o$), curvature ($\Omega_{ko}$), and jerk ($j_o$) parameters. Here, the subscript $o$ indicates the parameters are evaluated at the observer location. These parameters are defined in terms of the FLRW metric components by
\bea\label{eq:FLRWparams}
    H &\equiv \frac{\dot a}{a}, \quad\quad q \equiv - \frac{\ddot a}{aH^2}, \quad\quad
    j &\equiv \frac{\dot{\ddot a}}{a H^3}, \quad\quad \Omega_{k} \equiv - \frac{k}{a^2 H^2},
\eea
where $a$ is the FLRW scale factor, $k$ is the (constant) scalar curvature of spatial sections, and an over-dot represents a derivative with respect to time.

The parameters \eref{eq:FLRWparams} are derived without making any assumption on the form of the equations specifying the evolution of the FLRW scale factor $a(t)$, which makes it an attractive formalism to use to infer cosmological parameters. This means, for example, we do not need to assume a particular energy density of matter or dark energy to infer the parameters \eref{eq:FLRWparams} from low-redshift data. For this reason, such constraints are ``model-independent'' within the context of the FLRW metric tensor. 
We should note that using a cosmographic expansion with $z$ as the expansion parameter is strictly only valid for redshifts $z<1$, see Section~\ref{subsec:cg_apps} for further discussion on this.

\section{General cosmography}\label{sec:cg}

The coefficients \eref{eqs:dLi_FLRW} are derived assuming an FLRW metric tensor. 
Recently, \citet{Heinesen:2020b} derived these coefficients relaxing this assumption and thus arriving at a 
cosmographic relation for the luminosity distance which is fully general in terms of the metric tensor and field equations\footnote{Some assumptions have still been made, including the existence of a time-like congruence of observers and emitters as well as some regularity requirements for the Taylor series expansion, see \citet{Heinesen:2020b} for details.}. Within this formalism, the cosmological parameters are replaced by their ``effective'' namesakes which are dependent on both the observer's position and the direction of observation. 

The coefficients of the expansion \eref{eq:dL_expand} in the generalised cosmography are
\numparts\label{eq:dLexpand2}
\bea
    d_L^{(1)} &= \frac{1}{\Eu_o} \, , \qquad d_L^{(2)} =   \frac{1 - \mathfrak{Q}_o }{2 \Eu_o}  \ ,\\
    d_L^{(3)} &=  \frac{- 1 +  3 \mathfrak{Q}_o^2 + \mathfrak{Q}_o    -  \mathfrak{J}_o   + \mathfrak{R}_o }{ 6  \Eu_o}     \, , 
\eea\endnumparts
where a subscript $o$ again indicates that quantity is evaluated at the point of observation. The effective Hubble, deceleration, curvature, and jerk parameters are defined as
\numparts\label{eq:paramseff}
\bea
    \Eu &= - \frac{1}{E^2}     \frac{ {\rm d} E }{{\rm d} \lambda}  \, ,\label{eq:hdef}\\ %
    \mathfrak{Q}  &\equiv - 1 - \frac{1}{E} \frac{     \frac{ {\rm d} \Eu}{{\rm d} \lambda}    }{\Eu^2}   \, , \label{eq:Qdef} \\ %
    \mathfrak{R} &\equiv  1 +  \mathfrak{Q}  - \frac{1}{2 E^2} \frac{k^{\mu}k^\nu R_{\mu \nu} }{\Eu^2}   \, ,\label{eq:Rdef} \\
    \mathfrak{J}  &\equiv   \frac{1}{E^2} \frac{      \frac{  {\rm d^2} \Eu}{{\rm d} \lambda^2}    }{\Eu^3}  - 4  \mathfrak{Q}  - 3 \, , \label{eq:Jdef}
\eea\endnumparts
respectively. In the above, we are considering an observer moving with 4--velocity $u^\mu$ observing an incoming photon with 4--momentum $k^\mu$ which has travelled along a geodesic parametrised by $\lambda$. The energy of the photon, as seen by the observer, is $E\equiv - u_\mu k^\mu$, 
the derivative along the line of sight is $\frac{\rm d}{{\rm d}\lambda}=k^\mu \nabla_\mu$, and $R_{\mu\nu}$ is the Ricci tensor of the space-time. The set of parameters \eref{eq:paramseff} can be considered as inhomogeneous and anisotropic generalisations of the FLRW parameters \eref{eq:FLRWparams}. 
Each of the effective cosmological parameters can be expressed as multipole series expansions in the vector describing the direction of observation, $e^\mu$. This might represent, for example, the direction of an object in a given cosmological survey. Such an expansion allows us to pinpoint the anisotropic contributions to each effective cosmological parameter. For the effective Hubble parameter, for example, we have
\bea\label{eq:Eudecomp}
    \Eu (\boldsymbol{e}) = \frac{1}{3}\theta - e^\mu a_\mu + e^\mu e^\nu \sigma_{\mu\nu}, 
\eea
where $\theta$ %
is the volume expansion of the congruence defined by $u^\mu$, $a_\mu$ is its 4--acceleration, and $\sigma_{\mu\nu}$ is its shear tensor \citep[see MH21 or][for explicit definitions of these quantities]{Heinesen:2020b}. Similarly, the higher-order effective cosmological parameters can also be written as multipole decompositions, however, we refer the reader to \citet{Heinesen:2020b} for their explicit forms.

\subsection{Applications of cosmography}\label{subsec:cg_apps}

FLRW cosmography is perhaps most commonly adopted in the analysis of supernova Type Ia (\sn) as standardisable candles to infer the local Hubble expansion \citep[e.g.][]{Riess:2021,Freedman:2019}.
Any Taylor series expansion of luminosity distance using the redshift $z$ as a parameter is divergent for $z>1$ \citep{Cattoen:2007}. %
This result holds for both the FLRW and general cosmographies outlined above and we are therefore constrained to studying the low-redshift Universe with these methods. Alternatives have been proposed to be able to study high-redshift data with cosmography by, e.g., re-parameterising the redshift as $y=z/(1+z)$ \citep{Cattoen:2007} or using Pad\'e approximations in place of the Taylor expansion \citep{Gruber:2014}. Such formalisms have been applied to study observational samples with $z>1$ such as quasars, gamma-ray bursts, baryon acoustic oscillations, and \sne\ \citep[e.g.][]{Yang:2020,Lusso:2019,Capozziello:2019}. However, concerns have been raised regarding the convergence of these techniques at high redshift \citep{Banerjee:2021}. 

The general cosmographic formalism outlined in Section~\ref{sec:cg} contains a total of 61 independent degrees of freedom \citep[see][]{Heinesen:2020b} which is too many to constrain using current data. In \citet{Heinesen:hm2021}, the authors used the \textit{`quiet universe'} dust models --- which should provide a good description to the late-time Universe --- to bring this down to $\sim 30$ degrees of freedom. However, even this is still far too many to constrain with currently available data sets. 
In MH21, the authors found the effective Hubble and deceleration parameters of the general cosmography should be dominated by a quadrupole and dipole anisotropy, respectively. %
Focusing on the anisotropic signals we might expect to be dominant, and neglecting all others, is one way to drastically reduce the number of degrees of freedom in data analysis. 

Recently, several works have constrained a dipole anisotropy in the cosmographic deceleration parameter from \sn\ data by adding a dipole term on top of the usual FLRW cosmography \citep{Rahman:2022,Colin:2019b,Rubin:2020,Colin:2019}. 
\citet{Dhawan:2022} %
presented the first constraints of anisotropies in \sn\ data which were theoretically motivated by the simplified general cosmography \citep[MH21,][]{Heinesen:2020b}, including the first constraints on a quadrupole in the Hubble parameter from \sn. 
While we cannot claim a significant anisotropy in current data, %
new and improved data sets will allow for much stronger constraints on individual multipoles --- as well as more complete hierarchies of multipoles --- in near-future analyses \citep{Brout:2022}.

In Appendix~A of MH21, the authors showed that the general cosmography as expanded to third order in redshift %
may only accurately represent the true luminosity distance for narrow redshift intervals. 
Specifically, for smoothing scales of $100 h^{-1}$ Mpc, they predicted that the cosmographic luminosity distance should be correct to within $\sim 1\%$ for redshifts $z\sim$0.02--0.03. For larger smoothing scales of $200 h^{-1}$ Mpc, it was predicted to be accurate within $\sim 1\%$ for redshifts $z\sim$0.04--0.06. 
Determining the regime of applicability of the cosmographic luminosity distance is vital in ensuring accurate results when applying it to data, however, the above numbers are order-of-magnitude estimates. 
Methods to alleviate these potential issues could involve allowing for a rapid decay of the anisotropies with redshift \citep[as is considered in, e.g.][]{Dhawan:2022,Rahman:2022,Rubin:2020,Colin:2019}, or implementing some smoothing of observational data prior to the cosmological fit. %
A key part of this paper is to validate the regime of validity of the cosmographic luminosity distance, as estimated in MH21, using distances calculated with ray tracing.

\section{Simulations} \label{sec:sims}

We are interested in studying the regime of validity of the general cosmographic framework. Since this framework makes no assumption on the specific form of the metric tensor, we wish to maintain this generality in our analysis as much as possible. Numerical relativity has proven to be a viable tool for cosmological simulations without the need to make assumptions on the existence of a fictitious global ``background'' metric tensor \citep{giblin:2016a,Bentivegna:2016a,Macpherson:2017}. In particular, such simulations have proven useful for studies of general-relativistic effects on observables in the context of inhomogeneous cosmology \citep{giblin2016b,east:2018a,Tian:2021a,Macpherson:2021}. 

Here, we will use the same simulations as in MH21 in order to be able to directly compare their predictions with our results. 
The simulations were performed using the Einstein Toolkit\footnote{\url{https://einsteintoolkit.org}} \citep{Loffler:2012,Zilhao:2013}; an open-source NR code based on the Cactus\footnote{\url{https://www.cactuscode.org}} infrastructure. 
The ET has been adapted and used for cosmological simulations in a number of works \citep[e.g.][]{Bentivegna:2016a,Bentivegna:2017a,Macpherson:2017,Wang:2018} and has proven to be a valuable tool to study inhomogeneous cosmology in the nonlinear regime.
Our cosmological initial data was generated using \flrwsolver\footnote{\url{https://github.com/hayleyjm/FLRWSolver_public}} \citep{Macpherson:2017} under the assumption of linear perturbations atop a flat FLRW background space-time. The initial density fluctuations are assumed to be Gaussian random and are drawn from the matter power spectrum at $z=1000$ generated with the CLASS\footnote{\url{https://lesgourg.github.io/class_public/class.html}} code \citep{CLASS:2011,Lesgourgues2011}. The initial power spectrum is generated using $h=0.7$ (and otherwise default parameters), which need only be used to translate quoted length scales from the simulation and does not imply a particular Hubble expansion. 
The simulation evolution is matter dominated (no cosmological constant is included in Einstein's equations) and thus we compare our results to the equivalent FLRW model, namely, the Einstein-de Sitter (EdS) model. 
We adopt a continuous fluid description for the matter content, with pressure such that $P\ll \rho$, and use periodic boundary conditions \citep[see][for more specifics on the simulation setup]{Macpherson:2019}. 
We note that once the simulation starts there is no explicit constraint for the space-time to remain close to an FLRW background, however, we do find that spatial averages of the model universe on the $z\approx0$ slice agree with the EdS model to within a few percent \citep[see MH21 and also][]{Macpherson:2019}.

We use two simulations with cubic domain lengths $L=12.8 h^{-1}$ Gpc and $25.6 h^{-1}$ Gpc, both with numerical resolution $N=128$ (where the total resolution is $N^3$ grid cells). These domain size and resolution combinations were chosen in order to explicitly exclude any structure beneath $100$ and $200 h^{-1}$ Mpc, respectively, due to the requirements of the general cosmography in sampling only large-scale (expanding) space-time regions \citep[see][and MH21]{Heinesen:2020b}.

\section{Analysis software} \label{sec:analysis}

We use the new ray tracer built as a part of the \mesc\ post-processing analysis code \citep{Macpherson:2019} to calculate the luminosity distance and redshift of geodesics in our simulations. In Section~\ref{sec:mesc} we present some key details of the code, in Section~\ref{sec:metricconv} we introduce relevant conventions and definitions, and discuss the assumptions made in the code in Section~\ref{sec:assump}.

\subsection{\texttt{Mescaline}: Extracting interesting things from Cactus}\label{sec:mesc}

\texttt{Mescaline} is a post-processing analysis code written specifically to study the effects of nonlinear GR in cosmological simulations performed with the ET \citep{Macpherson:2019}.%
In \citet{Macpherson:2019} and \citet{Macpherson:2018} the authors used \mesc\ to study the averaged dynamics of inhomogeneous ET simulations on a variety of scales to address the backreaction problem for the first time in a realistic cosmic web. 
As mentioned above, MH21 used \mesc\ to study the anisotropies in the luminosity distance as predicted by the general cosmography outlined in Section~\ref{sec:cg}. All works using \mesc\ thus far have considered either spatial averaging or the study of local derivatives in the observer's vicinity to predict observable signatures.

In this paper, we make use of the new ray-tracing capabilities of \mesc\ for the first time. This new feature of the code advances the geodesic equation and the Jacobi matrix equation to calculate the angular diameter distance, luminosity distance, and redshift along a geodesic in the numerical space-time (see Section~\ref{sec:raytracing} for details of the equations evolved). In \ref{appx:tests} we show that the ray tracer matches known analytic solutions and prove its numerical convergence at the expected rate.

\subsection{Metric conventions and definitions}\label{sec:metricconv}

The primary application of \mesc\ is the analysis of NR simulations. Therefore, we adopt a 3+1 split of space-time, where the metric tensor $g_{\mu\nu}$ takes the form
\bea
    ds^2 &= g_{\mu\nu} dx^\mu dx^\nu, \\
    &= -\alp^2 dt^2 + \gam_{ij} dx^i dx^j,
\eea
where $x^\mu=(t,x^i)$ are the space-time coordinates of the simulation, $\alpha$ is the lapse function describing the spacing between spatial surfaces in time, $\gam_{\mu\nu}\equiv g_{\mu\nu}+ n_\mu n_\nu$ is the induced spatial metric of the surfaces, and we adopt a gauge condition for the shift vector of $\beta^i=0$ throughout. The normal vector describing the 3--dimensional spatial surfaces is defined as
\bea
    n_\mu \equiv - \alp \nabla_\mu x^0,
\eea
where $\nabla_\mu$ is the covariant derivative associated with $g_{\mu\nu}$, and we have $n_\mu = (-\alp,{\bf 0})$ and $n^\mu = (1/\alp,{\bf 0})$ for our chosen zero shift.

The extrinsic curvature, $K_{ij}$, describes the embedding of the spatial surfaces in space-time. It is defined as the covariant derivative of the normal vector projected onto the spatial surfaces. We can also relate it to the time derivative of the spatial metric as
\bea\label{eq:dtgamij}
    \pd_t \gam_{ij} = -2\alp K_{ij},    
\eea
where $\pd_\mu \equiv \pd/\pd x^\mu$.
The Christoffel symbols associated with the metric tensor $g_{\mu\nu}$ are defined as
\bea\label{eq:four_gammas}
    \Gamma^\mu_{\alp\beta} = \frac{1}{2}g^{\mu\nu} \left( \pd_\alp g_{\nu\beta} + \pd_\beta g_{\alp\nu} - \pd_\nu g_{\alp\beta} \right),
\eea
and the \textit{spatial} Christoffel symbols associated with $\gam_{ij}$ are
\bea\label{eq:spatial_gamma}
    {}^{(3)}\Gamma^i_{jk} = \frac{1}{2}\gam^{il} \left( \pd_j \gam_{lk} + \pd_k \gam_{jl} - \pd_l \gam_{jk} \right),
\eea
where in the special case of $\beta^i=0$ we have $\Gam^i_{jk}={}^{(3)}\Gam^i_{jk}$, and so we drop the superscript $(3)$ for the rest of this work. 
The Ricci tensor of the space-time is the contraction of the Riemann tensor, namely $R_{\mu\nu}\equiv R^\alp_{\ph{\alp}\mu\alp\nu}$, and $R\equiv g^{\mu\nu}R_{\mu\nu}$ is its trace. 

\texttt{Mescaline} calculates the spatial Christoffel symbols, the components of the 4--Ricci tensor and its trace, some relevant components of the Weyl tensor $C_{\mu\nu\alp\beta}$ (see Section~\ref{subsec:Weyl}) and its electric part in the fluid frame, %
the components of the 3--Ricci tensor ${}^{(3)}R_{ij}$ and its trace ${}^{(3)}R$, and the fluid rest-frame 3--curvature scalar \citep[see Section~4 of][]{Buchert:2020a}. %
The code also calculates kinematic and dynamical quantities in the rest-frame of the fluid, namely, the expansion scalar $\theta$, the shear tensor $\sigma_{\mu\nu}$, the vorticity tensor $\omega_{\mu\nu}$, and the 4--acceleration $a_\mu$. Following the averaging approach of \citet{Buchert:2020a}, \mesc\ also calculates \textit{fluid-intrinsic} spatial averages over a user-defined domain $\mathcal{D}$ and calculates the effective cosmological parameters including the kinematical backreaction, average curvature, volume of the domain $V_\mathcal{D}$, and effective scale factor, $a_\mathcal{D}\equiv[V_\mathcal{D}(t)/V_{\mathcal{D}}(t_{\rm ini})]^{1/3}$, where $t_{\rm ini}$ is the initial simulation time.

\subsection{Assumptions and input}\label{sec:assump}

\texttt{Mescaline} assumes a continuous fluid description of the matter content by taking input of a smooth density field, $\rho$, and velocity field with respect to the Eulerian observer, $v^i$, at all points on the grid. 
It also takes as input the spatial metric tensor of the hypersurfaces, $\gamma_{ij}$, the extrinsic curvature, $K_{ij}$, and the lapse function $\alpha$. It enforces the gauge condition $\beta^i=\pd_t \beta^i = 0$ throughout, however, the lapse and its time derivative are kept general (though the latter must be specified according to the gauge condition of the simulation). The code assumes a uniform Cartesian grid, i.e. $\Delta x = \Delta y = \Delta z$ for the grid spacing in each dimension $x^i = (x,y,z)$ when taking spatial derivatives. We adopt geometric units, $G=c=1$, throughout the code and enforce periodic boundary conditions in the calculation of spatial derivatives. 

The simulation frame is assumed to be generally separate from the frame of the fluid flow, i.e., the hypersurface normal $n^\mu$ is not equal to the fluid 4--velocity $u^\mu$. However, the observers we define in both the cosmography and ray-tracing calculations are {chosen} to be co-moving with the fluid flow by {defining} their 4--velocity to be that of the fluid at their location.

\section{Ray tracing}\label{sec:raytracing}

In this Section, we detail the ray-tracing calculation performed in \mesc. In Section~\ref{sec:geodesics} we introduce the geodesic equation and describe how it is evolved and in Section~\ref{sec:distances} we introduce the cosmological distances relevant in this work and describe the methods for calculating these distances along the geodesic.

\subsection{Geodesics}\label{sec:geodesics}
We consider the propagation of a light bundle, or ray, through the 3+1 space-time along null geodesics. 
The photon 4--momentum is
\bea
	k^\mu \equiv \frac{dx^\mu}{d\lam},
\eea
where $\lam$ is the affine parameter of the geodesic, the total derivative is $d/d\lam \equiv k^\mu \partial_\mu$, and $k^\mu$ satisfies the null condition $k_\mu k^\mu = 0$. The geodesic equation,
\bea \label{eq:geodesic}
	\frac{D k^\mu}{d\lam} \equiv \frac{dk^\mu}{d\lam} + \Gamma^\mu_{\alpha\beta}k^\alpha k^\beta = 0,
\eea
describes the evolution of $k^\mu$ along the geodesic in the space-time described by $g_{\mu\nu}$, where the covariant total derivative $D/d\lam \equiv k^\mu \nabla_\mu$. 
It is useful to decompose the photon 4--momentum in the rest frame of an observer with 4--velocity $u^\mu$ as
\bea\label{eq:kmu_decomp}
    k^\mu = E ( u^\mu - e^\mu ),
\eea
where $E=-k^\mu u_\mu$ is the energy of the photon as measured in the observer's frame. In the above, $e^\mu$ is a unit vector describing the direction of the incoming photon on the observer's sky, which
satisfies the constraints 
\bea\label{eq:emuconstraints}
    g_{\mu\nu}e^\mu e^\nu = 1, \quad e^\mu u_\mu = 0,
\eea
i.e., it is a space-like unit vector orthogonal to the observer's 4--velocity.

A photon feels the effect of the expansion of the space-time and curvature via changes in its energy, $E$, along the geodesic. 
This change in energy is measured as the photon redshift, which in general is 
defined as the ratio of the photon energy when it was emitted at the source, $s$, to when it was received at our detectors (at the observer), $o$, namely
\bea
    1 + z \equiv \frac{E_s}{E_o}.
\eea

The vector $k^\mu$ fully specifies the geodesic itself, and only depends on the space-time metric. The photon energy, and hence redshift, is defined once we choose an observer. In \mesc, we choose our observers to be co-moving with the fluid flow. In practise, this means that each observer's $u^\mu$ coincides with the fluid 4--velocity at their location in the simulation. 
Usually, after correcting for local peculiar velocities (e.g., our motion around the Sun and the galaxy or the peculiar motion of a nearby object) we assume our cosmological measurements lie in the frame co-moving with the cosmic expansion. This frame might be considered as ``co-moving with the fluid flow''. However, connecting our observations to a particular frame in the simulation is not straightforward. 
See Section~\ref{sec:observers} for a discussion on this topic.

\subsubsection{Evolving the geodesic equation.}

We wish to numerically advance the geodesic equation \eref{eq:geodesic} to trace the path of photons through the simulated space-time. %
We can freely re-parametrise the geodesic in terms of the simulation coordinate time, $t$, as follows: 
\bea
	\frac{dk^\mu}{dt} & = \frac{d\lam}{dt} \frac{dk^\mu}{d\lam}, \\
    & =  \frac{1}{k^0} \frac{dk^\mu}{d\lam}, \label{eq:geodesic_dt}
\eea
so long as the directional aspect of the derivative $d/d\lam$ is preserved. Namely, we must ensure that we keep track of the change in position of the geodesic in time. 
In the above formulation, we are re-casting the directional derivative along the geodesic to be spaced in terms of coordinate time, $t$, instead of affine parameter, $\lam$. 
Next, substituting the geodesic equation \eref{eq:geodesic} into the right hand side of \eref{eq:geodesic_dt}
we arrive at the following system of equations
\numparts\label{eq:kmuevol_step1}
\bea
	\frac{dk^0}{dt} &= - \Gamma^0_{00} k^0 - \Gamma^0_{0j}k^j - k^j \Gamma^0_{j0} - \frac{k^k}{k^0} \Gamma^0_{kj}k^j, \\
	\frac{dk^i}{dt} &= - \Gamma^i_{00} k^0 - \Gamma^i_{0j}k^j - k^j \Gamma^i_{j0} - \frac{k^k}{k^0} \Gamma^i_{kj}k^j, \\
	\frac{dx^i}{dt} &= \frac{k^i}{k^0}, \label{eq:dxidt}
\eea\endnumparts
where the last equation describes the path of the geodesic through the simulation domain.
The time components of the Christoffel symbols appearing in the equations above, in terms of 3+1 variables, are
\bea
	\Gamma^0_{00} &= \frac{1}{\alpha} \partial_t \alpha,  \quad \Gamma^0_{j0} = \frac{1}{\alpha} \partial_j \alpha,  \quad
	\Gamma^0_{jk} = -\frac{1}{\alpha} K_{jk}, \nonumber\\ 
	\Gamma^i_{00} &= \alpha \gamma^{ij} \partial_j \alpha,  \quad
	\Gamma^i_{0j} = -\alpha \gamma^{ik} K_{jk},  \label{eqs:time_christ}
\eea%
and $\Gamma^i_{jk}$ is calculated using spatial derivatives of the metric tensor as defined in \eref{eq:spatial_gamma}. 
Substituting these into \eref{eq:kmuevol_step1} we arrive at
\numparts\label{eqs:kmuevol}
\bea
	\frac{dk^0}{dt} &= - \frac{k^0}{\alp}\pd_t \alp - \frac{2}{\alp}k^j\pd_j\alp + \frac{k^k}{\alp k^0} K_{kj}k^j, \\
	\frac{dk^i}{dt} &= - \alp k^0 \gam^{ij}\pd_j\alp  + 2 \alp K^i_{\ph{i}j} k^j - \frac{k^k}{k^0} \Gamma^i_{kj}k^j, \\
	\frac{dx^i}{dt} &= \frac{k^i}{k^0},
\eea\endnumparts
which is the system solved in \mesc\ to advance the geodesic through the simulated space-time. 
We describe the process of choosing initial data for $k^\mu$ in Section~\ref{sec:initial_data}.

\subsection{Distances}\label{sec:distances}

The geodesic equation tells us how the energy of one photon traverses through a given space-time. This is not enough to study observables in cosmology, since many of our observations consider distances. 
The most relevant distance for this work is the luminosity distance $D_L$, which can be defined as the ratio of the intrinsic luminosity of a source, $L$, to the flux received on Earth, $F$, namely
\bea
    D_L = \sqrt{ \frac{L}{4\pi F}}.
\eea
The luminosity distance can also be defined in terms of the distance modulus $\mu \equiv m-M = 5 {\rm log}_{10}\left(\frac{D_L}{10{\rm pc}}\right)$, where $m$ is the apparent magnitude of a source and $M$ its absolute magnitude. 
Another distance we are interested in is the angular diameter distance $D_A$, which is defined as
\bea\label{eq:DAdef}
    D_A^2 \equiv \frac{A_s}{\Omega_o},
\eea
where $A_s$ is the physical area of a source, and $\Omega_o$ is its observed angular extent. Assuming conservation of photons, these two distances are related by Etherington's reciprocity relation \citep[which holds in any space-time and under any theory of gravity, see, e.g.][]{Ellis:2007}%
\bea\label{eq:dLdA}
    D_L = (1+z)^2 D_A.
\eea
To calculate the angular-diameter distance we must track the evolution of a small bundle of photons, or a light beam, as they traverse the space-time geometry. This beam might be thought of as the collection of photons travelling from an extended source towards our telescopes. 
We use a separation vector, $\xi^\mu$, to describe the behaviour of two rays next to one another in the bundle in order to eventually relate the physical area of the source to its observed angular extent. 
Assuming the beam is narrow, we can track the evolution of the bundle using the evolution of $\xi^\mu$ along the geodesic, which gives rise to the geodesic deviation equation.
We refer the reader to Chapter~2 of \citet{Fleury:2015a} for an in-depth introduction to light propagation and light beams, including the derivation of the geodesic deviation equation.

\subsubsection{Screen space.}\label{subsec:screen}

The separation vector $\xi^\mu$ contains information about our infinitesimal ray bundle as it traverses through the space-time along the geodesic. We are interested in further relating this information to observable quantities.
For this purpose, it is useful to introduce a 2--dimensional plane (located in the 4--dimensional space-time) on which we project our light bundle. This is known as the ``screen space'', and is
defined by the set of %
basis vectors $s^\mu_A$, for $A\in[1,2]$. We can think of this 2--dimensional space as a flat plate that an observer holds perpendicular to the light bundle, thus projecting the light onto the ``screen'' (much like we do with our telescopes), at a point on the geodesic in order to measure the separation of rays within the bundle.
The equations we will solve (which we introduce in the next section) describe how the separation between light rays projected onto this screen evolves as the bundle travels along the geodesic. They describe how the curvature of space-time influences the shape and size of the beam throughout its journey. 

The vectors $s^\mu_A$ form an orthonormal basis of the screen space within the space-time defined by $g_{\mu\nu}$, and must satisfy the constraints
\bea\label{eq:smuconstraints}
	s^\mu_A u_\mu = s^\mu_A e_\mu = 0, \qquad s_\mu^A s_B^\mu = \delta^A_B,
\eea
where %
$\delta^A_B$ is the Kronecker delta. In words, the screen basis is orthogonal to the observers 4--velocity $u^\mu$, and the direction of observation $e^\mu$, and thus is also orthogonal to the direction of propagation of the ray bundle (i.e. we will also have $s^\mu_A k_\mu=0$ via \eref{eq:kmu_decomp}). They are also orthogonal to each other and normalised to have unit length. 

The screen vectors are essential for distance calculations along the geodesic, since the beams morphology is calculated from the separation of geodesics in screen space. 
The basis $s^\mu_A$ must therefore also be propagated along the geodesic, however, since they are an arbitrary basis they should not affect the physical evolution along the bundle. In \ref{appx:screen_prop}, we provide details of how we propagate the screen vectors and show that our physical results remain unchanged with different initial $s^\mu_A$.

\subsubsection{Jacobi Matrix.}\label{subsec:jacobi}

Now that we are familiar with the concept of screen space, we will briefly describe the connection of the separation vector to observables. %
We will only briefly touch on the derivation of the equations involved, mainly for interest and physical intuition of the equations we solve in \mesc.

Projecting the separation vector $\xi^\mu$ into the screen space gives 
the separation of rays in the bundle in the 2--dimensional plane, namely $\xi_A\equiv \xi_\mu s^\mu_A$. In combination with the geodesic deviation equation, $\xi_A$ will allow us to track intrinsic properties of the beam and relate these to quantities measured by observers in the space-time.

After this projection, the geodesic deviation equation gives rise to the Sachs vector equation \citep[see][]{Fleury:2015a}
\bea\label{eq:sachsvector}
    \frac{{d^2}\xi^A}{{d}\lam^2} = \mathcal{R}^A_{\ph{A}B}\xi^B,
\eea
where we have introduced the optical tidal matrix 
\begin{equation}\label{eq:RAB_def}
	\mathcal{R}_{AB} = 
    \pmatrix{%
	\mathscr{R} & 0\cr %
	0 & \mathscr{R}}
	+
    \pmatrix{%
	-{\rm Re}(\mathscr{W}) & {\rm Im}(\mathscr{W})\cr %
	{\rm Im}(\mathscr{W}) & {\rm Re}(\mathscr{W})},
\end{equation}
and the positioning of the indices $A,B$ does not matter since they are raised and lowered with $\delta^A_B$, however, summation is still implied over repeated indices. In \eref{eq:RAB_def}, the Ricci lensing scalar is
\bea
	\mathscr{R} &\equiv -\frac{1}{2} R_{\mu\nu}k^\mu k^\nu,   %
\eea
which describes the focusing of the light beam due to matter in its path. The Weyl lensing scalar is
\bea
	\mathscr{W} &\equiv -\frac{1}{2} C_{\mu\nu\alp\beta} \sigma^\mu k^\nu k^\alp \sigma^\beta, \label{eq:curlyW}
\eea
where $\sigma^\mu\equiv s^\mu_1 - {\rm i} s^\mu_2$ (with ${\rm i}^2=-1$), which describes the \textit{shearing} of the light beam due to nearby structures outside of the beam itself. This is natural given that the inherent nature of the Weyl tensor describes the distortion of bodies (which in this case is the light beam) due to external tidal forces. We discuss the method for calculating $\mathscr{W}$ in Section~\ref{subsec:Weyl} below.

The linearity of \eref{eq:sachsvector} implies that there exists a direct mapping between any solution and its initial conditions. If we consider a beam of light that converges at an observer at point $o$, and integrate \eref{eq:sachsvector} outwards to a source at point $s$, we can show that the physical separation of two rays in the bundle at the source, $\xi^A_s$, is
\bea\label{eq:jacobidef}
    \xi^A_s = \mathcal{D}^A_{\ph{A}B}(s \leftarrow o) \frac{{d}\xi^B}{{d}\lam}\bigg|_o,
\eea
where $s \leftarrow o$ indicates the direction of integration. 
The derivative on the right hand side can be written as \citep[see][]{Fleury:2015a}
\bea
    \frac{{d}\xi^B}{{d}\lam}\bigg|_o=-E_o \theta^B_o,
\eea
where $\theta^B_o$ is the angular separation of the rays on the observer's sky. Therefore, in \eref{eq:jacobidef}, $\mathcal{D}^A_{\ph{A}B}$ represents the Jacobi matrix mapping of the physical separation of rays in the bundle at the source, $\xi^A_s$, to the observed angular separation of the same rays at the observer, $\theta^B_o$. This matrix therefore contains all information relating the physical attributes of a source to how it appears to an observer.  
Especially relevant to this work is its determinant, %
which is the ratio of the physical area of the source to its observed angular extent, 
\bea
    E_o \,{\rm det}|\mathcal{D}| = \frac{A_s}{\Omega_o}, %
\eea
which is the definition of the angular diameter distance, $D_A$, as given in \eref{eq:DAdef}. 
The Jacobi matrix can also be decomposed in terms of quantities describing image shear and rotation due to gravitational lensing \citep[see Section~2.2.2 of][]{Fleury:2015a}.
Evolving the Jacobi matrix equation directly thus also allows for straightforward analysis of weak lensing observables using the same data.

We note that evolving a bundle of geodesics via the Jacobi matrix equation (or any of its decompositions) is not the only viable method to calculate distances in simulation data. One might instead choose to evolve the geodesic equation \eref{eq:geodesic} individually for a set of neighbouring light rays and calculate their separations $\xi_A$ along the way \citep[e.g.][]{Fluke:1999,Fluke:2011,Bentivegna:2017b,AkterEma2021,Breton:2022}.

\subsubsection{Evolving the Jacobi matrix equation.}

Interpreting the Jacobi matrix as relating observations of a source to its physical properties relies on the evolution of the Sachs vector equation from an observer out to the source. 
We can obtain the evolution of the Jacobi matrix along the geodesic by taking the second derivative of \eref{eq:jacobidef} and inserting the Sachs vector equation \eref{eq:sachsvector}, which gives its evolution in terms of the optical tidal matrix \citep{Fleury:2015a}
\bea\label{eq:jacobis}
	\frac{d^2}{d\lam^2} \mathcal{D}^A_{\ph{A}B} = \mathcal{R}^A_{\ph{A}C} \mathcal{D}^C_{\ph{C}B}.
\eea
As a next step, we might decompose the Jacobi matrix to define the \textit{Sachs optical scalars} $\theta_S$ and $\sigma_S$\footnote{We add a subscript $S$ here to avoid confusion with the kinematic fluid expansion and shear, however, these scalars are usually referred to without this subscript.}, which describe the expansion and shear rates of the beam, respectively \citep{Sachs:1961}. The expansion rate $\theta_S$ is singular at the point of observation, $\lam=\lam_o$, and so it is not ideal for numerical integration. It is therefore common for ray-tracing in numerical data to cast the evolution equations for the optical scalars into a different form to instead evolve, e.g., the area of the beam \citep[e.g.][]{giblin2016b} or the angular diameter distance directly \citep[e.g.][]{Lepori:2020}. See also \citet{Grasso:2021,Grasso:2022} for light propagation methods using bi-local geodesic operators \citep{Grasso:2019}.
However, here we do not take either of these routes. Instead, we directly integrate the evolution equation \eref{eq:jacobis} 
to obtain the full Jacobi matrix along the geodesic. 

We now need to write \eref{eq:jacobis} in a form suitable for numerical integration. First, as with the geodesic equation, we will re-cast the derivative with respect to $\lam$ into a derivative with respect to the time coordinate, $t$. Thus, we can write
\bea
	\frac{d^2}{d\lam^2} &= \frac{d}{d\lam}\left(k^0 \frac{d}{d t}\right) \\
	&= \frac{d k^0}{d \lam}\frac{d}{d t} + k^0 \frac{d}{d \lam}\frac{d}{d t} \\
	&= - \Gam^0_{\alp\beta}k^\alp k^\beta \frac{d}{d t} + \left(k^0\right)^2 \frac{d^2}{d t^2}
\eea
where to get to the last line we have used the geodesic equation \eref{eq:geodesic}. Therefore, \eref{eq:jacobis} becomes
\bea\label{eq:jacobi_step2}
	- \Gam^0_{\alp\beta}k^\alp k^\beta \frac{d}{dt}\mathcal{D}^A_{\ph{A}B} + (k^0)^2 \frac{d^2}{dt^2} \mathcal{D}^A_{\ph{A}B} = \mathcal{R}^A_{\ph{A}C} \mathcal{D}^C_{\ph{C}B},
\eea
and to simplify the system we define
\bea
	\mathcal{P}^A_{\ph{A}B} \equiv \frac{d}{dt}\mathcal{D}^A_{\ph{A}B},
\eea
such that we can re-write \eref{eq:jacobi_step2} to arrive at the following system of evolution equations
\bea
	\frac{d}{dt}\mathcal{P}^A_{\ph{A}B} &= \frac{1}{\left(k^0\right)^2}\Gam^0_{\alp\beta}k^\alp k^\beta \mathcal{P}^A_{\ph{A}B} + \mathcal{R}^A_{\ph{A}C} \mathcal{D}^C_{\ph{C}B},\label{eq:dtPAB} \\
	\frac{d}{dt} \mathcal{D}^A_{\ph{A}B} &= \mathcal{P}^A_{\ph{A}B}.
\eea
Expanding the sum $\Gamma^0_{\alp\beta}k^\alp k^\beta$ in \eref{eq:dtPAB} for a 3+1 metric, we arrive at the full system of evolution equations
\numparts\label{eqs:jacobievol}
\bea
	\frac{d}{dt}\mathcal{P}^A_{\ph{A}B} &= \bigg[ \,\frac{2}{\alp k^0} k^i \pd_i\alp - \frac{1}{\alp(k^0)^2}K_{ij}k^i k^j + \frac{1}{\alp}\pd_t\alp\, \bigg]\, \mathcal{P}^A_{\ph{A}B} \nonumber\\
	&\quad\quad + \frac{1}{(k^0)^2}\mathcal{R}^A_{\ph{A}C} \mathcal{D}^C_{\ph{C}B}, \\
	\frac{d}{dt} \mathcal{D}^A_{\ph{A}B} &= \mathcal{P}^A_{\ph{A}B},
\eea\endnumparts
and the initial conditions at the point of observation are \citep{Fleury:2015a}
\numparts\label{eq:DABini}
\bea
	\mathcal{D}^A_{\ph{A}B}(\lam_o) &= 0, \\
	\mathcal{P}^A_{\ph{A}B}(\lam_o) &= \delta^A_{\ph{A}B}. %
\eea\endnumparts

The system of equations \eref{eqs:jacobievol} combined with the evolution equations for $k^\mu$ in \eref{eqs:kmuevol} and the screen vector evolution in \ref{appx:screen_prop} fully specifies the evolution of the Jacobi matrix along the geodesic. From this system we can extract the angular diameter distance and the redshift at all points along the geodesic for a given numerical metric tensor. %

\subsubsection{Weyl tensor.}\label{subsec:Weyl}

The Weyl tensor appears in the optical tidal matrix \eref{eq:RAB_def} in the Weyl lensing scalar \eref{eq:curlyW}.
It is therefore an important part of the evolution of the Jacobi matrix. The Weyl tensor is defined as the trace-free part of the Riemann tensor, namely,
\bea
\label{eq:Weyldef}
	C_{\mu\nu\alp\beta} & \equiv R_{\mu\nu\alp\beta} + \frac{1}{2} \bigg[ g_{\mu\beta}R_{\nu\alp} + g_{\nu\alp}R_{\mu\beta} - g_{\mu\alp}R_{\nu\beta} - g_{\nu\beta}R_{\mu\alp}\bigg] \nonumber\\
	&\quad\quad + \frac{1}{6} \bigg( g_{\mu\alp} g_{\nu\beta} - g_{\mu\beta} g_{\nu\alp} \bigg) R. 
\eea

To calculate the Weyl curvature scalar $\mathscr{W}$, we adopt a method similar to that described in Appendix~B of \citet{giblin2016b}. 
We begin by collapsing the space-time indices in \eref{eq:curlyW} into a new set of indices, namely
\bea\label{eq:scrW_step0}
	\mathscr{W} = -\frac{1}{2} C_{\Sigma \Pi} X^\Sigma Y^\Pi,
\eea
with $X^\Si\equiv \sigma^\mu k^\nu$ and $Y^\Pi\equiv k^\alp \sigma^\beta$, 
and the index $\Si$ therefore takes values over all combinations of $\mu\nu$, and $\Pi$ takes values over all combinations of $\alpha\beta$. 
Substituting the definition of the Weyl tensor into \eref{eq:scrW_step0}, all terms in brackets in \eref{eq:Weyldef} are zero either due to the constraints on the screen vectors \eref{eq:smuconstraints} (i.e. after contracting with $\sigma^\mu\equiv s^\mu_1 - {\rm i}s^\mu_2$), or because they cancel with one another. The expression therefore simplifies to
\bea\label{eq:scrW_step1}
	\mathscr{W} = -\frac{1}{2} R_{\Sigma \Pi} X^\Si Y^\Pi,
\eea
where $R_{\Sigma \Pi}\equiv R_{\mu\nu\alp\beta}$.
This expression is symmetric in changes $\Si\leftrightarrow\Pi$ due to the symmetries in the Riemann tensor. %
All twelve possible values for the indices are $\Si,\Pi \in$ [01, 02, 03, 12, 13, 23, 10, 20, 30, 21, 31, 32], because combinations with repeated indices in either $\Si$ or $\Pi$ (i.e., $\Si=00,11$, etc.) have zero Weyl tensor component. We define the `forward' indices as $\Si_f,\Pi_f \in$ [01, 02, 03, 12, 13, 23], and the `backward' indices as their reverse values $\Si_b,\Pi_b \in$ [10, 20, 30, 21, 31, 32]. The Riemann tensor $R_{\Si\Pi}$ is antisymmetric in exchanges $\Si_f \rightarrow \Si_b$ or $\Pi_f \rightarrow \Pi_b$. Expanding the sum in index $\Si$ in \eref{eq:scrW_step1} therefore gives
\bea
	R_{\Si \Pi} X^\Si Y^\Pi &= R_{\Si_f \Pi} X^{\Si_f} Y^\Pi + R_{\Si_b \Pi} X^{\Si_b} Y^\Pi, \\
	&= R_{\Si_f \Pi} (X^{\Si_f} - X^{\Si_b})Y^\Pi, \\
	&= 2 R_{\Si_f \Pi} X^{[\Si_f]} Y^\Pi, \label{eq:scrW_step2}
\eea
since $R_{\Si_b \Pi}=-R_{\Si_f \Pi}$ and we have denoted the antisymmetric part of $X^{\Si_f}$ as $X^{[\Si_f]}\equiv \frac{1}{2}(X^{\Si_f} - X^{\Si_b})$. Expanding the $\Pi$ index in \eref{eq:scrW_step2} then gives
\bea
	2 R_{\Si_f \Pi} X^{[\Si_f]} Y^\Pi &= 2 R_{\Si_f \Pi_f} X^{[\Si_f]} Y^{\Pi_f} + 2R_{\Si_f \Pi_b} X^{[\Si_f]} Y^{\Pi_b},\nonumber \\
	&= 4 R_{\Si_f \Pi_f} X^{[\Si_f]} Y^{[\Pi_f]},
\eea
following similar steps. 
The Weyl curvature scalar finally becomes
\bea\label{eq:scrW} 
	\mathscr{W} =-2 R_{\Si\Pi} X^{[\Si]} Y^{[\Pi]},
\eea
where we have dropped the subscript $f$ in the indices for brevity. %
We next need to split this expression into its real and imaginary parts to calculate the tidal matrix components in \eref{eq:RAB_def}. The imaginary part comes from the vector $\sigma^\mu$, and therefore is in $X^\Si$ and $Y^\Pi$. We split these into 
\numparts\label{eq:XAsplit}
\bea
    X^\Si &= X^\Si_1 - {\rm i} X^\Si_2, \\
    Y^\Pi &= Y^\Pi_1 - {\rm i} Y^\Pi_2,
\eea\endnumparts
where
\bea
	X^\Si_1 &\equiv s^\mu_1 k^\nu, \quad X^\Si_2 \equiv s^\mu_2 k^\nu, \\
	Y^\Pi_1 &\equiv k^\alp s^\beta_1, \quad Y^\Pi_2 \equiv k^\alp s^\beta_2.
\eea
Substituting \eref{eq:XAsplit} into \eref{eq:scrW} gives
\bea
	{\rm Re}(\mathscr{W}) &= -2 R_{\Si\Pi} \left( X^{[\Si]}_1 Y^{[\Pi]}_1- X^{[\Si]}_2 Y^{[\Pi]}_2  \right), \\
	{\rm Im}(\mathscr{W}) &= 2 R_{\Si\Pi} \left(X^{[\Si]}_1 Y^{[\Pi]}_2 + X^{[\Si]}_2 Y^{[\Pi]}_1\right),
\eea
with $\Si,\Pi\in[01,02,03,12,13,23]$.
This calculation of the Weyl scalar is completely general for any metric $g_{\mu\nu}$, and only depends on the constraints on the screen vectors \eref{eq:smuconstraints} and the null condition for the photon 4--momentum (i.e., when either set of vectors are contracted with $g_{\mu\nu}$) as well as some symmetries of the Weyl tensor. 

The components of the Riemann tensor we need %
for these values of $\Si,\Pi$ can be written in terms of 3+1 variables, for zero shift vector, as
\numparts\bea
	R_{0i0j} &= \alp \pd_0 K_{ij} + \alp \pd_i \pd_j \alp - \alp \pd_k (\alp) \Gam^k_{ij} + \alp^2 K_{kj} \gam^{kl} K_{li}, \\
	R_{0ijk} &= \alp \pd_j K_{ik} - \alp\pd_k K_{ij} + \alp K_{lj} \Gam^l_{ik} - \alp K_{lk} \Gam^l_{ij}, \\
	R_{ijkl} &= \gam_{im} \left[ \pd_k \Gam^m_{jl} - \pd_l \Gam^m_{jk} + \Gam^m_{nk}\Gam^n_{jl} - \Gam^m_{nl}\Gam^n_{jk}\right] + K_{ik} K_{jl} - K_{il}K_{jk}.
\eea\endnumparts

\subsection{Initial data}\label{sec:initial_data}

To evolve our system and trace the path of the geodesic back in time, we first need to specify initial data for $k^\mu, x^\mu, \mathcal{D}^A_{\ph{A}B}$, $\mathcal{P}^A_{\ph{A}B}$, and the screen vectors $s^\mu_A$ for each geodesic at the observer position. %
The initial data for $x^\mu$ is just the position of the observer and the initial data for the Jacobi matrix was given in \eref{eq:DABini}. 
The initial $k^\mu$ will be different for each geodesic and is dependent on what kind of sky sampling we are interested in. In general, we need to translate a set of lines of sight as seen by the observer, e.g. the coordinates of a particular survey, into an initial set of $k^\mu$. From the decomposition \eref{eq:kmu_decomp}, the photon 4--momentum $k^\mu$ is fully specified by the observer 4--velocity, $u^\mu$, and the direction of the source on the sky, $e^\mu$, (up to a scaling by the energy of the photon, $E$) both of which are arbitrary so long as the relevant constraints are satisfied for $e^\mu$. We take our observers to be co-moving with the fluid flow and thus $u^\mu$ is specified using the 4--velocity of the fluid at the observers location.
To set the initial $e^\mu$, we first write the spatial components as
\bea\label{eq:eimi}
	e^i = D \, m^i , 
\eea
where $m^i$ is some spatial direction vector and $D$ is a scaling factor to be determined from the constraints. In the frame of the observer, $e^\mu$ is purely spatial and thus $e^0=0$. However, the simulation is not necessarily performed in the rest frame of the observer and therefore the coordinates $\{t,x^i\}$ are in general not adapted to the fluid flow. We will in general have $e^0\neq 0$, and we must also determine this from the constraints on $e^\mu$.
We take the vector $m^i$ to represent the $\{x,y,z\}$ Cartesian coordinates of the direction of an incoming geodesic on the observer's sky. 
Given an initial $m^i$, we determine $D$ and $e^0$ by substituting \eref{eq:eimi} into the constraints \eref{eq:emuconstraints}, which gives
\numparts\label{eq:e0_D}
\bea
	e^0 &= \frac{\pm m^i u_i}{\sqrt{u_0^2 \gam_{ij} m^i m^j - \alp^2 m^i m^j u_i u_j}}, \\
	D &= \frac{\mp u_0}{\sqrt{u_0^2 \gam_{ij} m^i m^j - \alp^2 m^i m^j u_i u_j}},
\eea\endnumparts
where $\alp$, $\gam_{ij}$, and $u_i$ are evaluated at the observer's position in the simulation \citep[see also Appendix~B3 of][]{Macpherson:2021}.

\texttt{Mescaline} takes a set of directions $m^i$ as defined \textit{in the observer's frame} representing, e.g., a mock supernova survey or an even coverage in the direction of \texttt{HEALPix}\footnote{\url{https://healpix.sourceforge.io}} indices \citep[see][]{Gorski:2005}.
To ensure our initial data is defined in the same frame used to advance the geodesics, we first transform these $m^i$ into the simulation frame prior to calculating $e^\mu$ (see \ref{appx:fermi}). %
After determining the components of $e^\mu$ using \eref{eq:e0_D}, we then use the decomposition \eref{eq:kmu_decomp}, with an arbitrary choice of the photon energy at the observer $E\rightarrow E_o$ (since we are interested in the \textit{ratio} of the energies at observer and source), and the value of the 4--velocity at the observer's position, to determine the initial $k^\mu$.

We detail the process of generating initial data for the screen vectors in \ref{appx:screen_init}. Their initial values are built from the constraints \eref{eq:smuconstraints} after specifying the three remaining degrees of freedom arbitrarily. This choice does not affect the final physical result, which we show in \ref{appx:screen_init_test}.

\subsection{Time stepping, interpolation, and tests}

The equations are advanced in time using a Runge-Kutta $2^{\rm nd}$ order (RK2, or Heun) method, and all spatial derivatives are calculated using fourth-order accurate stencils.
The evolution tracks the propagation of light beams through a discretised Cartesian grid at the specific coordinate time intervals defined by the constant-$t$ surfaces of the simulation output. %
The position of the light beam at each time step will not necessarily be positioned on a spatial grid cell. We therefore need to interpolate the quantities on the right hand side of the evolution equations to the position of the beam at each time step. 
In \mesc, we adopt a $5^{\rm th}$ order spatial, 3--dimensional $B$-spline interpolation 
(see \ref{appx:interp}).

In \ref{appx:tests} we present a variety of tests of the numerical accuracy of the \mesc\ ray tracer. We test the redshift and distance calculations for analytic EdS and linearly-perturbed EdS space-time metrics and confirm the error converges at the expected rate. We also test the calculation on a set of controlled-mode NR simulations and perform a Richardson extrapolation to estimate the error in our calculations for simulation data. Lastly, we test the violation in the null condition $k^\mu k_\mu = 0$ --- which is not explicitly enforced anywhere in \mesc\ and therefore serves as an additional numerical test --- and confirm it reduces with increasing resolution and is thus dominated by numerical error. Based on these tests, we conclude the \mesc\ ray tracer is accurate and as precise as possible for a second-order accurate integration scheme. 

Each step of the ray-tracing calculation is explicitly propagating the light bundle onto the next constant-$t$ hypersurface of the simulation, and therefore each step for all observers is associated with the same coordinate time. However, due to the inhomogeneous nature of the simulated space-time, for each observer each step will correspond to a different redshift. Further, not all lines of sight will have the same redshift at each time step. To isolate the anisotropies in $D_L$ itself, we can interpolate each line of sight to the same $z$ value after the analysis is done. In this work, we use the mean value of $z$ across the sky for each observer at each step and linearly interpolate each line of sight to find $D_L$ at this redshift.

\subsection{Observers} \label{sec:observers}

Quantities calculated via ray tracing are purely observable and are thus directly comparable to distances and redshifts from cosmological surveys. However, they are dependent on our choice of observer and so this choice requires some discussion. 
For the framework presented here, the observers are moving with some general 4--velocity $u^\mu$ and in \mesc\ we have further assumed that this coincides with the fluid 4--velocity.
This is a common choice in both analytic and numerical calculations of observables which 
might manifest either by explicitly taking both observer and emitter to be co-moving with the fluid flow 
\citep[e.g.][]{Sanghai:2017,Ben-Dayan:2012,Rasanen:2009,Bonvin:2006}, 
or via the use of the synchronous co-moving gauge with observers at rest with respect to the coordinates 
\citep[e.g.][]{Grasso:2021,giblin2016b,Li:2007,Kolb:2006,Buchert:2000}. 

Ideally, we want to mimic real cosmological observations as closely as possible with our ray-tracing data. 
Relating the frame which is co-moving with the fluid flow within a simulation to observational analysis is perhaps not straightforward. 
For example, in analysis of observations we might assume that the rest frame of the CMB is the frame in which the Universe is well-described by the FLRW models. 
The redshift of distant sources will be dominated by cosmic expansion, whereas low-redshift sources can have a significant contribution from peculiar velocities.  
For the latter, we might apply corrections to our observed redshifts in an attempt to alleviate this and end up with a purely cosmological redshift \citep[see, e.g.][]{Peterson:2021}. 
Additionally, we as observers are also moving with respect to this global cosmological frame and this must also be taken into account from observed redshifts. Our motion is inferred from the kinematic dipole we observe in the CMB radiation \citep{PlanckDipole:2014a}, %
though recently the purely kinematic origin of this dipole has been called into question \citep[e.g.][]{Siewert:2021,Secrest2021,Colin:2019} \citep[see also][]{Dalang:2022}.

The use of the CMB frame could be motivated by the fact that the early Universe was very close to homogeneous and isotropic, and therefore close to the reference FLRW model which best fits the CMB radiation. 
The simplest way to define the rest frame of the CMB in NR simulations is to use the dipole that each observer measures in their CMB map. However, this would require ray-tracing the full sky to very high redshift which is beyond the computational capabilities of this work. 
Since our simulations start close to an FLRW model by construction, we might choose the simulation hypersurface to coincide with the ``CMB frame'' for all observers. In fact, the perturbations to the metric tensor itself remain small and our hypersurfaces thus remain close to the FLRW expansion chosen on the initial slice. However, large perturbations in the matter develop during the simulation, and so the frame co-moving with the fluid is not necessarily close to this FLRW model on all scales. We thus might consider using the observer's peculiar velocity with respect to the Eulerian grid (i.e., the hypersurfaces) to boost their measured redshifts to this fictitious ``CMB frame'' for analysis. 
We proceed using observers who are co-moving with the fluid flow for the remainder of this work. However, in \ref{appx:CMBframe} we outline the process by which we can correct our simulation data using a single point-wise boost to the hypersurface frame. We also present some of our main results after applying this boost for comparison. 

For all calculations, observers are placed on the spatial hypersurface of the simulation closest to $z=0$. This slice is chosen using the effective scale factor of the simulation, $a_\mathcal{D}$, and the initial redshift of the evolution. Observer positions are chosen (pseudo-)randomly in space across the redshift zero slice.

\section{Cosmography calculation}\label{sec:cosmographic_calc}

We calculate the luminosity distance as predicted by the general cosmography given in Section~\ref{sec:cg} using the coefficients \eref{eq:dLexpand2}, using the effective cosmological parameters as calculated and presented in MH21. We will briefly summarise the method used to calculate these parameters in this section.

In MH21, the authors first calculated the effective Hubble parameter, using its multipole decomposition given in \eref{eq:Eudecomp} %
for each observer. 
The first step is to evaluate the volume expansion $\theta$, the 4--acceleration $a^\mu$, and the shear tensor $\sigma_{\mu\nu}$ at the observer's location (explicit definitions of these quantities are given in Appendix~B of MH21). 
We then use these together with the unique $e^\mu$ for each line of sight (via the same process outlined in Section~\ref{sec:initial_data}) to calculate $\Eu(\boldsymbol{e})$ across each observer's sky. 
The effective deceleration parameter $\mathfrak{Q}$ in \eref{eq:Qdef} and jerk $\mathfrak{J}$ in \eref{eq:Jdef} are then calculated using the first and second derivatives of $\Eu$, respectively, along the direction of the incoming null ray (see Appendix~B of MH21 for the exact expressions of these derivatives). 
Finally, the effective curvature parameter $\mathfrak{R}$ in \eref{eq:Rdef} is calculated using the 4--Ricci tensor, at the observer's location, contracted with $k^\mu$ for each line of sight ($k^\mu$ is also calculated via the method given in Section~\ref{sec:initial_data}).
All calculations were also done using \mesc, though via a separate set of routines to the ray-tracing code.

In this work, we use the calculations of $\Eu,\mathfrak{Q},\mathfrak{R}$, and $\mathfrak{J}$ from MH21 and substitute them into the Taylor series expansion \eref{eq:dL_expand} to find the cosmographic $d_L(z)$ across the sky for all 100 observers (correct to third order in redshift). 
We calculate $d_L(z)$ for a set of redshift values for each observer which coincide with the mean $z$ across the sky as output from the ray tracer (this is the same value which we interpolate $D_L(z)$ to create the ray-traced sky maps).

\section{Results and discussion}\label{sec:results}

In Section~\ref{sec:MH21sims}, we compare the $D_L$ from ray tracing to the cosmographic $d_L$ for the simulations presented in MH21. In Section~\ref{sec:cut200} we assess the accuracy of the cosmographic prediction in a simulation containing more small-scale structure, but which still contains relatively low density contrasts. In Section~\ref{sec:cut100}, we further investigate the anisotropy in a simulation with as much small-scale structure as we can reliably resolve with the simulation software we use. %

In our results presented below, we often present the luminosity distances normalised by the relevant FLRW luminosity distance. For our matter-dominated simulations, this is the EdS model which has cosmological parameters $\Omega_\Lambda=\Omega_k=0$ and $\Omega_m=1$. This is the model around which we apply small perturbations to set our initial data. We have 
\bea
    d_L({\rm EdS}) = \frac{2}{H_0} \left(1 + z - \sqrt{1+z} \right),
\eea
where $H_0$ is the Hubble constant. We take the globally-averaged Hubble parameter in the simulation $\mathcal{H}_{\rm all}\equiv \langle \theta \rangle_{\rm all}/3$ as $H_0$. Here, $\langle\theta\rangle_{\rm all}\equiv \int_{\rm all} \sqrt{\gamma}\,\theta\, d^3X / V_{\rm all}$ is the volume average over the entire $z=0$ slice, with $V_{\rm all}=\int_{\rm all} \sqrt{\gamma}\,d^3 X$ is the volume of the slice and ${\gamma}$ is the determinant of the spatial metric $\gamma_{ij}$. As mentioned in Section~\ref{sec:sims} above and in MH21, we find this Hubble rate to coincide with the FLRW model prediction of $H_0=100 h$ km/s/Mpc (approximately 45 km/s/Mpc for the EdS model) to within one percent for all simulations we use here. 

Throughout our results, we generally distinguish between the ray-traced luminosity distance and the cosmographic prediction for the luminosity distance (correct to third order in redshift) using $D_L$ for the former and $d_L$ for the latter.

\begin{figure*}[ht]
    \includegraphics[width=\textwidth]{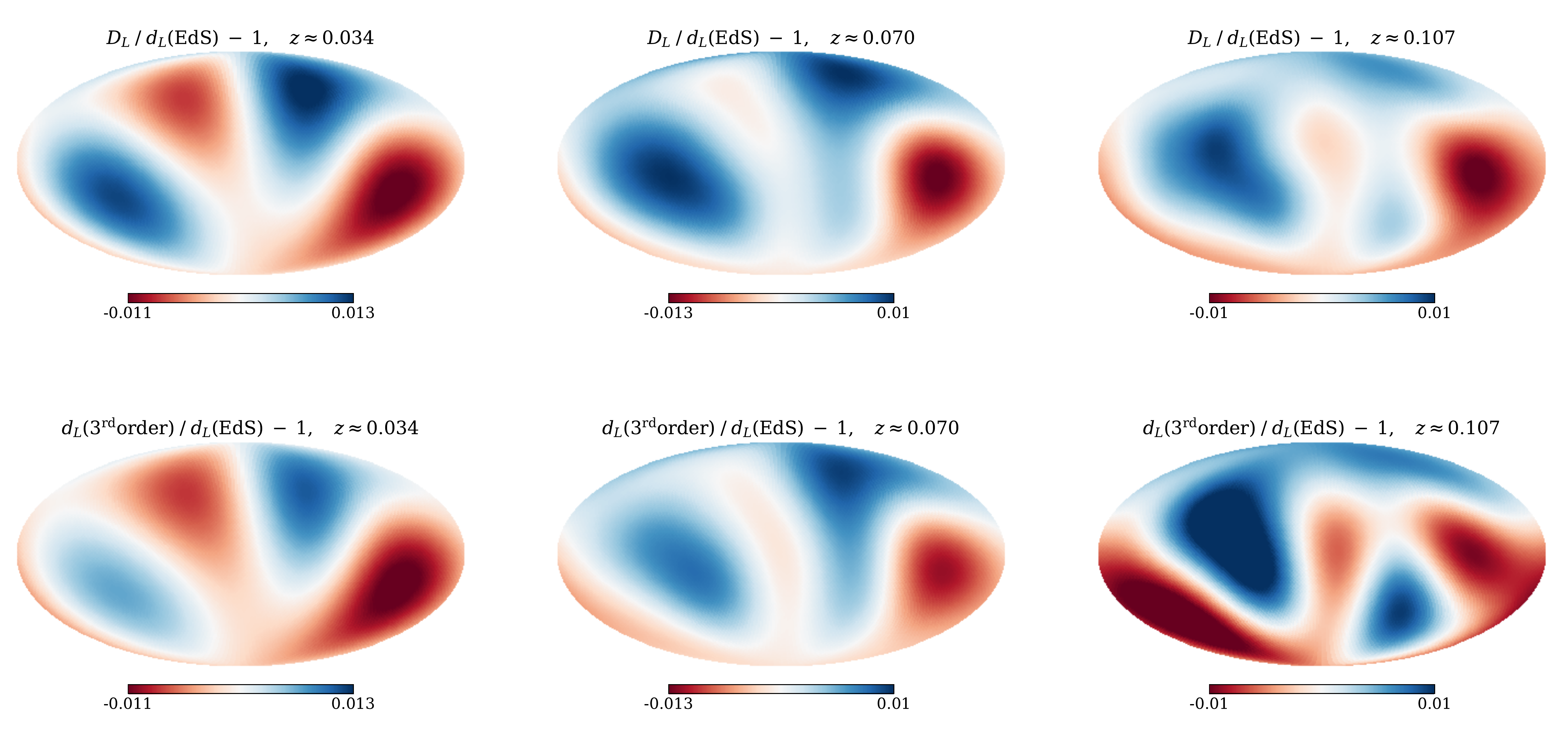}
    \caption{Sky maps of the luminosity distance for one observer in the simulation with $200\,h^{-1}$ Mpc strict smoothing scale. Top row shows the ray-traced luminosity distance $D_L$ at three redshifts $z\approx 0.034, 0.07,$ and $0.107$ (left to right, respectively), normalised by the EdS luminosity distance at that same redshift. Bottom row shows the cosmographic luminosity distance $d_L$ correct to third order in $z$ for the same three redshifts, also normalised by the EdS distance.}
    \label{fig:skymaps}
\end{figure*}

\subsection{Strictly large-scale simulations}\label{sec:MH21sims}

Here we compare the cosmographic predictions based on the simulations presented in MH21 to our new calculations using ray tracing in full GR. These simulations enforce \textit{strict} smoothing scales by explicitly neglecting all structures beneath a chosen scale. There are two simulations with resolution $N=128$ and smoothing scales of 100~$h^{-1}$ Mpc and 200~$h^{-1}$ Mpc. 

\begin{figure*}
    \includegraphics[width=\textwidth]{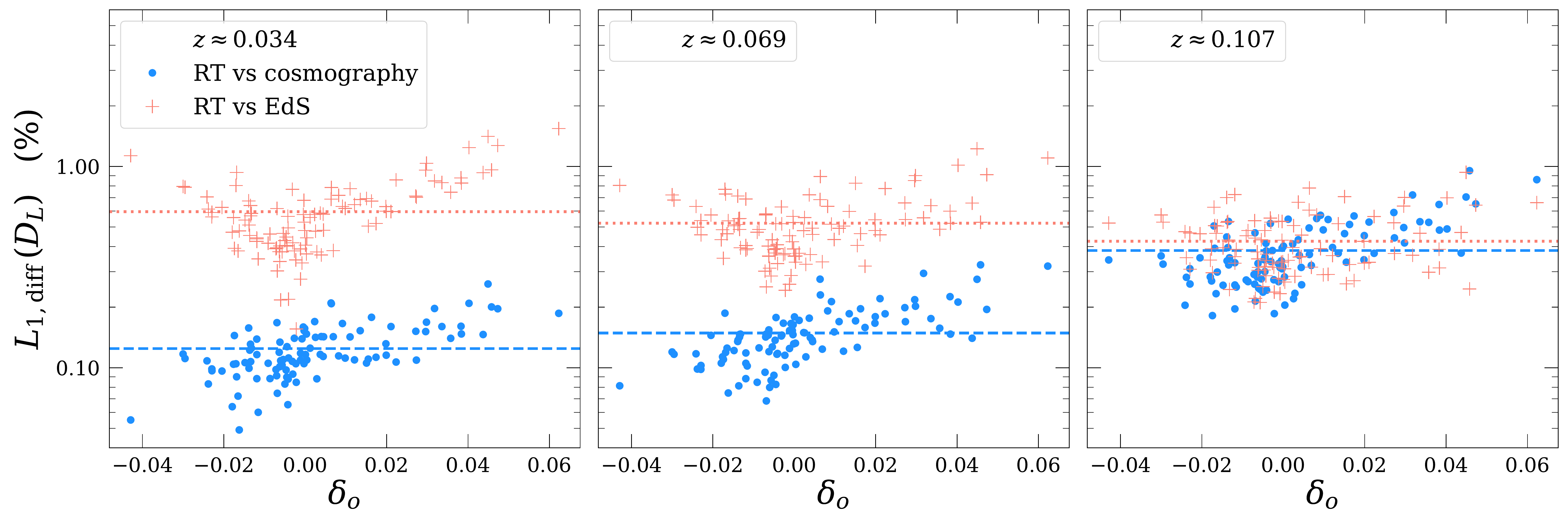}
    \caption{Percentage $L_1$ difference between the ray-traced $D_L$ and the cosmographic $d_L$ (blue circles, dashed line average), and the difference between the ray-traced $D_L$ and the EdS $d_L$ (red crosses, dotted line average). We show the differences for 100 observers at three redshift slices (panels) as a function of the density contrast at the observer, $\delta_o$. All observers are randomly placed in the simulation with a strict $200\, h^{-1}$ Mpc smoothing scale.}
    \label{fig:dLdiffs_200Mpc}
\end{figure*}
Figure~\ref{fig:skymaps} shows a comparison of the ray-traced luminosity distance $D_L$ (top row) to the cosmographic $d_L$ (bottom row) for one observer in the simulation with a strict 200~$h^{-1}$ Mpc smoothing scale. Panels, left to right, show three redshift slices $z\approx 0.034, 0.07,$ and $0.107$, respectively. All distances are normalised by the EdS luminosity distance for that redshift.
By eye, the cosmographic $d_L$ appears to give a very good approximation to the exact $D_L$ at least until $z \approx 0.07$.
To quantify the accuracy of the approximation, we calculate the difference between the two distances along each line of sight for all 100 observers at three redshifts slices.
Specifically, we calculate the ``$L_1$ difference'', defined as
\bea\label{eq:L1diff}
    L_{1,{\rm diff}}(D_L) \equiv \frac{1}{N_{\rm LOS}} \sum_{i=1}^{N_{\rm LOS}} \left|\frac{D^i_L}{d^i_L}-1\right|,
\eea
where $D^i_L$ is the value of $D_L$ for line of sight $i$ (with $N_{\rm LOS}$ total lines of sight), and we take either $d_L=d_L(3^{\rm rd} {\rm order})$ or $d_L=d_L({\rm EdS})$ to compare $D_L$ to both the cosmographic prediction and that of the EdS model, respectively.

Figure~\ref{fig:dLdiffs_200Mpc} shows the percentage $L_1$ difference between the cosmographic $d_L$ and the ray-traced $D_L$ (blue circles) for 100 observers as a function of their local density contrast, $\delta_o\equiv\rho_o/\langle\rho\rangle_{\rm all} - 1$. 
The red crosses show the difference between $D_L$ and the EdS distance for the same observers. Horizontal lines show the mean over all observers for the relevant case. 
Panels, left to right, show three redshift slices of $z\approx 0.034, 0.069$, and 0.107, respectively. We note the mean redshift on the slice differs slightly between observers. 
All observers shown here are placed randomly in the simulation with a strict $200\, h^{-1}$ Mpc smoothing scale. 
The cosmographic distance gives a sub-percent accurate representation of $D_L$ at all three redshifts for all observers. For $z\lesssim 0.07$, the luminosity distance is captured to within $\sim 0.2\%$ for all observers. Due to the very low density contrasts in the simulation, the difference between $D_L$ and the EdS distance is always $\sim 1\%$. However, the cosmographic $d_L$ still performs better than the EdS relation for $z\lesssim 0.1$. 

In Appendix~A of MH21, the authors predicted that the cosmography (as truncated at third order in redshift) should represent the exact luminosity distance to within $1\%$ at $z\sim$0.04--0.06 for a \emph{strict} smoothing scale of $200 h^{-1}$ Mpc (as predicted based on this exact simulation). We find the cosmographic prediction to actually perform \textit{better} than this prediction, matching within 1\% even out to $z\approx0.1$ for all 100 observers.

The results so far have been based on the simulation in MH21 with strictly no structure beneath 200 $h^{-1}$ Mpc. Next, we will consider the second simulation presented in MH21: which instead contains no structure beneath 100 $h^{-1}$ Mpc, and is thus slightly less smooth. 
\begin{figure*}
    \includegraphics[width=\textwidth]{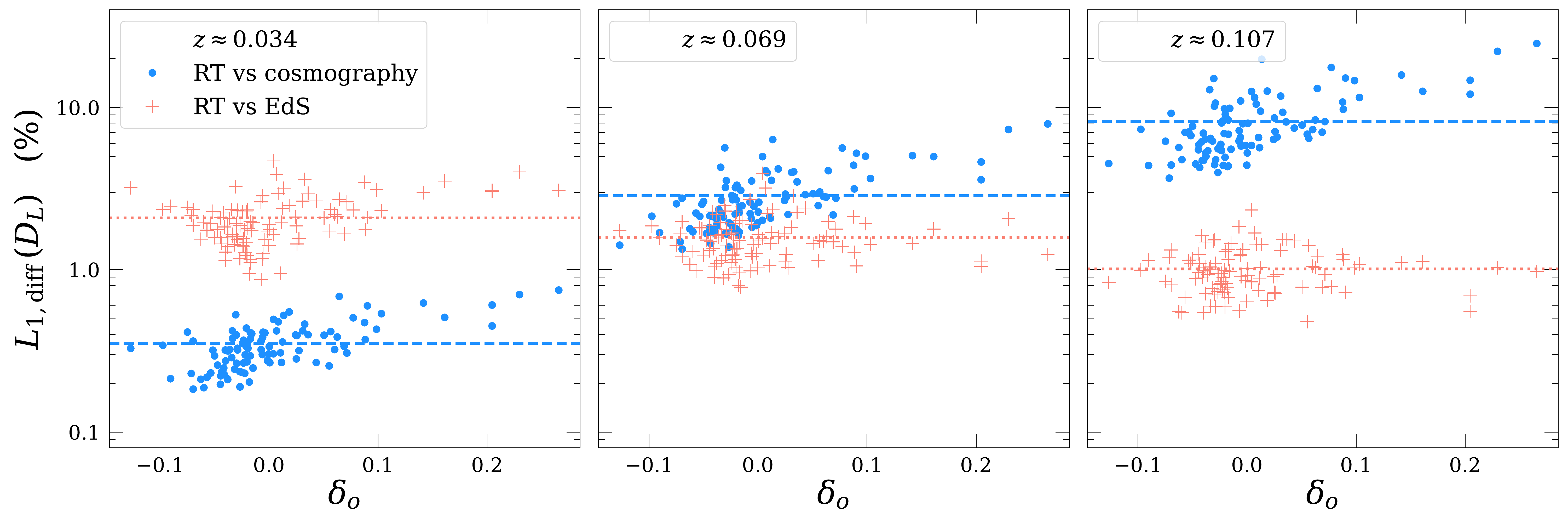}
    \caption{Same as Figure~\ref{fig:dLdiffs_200Mpc} except for a different set of 100 observers in the simulation with the strict $100\, h^{-1}$ Mpc smoothing scale.}
    \label{fig:dLdiffs_100Mpc}
\end{figure*}
Figure~\ref{fig:dLdiffs_100Mpc} shows the percentage $L_1$ difference between $D_L$ and the cosmographic $d_L$ (blue circles, dashed line average) for a set of 100 observers in the simulation with %
a strict smoothing scale of $100 h^{-1}$ Mpc. Again, we show differences as a function of the local density contrast at the observer's location. We note the higher density contrasts on the x-axis with respect to Figure~\ref{fig:dLdiffs_200Mpc}, since this simulation has a smaller smoothing scale. Red crosses again show the difference between $D_L$ and the EdS distance for the same three redshift slices as in Figure~\ref{fig:dLdiffs_200Mpc}. 
Here we see the accuracy of the cosmography is better than 1\% for all observers only to $z\approx0.034$, reaching a few percent at $z\approx0.069$ and eventually $\sim 10$ percent at $z\approx 0.107$. We notice the ray-traced distance $D_L$ is always within a few percent of the EdS distance for all observers out to $z\approx 0.1$. 
The cosmographic distance gives about an order-of-magnitude better prediction than the EdS relation at low redshift, due to the incorporation of anisotropies induced by the smaller-scale structures. 
However, as we approach higher redshifts the $>\mathcal{O}(z^3)$ terms become important and the cosmographic relation (as truncated at third order) over-estimates the anisotropy at $z\gtrsim0.07$ (which can also be seen in the right-most panel in Figure~\ref{fig:skymaps} for the case of the $200 h^{-1}$ Mpc smoothing scale). 

The prediction from MH21 for a 100~$h^{-1}$ Mpc smoothing scale was that the cosmographic $d_L$ should be accurate to within 1\% for redshifts of $z\sim$0.02--0.03. We find our results are consistent with, and even slightly better than, this prediction. 

For both simulations we discussed in this section, the ray-traced distance is within a few percent of the EdS distance for all redshifts we analysed. However, we stress again that \textit{all structure} beneath 100~$h^{-1}$ Mpc or 200~$h^{-1}$ Mpc has been strictly ignored (and thus cannot affect the evolution of larger scales at all), which is not a realistic universe scenario. 
In the next section, we perform a similar comparison for a simulation with more small-scale structure to demonstrate the effects of local inhomogeneities on the ability of the cosmography (and the EdS model) to capture distances.

\begin{figure}
    \centering
    \includegraphics[width=0.5\columnwidth]{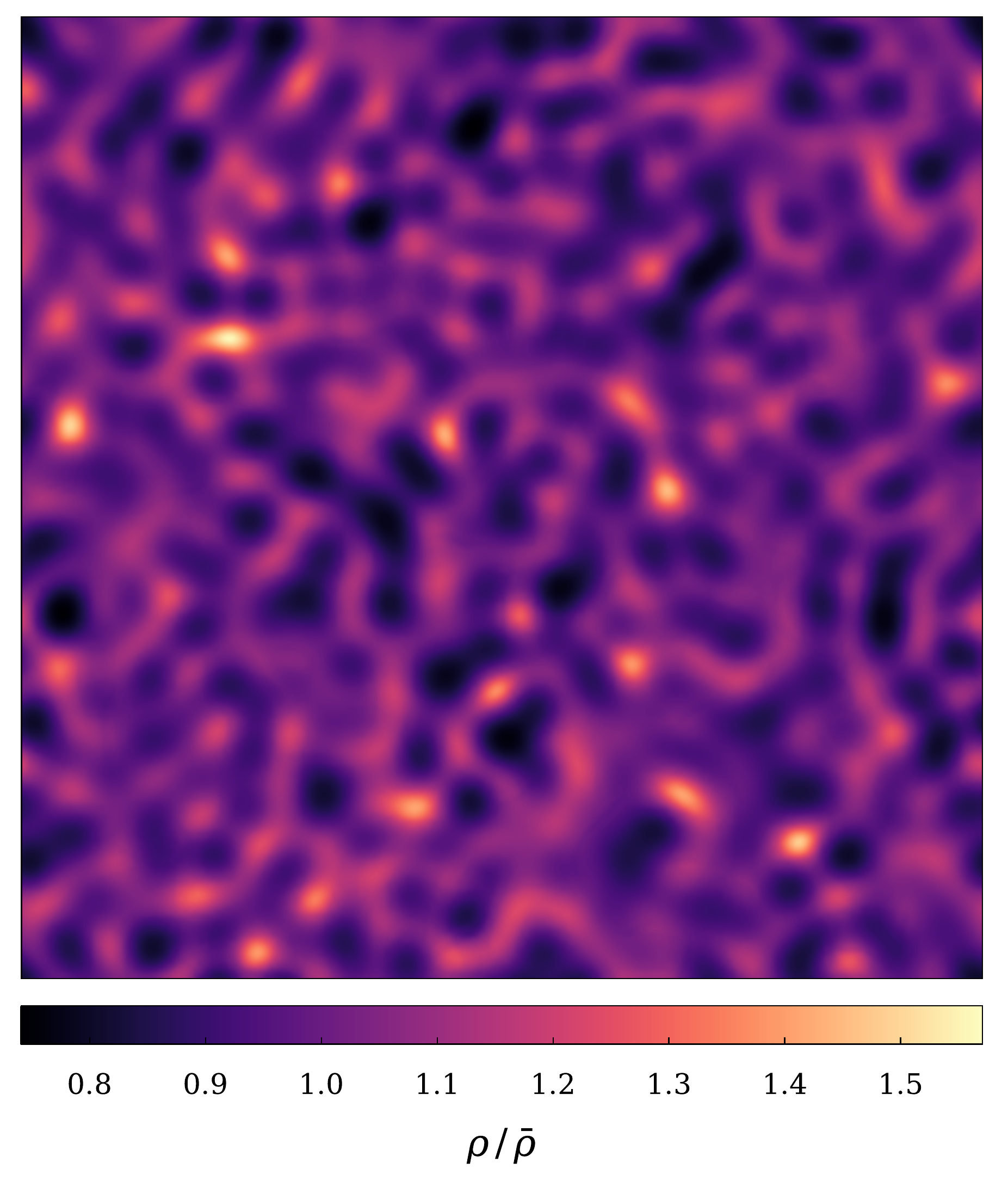}
    \caption{2--dimensional slice of the density field at $z\approx 0$ in a simulation with box size 3$h^{-1}$Gpc and resolution $N=256$. All modes beneath $\sim 200 h^{-1}$Mpc are removed from the initial data. We show the matter density normalised by the average over the whole domain, $\rho/\bar{\rho}$.}
    \label{fig:rho_cut200}
\end{figure}

\subsection{A more realistic model universe}\label{sec:cut200}

In this Section, we will assess the ability of the cosmographic luminosity distance to capture the ray-traced distance in a more realistic model universe. 
We use a simulation with box length $L=3h^{-1}$Gpc and resolution $N=256$, which was also performed using the ET with initial data from \flrwsolver\ generated in a manner identical to that described in Section~\ref{sec:sims}. 
However, for this simulation we have removed some small-scale structure from the initial power spectrum to impose a ``soft'' smoothing scale. This means that initially our simulations contain \textit{no} structure beneath a certain scale. However, contrary to the previous section, we do not strictly prevent structure beneath this scale from forming later in the simulation. 
In this section, we will study a simulation with all power below $\sim 200\,h^{-1}$Mpc initially removed. This implies that the minimum mode in the initial data is sampled by about 17 grid cells. %
In practise we remove this small-scale power by choosing a scale $k_{\rm cut}=2\pi/\lambda_{\rm cut}$ with $\lambda_{\rm cut}=200 h^{-1}$Mpc and set $P(k>k_{\rm cut})=0$, with $P(k\leq k_{\rm cut})$ left as the power spectrum output from CLASS for $k\leq k_{\rm cut}$.
In Section~\ref{sec:cut100} below, we will study an identical simulation (same domain size, resolution, and random realisation) with a slightly lower cut of $\lambda_{\rm cut}=100 h^{-1}$Mpc. 
\begin{figure*}
    \includegraphics[width=\textwidth]{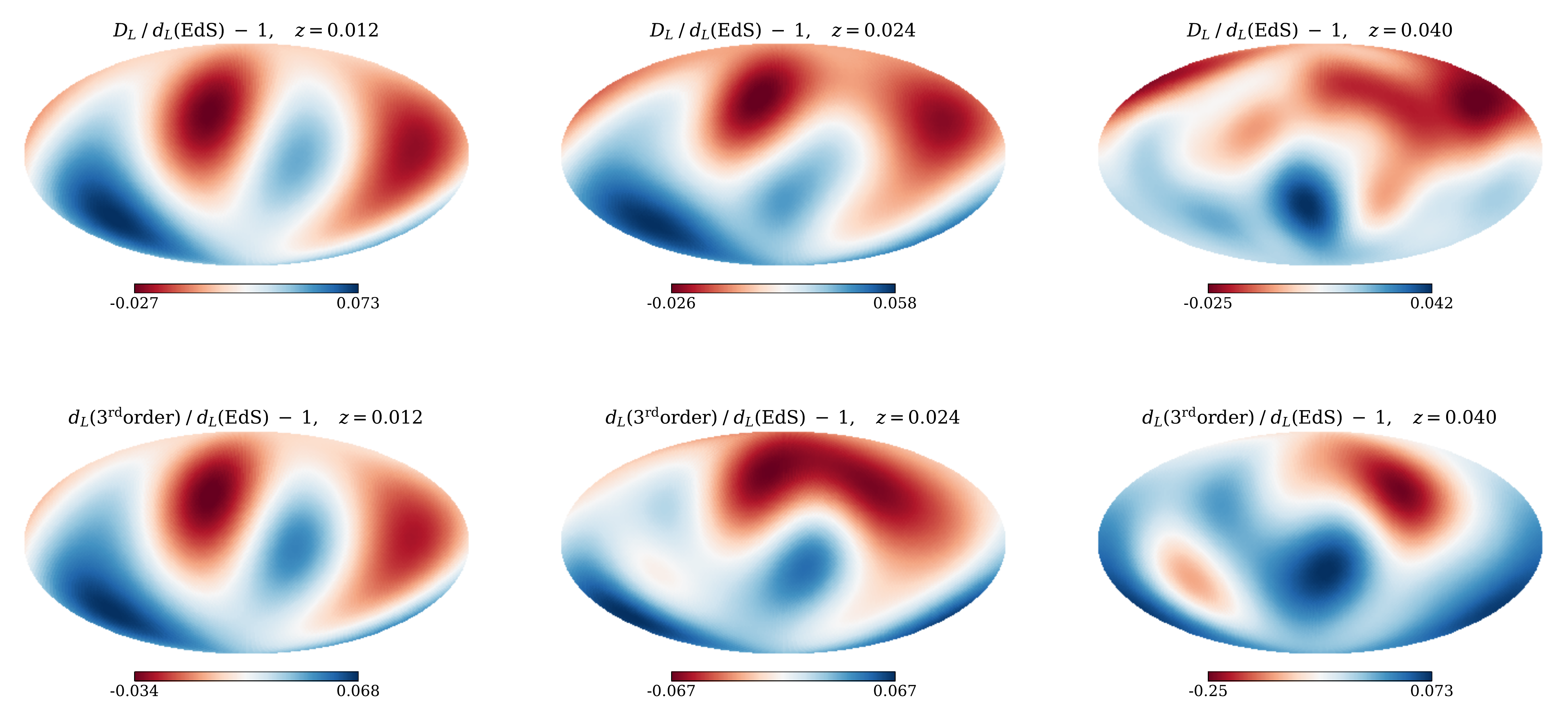}
    \caption{Sky maps of the luminosity distance for one observer in a simulation with all structure beneath $200\,h^{-1}$ Mpc removed from the initial data. Top row shows the ray-traced luminosity distance $D_L$ at three redshifts $z=0.012, 0.024,$ and $0.04$ (left to right, respectively), normalised by the EdS distance. Bottom row shows the cosmographic luminosity distance $d_L$ correct to third order for the same redshifts, also normalised by the EdS distance. Note that the colour bars between the top and bottom rows differ.}
    \label{fig:skymaps_cut200}
\end{figure*}
Figure~\ref{fig:rho_cut200} shows a 2--dimensional slice of the density field (relative to the mean over the whole slice) at redshift $z\approx0$ for the simulation with initial $\lambda_{\rm cut}=200 h^{-1}$Mpc. 
Here we see $\sim 10$--20\% typical under-dense regions and $\sim 30$--50\% typical over-dense regions. Since this simulation is matter-dominated, typical density contrasts on a given scale will in general be larger than in the \lcdm\ model due to the enhanced growth rate of structure. However, as also discussed in MH21, we expect the qualitative anisotropic signatures to be the same as in a \lcdm\ model, though with generally larger amplitudes. Again, while all structures beneath $\sim 200\, h^{-1}$ Mpc were removed from the initial data, the initial perturbations do collapse to scales smaller than this by the end of the simulation. It is thus difficult to refer to a specific ``smoothing scale'' at $z\approx 0$, and so we refer to the simulations by their initial smoothing scales instead. 

We place 25 observers and perform the ray tracing to obtain $D_L$ to $z\approx 0.1$. We use fewer observers for this simulation, relative to those in Section~\ref{sec:MH21sims}, since it is higher resolution and thus the calculations are more expensive. 
We also calculate the cosmographic $d_L$ for the same set of observers using the same process as outlined in Section~\ref{sec:cosmographic_calc} and in MH21. 
For both cases, we calculate distances for $12\times N^2_{\rm side}$ lines of sight in directions of \texttt{HEALPix} indices with $N_{\rm side}=32$.

Figure~\ref{fig:skymaps_cut200} shows a comparison of the ray-traced $D_L$ (top panels) and the cosmographic $d_L$ (bottom panels) at three redshift slices $z=0.012, 0.024$, and 0.04 (left to right rows, respectively) for one observer in the simulation with initial $\lambda_{\rm cut}=200 h^{-1}$ Mpc. We note the lower redshift slices here relative to the previous section, as the presence of more small-scale structure spoils the ability of the cosmography to capture $D_L$ to higher redshift. We also point out that the colour bars between the top and bottom rows of Figure~\ref{fig:skymaps_cut200} are different for $D_L$ and $d_L$. 
Here we notice that the cosmography reproduces the anisotropic signatures for this observer quite well to $z\approx 0.024$, however, the amplitudes of these anisotropies are mostly over estimated. We also notice that at the lowest redshift $z\approx 0.012$ (left-most panels), we have a quadrupole dominating the anisotropy in the luminosity distance, while as we move to higher redshift of $z\approx 0.04$, we see some higher-order multipoles mixed in as well as a discernable dipole signature. 
\begin{figure*}
    \includegraphics[width=\textwidth]{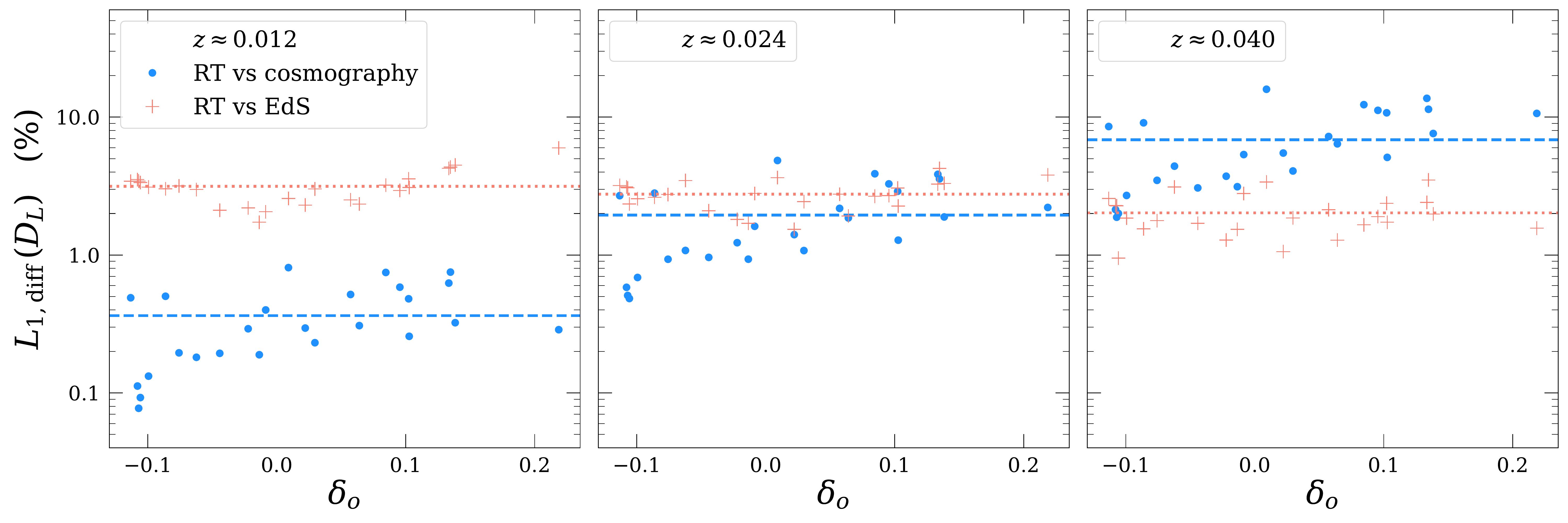}
    \caption{Percentage difference between the ray-traced $D_L$ and the cosmographic $d_L$ (blue circles, dashed line average), and the ray-traced $D_L$ and the EdS $d_L$ (red crosses, dotted line average), at three redshift slices (panels). We show the difference for 25 observers as a function of the density contrast at their location, $\delta_o$. The observers are placed in the simulation with all power below $200\, h^{-1}$Mpc removed from the initial data.}
    \label{fig:dLdiffs_cut200}
\end{figure*}
Figure~\ref{fig:dLdiffs_cut200} shows the percentage $L_1$ difference between the ray-traced $D_L$ and the cosmographic $d_L$ (blue circles with average as a dashed line) and the ray traced $D_L$ and the EdS $d_L$ (red crosses with average as a dotted line) for this case. 
For $z\approx 0.012$, the cosmography approximates the ray-traced $D_L$ to within 1\% for all observers and to within 0.5\% for $\sim 70\%$ of observers. The ray-traced distance is consistent with the EdS distance to within $\sim 2$--3\% for most observers for $z\lesssim 0.04$. 
For redshift $z\approx0.024$ the cosmography provides a $\sim 2\%$ accurate prediction for $D_L$ on average across all observers. At $z\approx0.04$, this has increased to $\sim 7\%$. 

Even though we have included smaller scales than the simulations presented in Section~\ref{sec:MH21sims}, we are still neglecting a lot of small-scale structure in the simulations analysed in this section. 
Due to the fluid approximation adopted in the code we use, we can only reliably simulate physical scales above those of the largest galaxy clusters ($\sim$8--10~$h^{-1}$ Mpc). In the next section, we will analyse a simulation sampling down to these smallest physically-reliable scales. 

\subsection{An even more realistic model universe}\label{sec:cut100}

Here we assess the level of anisotropy in a simulation with even more small-scale structure than that presented in the previous section. Namely, we remove all modes beneath $\sim 100\,h^{-1}$~Mpc from the initial data (such that the minimum mode in the initial data is sampled by about 8 grid cells). The power spectrum and random seed used to generate the initial data (and all parameters associated with setting initial data and evolution of the simulation) are identical to the simulation in the previous section, and \textit{only} the initial $k_{\rm cut}$ differs between the two. 
\begin{figure}
    \centering
    \includegraphics[width=0.5\columnwidth]{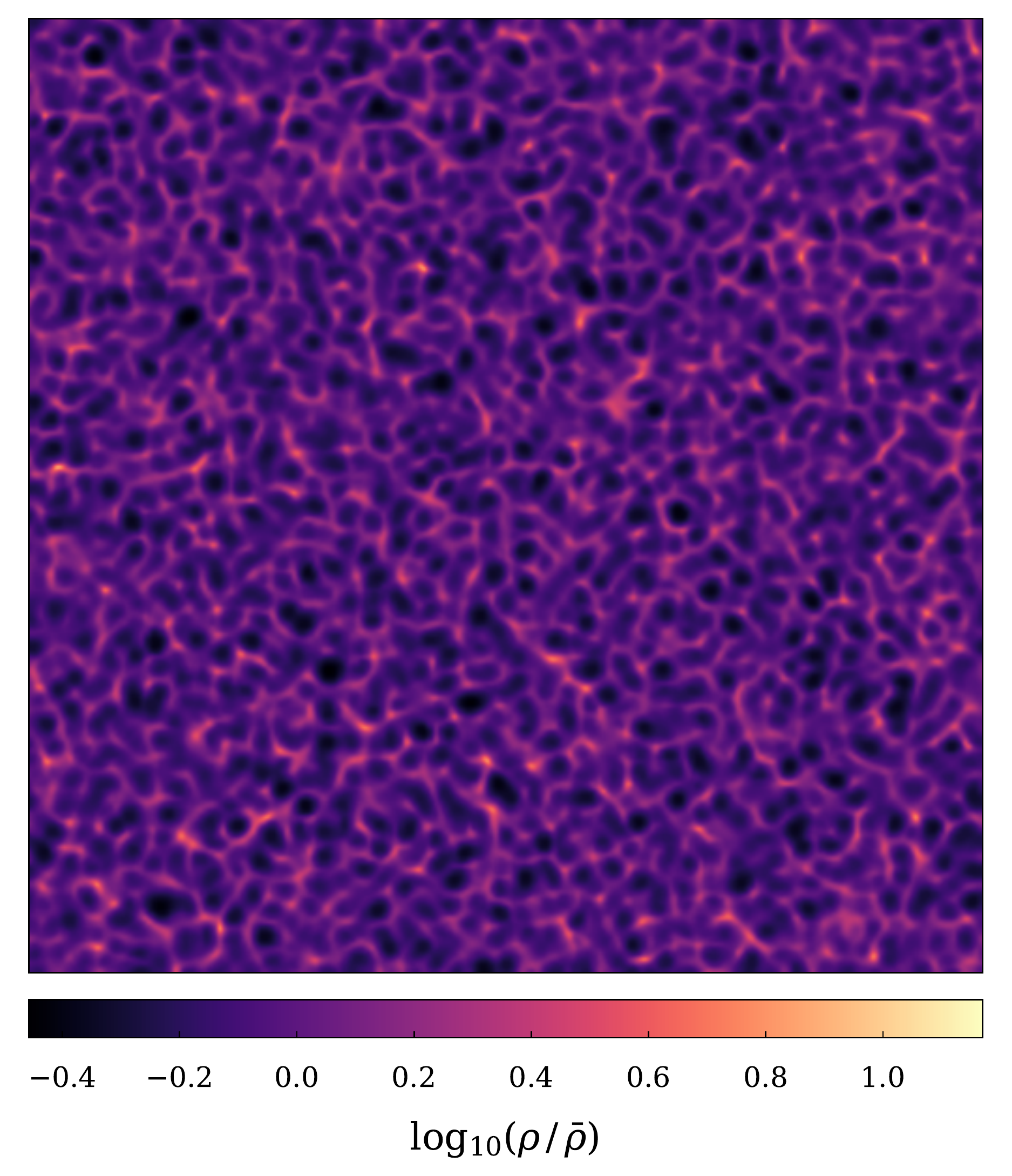}
    \caption{2--dimensional slice of the density field at $z\approx 0$ in a 3~$h^{-1}$~Gpc simulation with all modes beneath $\sim 100 \,h^{-1}$Mpc removed from the initial data. We show the logarithm of the matter density normalised by the average over the whole domain, $\rho/\bar{\rho}$.}
    \label{fig:rho_cut100}
\end{figure}
Figure~\ref{fig:rho_cut100} shows a 2--dimensional slice of the logarithm of the matter density field in the simulation used in this section on the spatial slice at redshift $z\approx 0$. Comparing to Figure~\ref{fig:rho_cut200}, we can see the increase in small-scale structure and larger density contrasts by about an order of magnitude. 

In the previous section, we found that the cosmographic prediction was inaccurate at the $\sim 10\%$ level beyond $z\approx 0.04$. We therefore expect it to provide an even less accurate prediction for this simulation for these low redshifts. 
Further, the validity of the generalised cosmography relies on the condition $\mathfrak{H}\neq 0$ at all points along the geodesic \citep[see][for more details and a discussion of the implications of this assumption]{Heinesen:2020b}. Thus, the cosmography applies only to large-scale cosmological models without collapsing structures. The simulation shown in Figure~\ref{fig:rho_cut100} contains non-linear structure, and thus collapsing regions. We therefore expect $\mathfrak{H}$ might change sign along some lines of sight, and the validity of the cosmographic prediction breaks down. 
Accurately predicting the anisotropic features in this simulation using the general cosmography therefore must involve some smoothing of small-scale structures in order to build a large-scale model from this simulation, which is beyond the scope of this work. In \citet[][in prep.]{MacHein:2023}, we use these same simulations to assess the ability of the general cosmography to predict the dipolar signature in $D_L$ for various smoothing scales. 
Studying even smaller scales than those sampled in Figure~\ref{fig:rho_cut100} --- eventually reaching a sampling which mimics our true observations --- will require going beyond the fluid approximation and using an N-body particle cosmological code such as \textit{gevolution} \citep{Adamek:2013}. 

\begin{figure}
    \centering
    \includegraphics[width=\columnwidth]{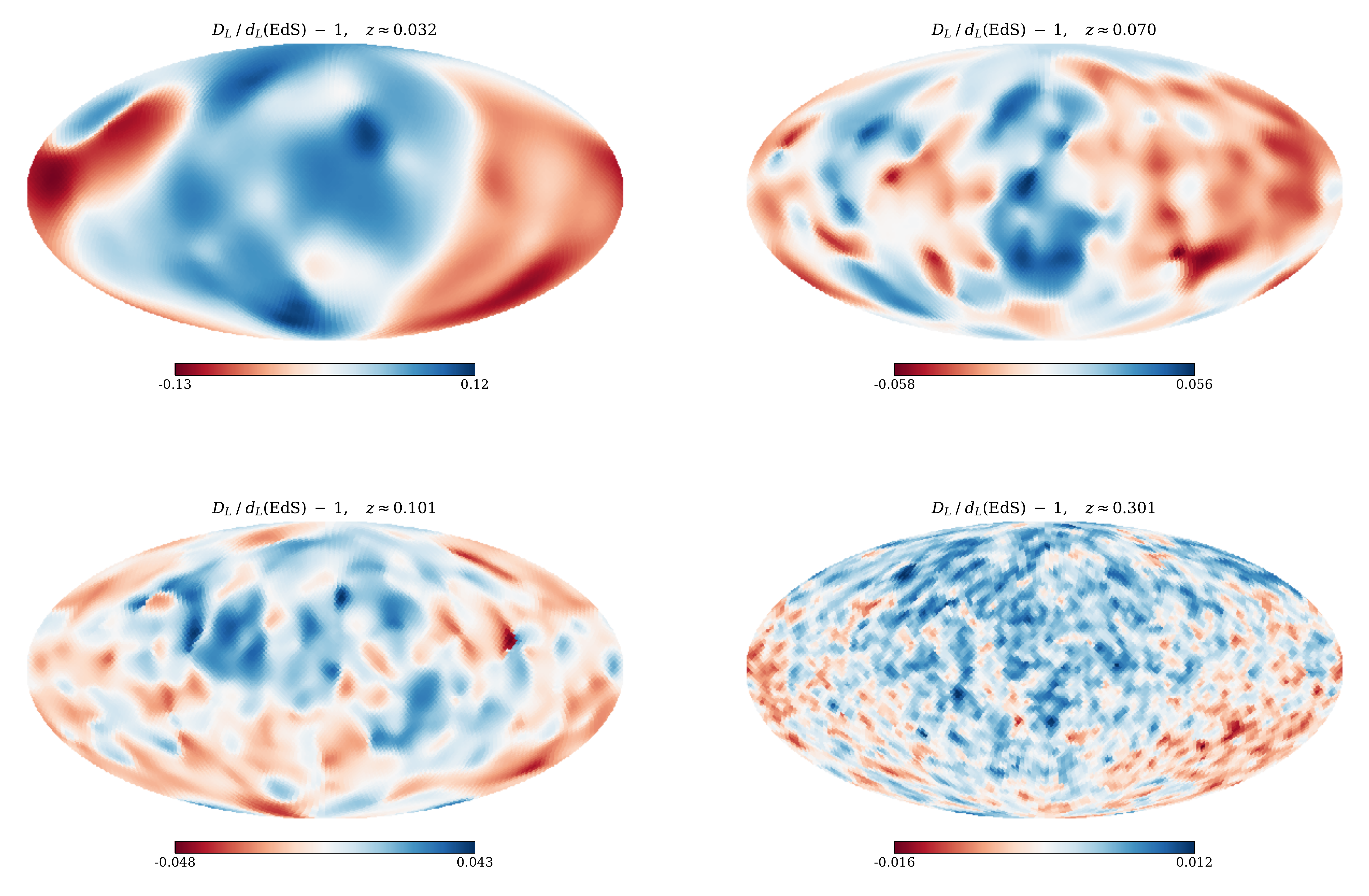}
    \caption{Sky maps of the ray-traced luminosity distance $D_L$, relative to the EdS prediction, on four uniform redshift slices $z\approx 0.031, 0.071, 0.101$, and 0.301 (top left to bottom right, respectively) for one observer in the simulation shown in Figure~\ref{fig:rho_cut100}.}
    \label{fig:dLmaps_cut100}
\end{figure}
Instead of directly comparing to the cosmographic prediction, we will present and assess 
how the ray-traced luminosity distance varies across the sky in this simulations. 
We defer an involved comparison with cosmographic predictions for specific multipole signatures to future work \citep[][in prep.]{MacHein:2023}. 

We place 25 observers at the same locations as for the simulation in Section~\ref{sec:cut200}, and ray trace for the same lines of sight. 
Figure~\ref{fig:dLmaps_cut100} shows all-sky maps of the ray-traced luminosity distance $D_L$, relative to EdS, for one observer in the simulation with structure beneath $\sim 100\,h^{-1}$ Mpc cut out of the initial data (as shown in Figure~\ref{fig:rho_cut100}). We show four uniform redshift slices at $z\approx 0.031, 0.071, 0.101$, and 0.301 in the top-left to bottom-right panels, respectively.%
The maps here are just for one observer's sky. The local environment of all observers will yield different anisotropic signatures, since the effects themselves arise from local kinematics such as differential expansion and shear. %
\begin{figure*}
    \centering
    \includegraphics[width=\textwidth]{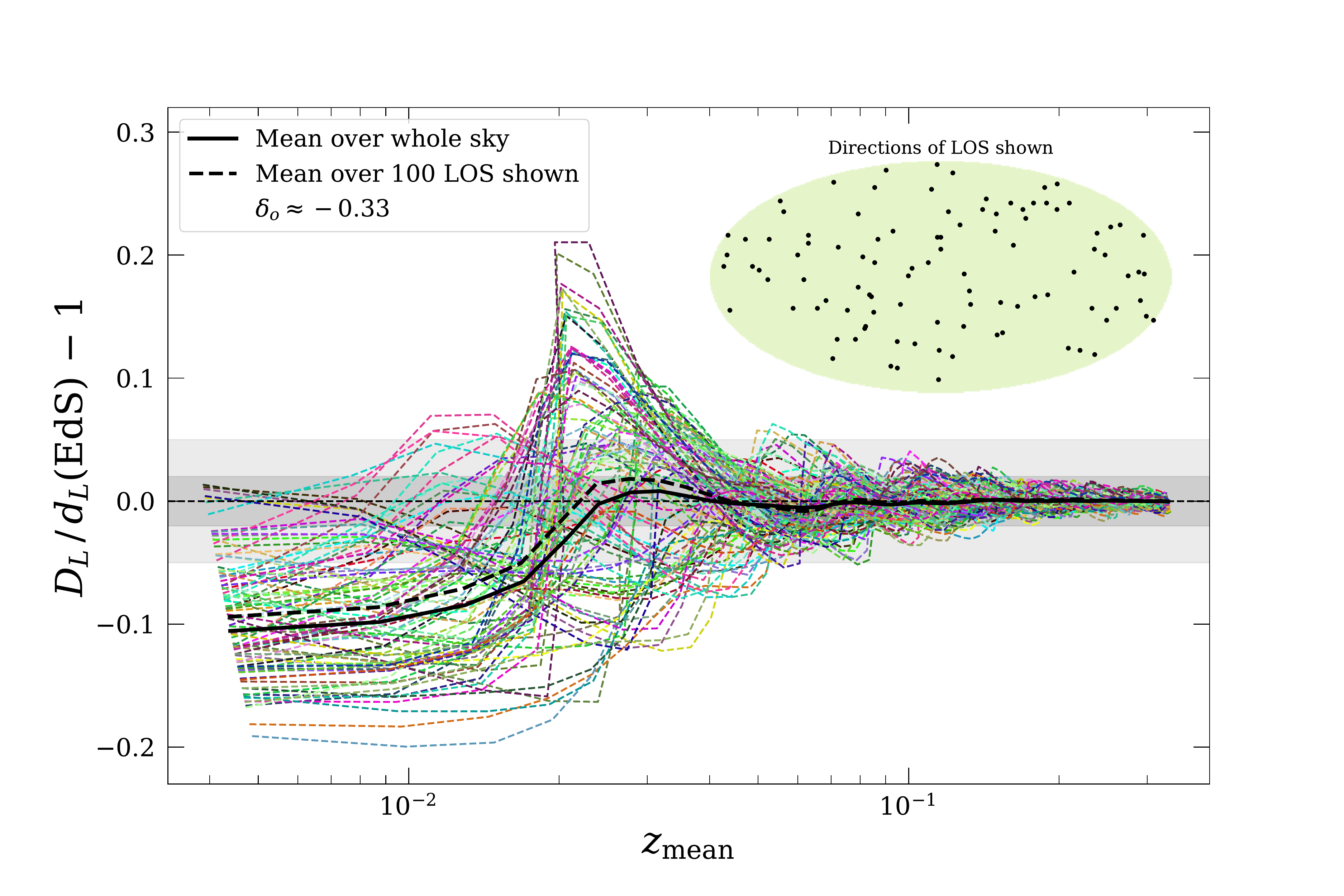}
    \caption{Deviation in ray-traced luminosity distance from the EdS prediction as a function of redshift. We show 100 lines of sight for the observer shown in Figure~\ref{fig:dLmaps_cut100} (coloured dashed curves), as well as the mean over the whole sky (thick solid black curve) and the mean over the 100 lines of sight shown (thick dashed black curve). The inset shows the directions of the lines of sight shown on the observer's sky, and the dark and light grey shaded regions show $\pm 2$\% and $\pm 5$\% deviations, respectively.}
    \label{fig:hubble_cut100}
\end{figure*}
Figure~\ref{fig:hubble_cut100} shows the ray-traced luminosity distance as a function of redshift for a set of lines of sight for the same observer as shown in Figure~\ref{fig:dLmaps_cut100}. We show the deviation of $D_L$ from the EdS distance for 100 randomly chosen lines of sight (coloured dashed curves), the mean over the whole sky (thick solid black curve), and the mean over the 100 lines of sight shown here (thick dashed black curve). The inset shows a Mollweide projection of the directions of the 100 lines of sight shown here on the observer's sky. The dark and light grey shaded regions show a $\pm 2$\% and $\pm 5$\% variation from EdS, respectively. This observer is located in a region with density contrast of $\delta_o \approx -0.33$, i.e., a $\sim 30\%$ under-density relative to the average over the spatial slice at $z\approx 0$. This results in a higher expansion in the observer's vicinity relative to the ``global'' expansion, which presents as smaller luminosity distances locally for $z\lesssim 0.02$ in Figure~\ref{fig:hubble_cut100}. 

\begin{figure}
    \centering
    \includegraphics[width=0.9\columnwidth]{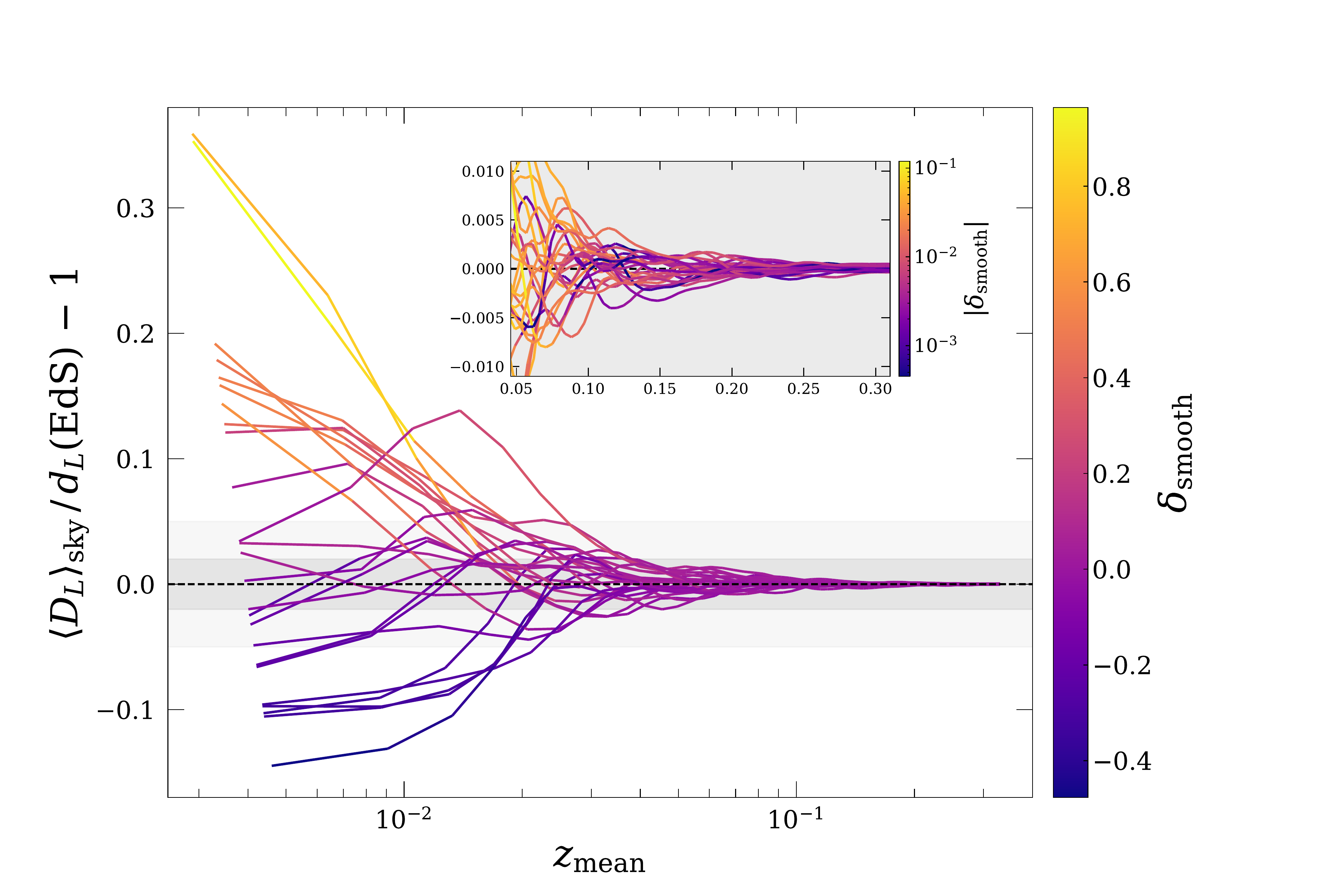}
    \caption{Mean luminosity distance relative to the EdS prediction as a function of redshift. We show the all-sky average $D_L$ for 25 observers in the simulation shown in Figure~\ref{fig:rho_cut100}. Each curve is coloured according to that observer's local density contrast as smoothed to the scale corresponding to $z_{\rm mean}$, namely, $\delta_{\rm smooth}$. 
    The dark and light shaded regions show 2\% and 5\% deviations from EdS, respectively. The inset shows a zoom-in of the redshift range $z_{\rm mean}=0.05$--0.3, with $|\delta_{\rm smooth}|$ on a log scale.}
    \label{fig:hubble_mean_cut100}
\end{figure}
Figure~\ref{fig:hubble_mean_cut100} shows the all-sky average luminosity distance, $\langle D_L\rangle_{\rm sky}=1/N_{\rm LOS} \sum_i D^i_L$, relative to the EdS distance, as a function of redshift (mean $z$ across the slice) for all 25 observers. This corresponds to the equivalent of the thick solid black curve in Figure~\ref{fig:hubble_cut100} for all observers. The dark and light grey shaded regions again show $\pm 2$\% and $\pm 5$\% variation from EdS, respectively. 
Each line segment is coloured according to the local density environment nearby that observer for that redshift scale, $\delta_{\rm smooth}$. Specifically, for each value of $z_{\rm mean}$ we use a Gaussian kernel to 
smooth the density field over the corresponding spatial scale, namely, %
\begin{equation}\label{eq:delta_smooth}
    \delta_{\rm smooth} \equiv \int \delta(x^i) \times W(x^i,\sigma) \, dV,
\end{equation}
where $\delta(x^i)$ is the density contrast defined on the $z\approx 0$ simulation hypersurface, and $W(x^i,\sigma)$ is a normalised Gaussian kernel centred on the observer's position with width $\sigma = d_{C,{\rm EdS}}(z_{\rm mean}) / 3$. Here, $d_{C,{\rm EdS}}(z_{\rm mean})$ is the co-moving distance in the EdS model at redshift $z_{\rm mean}$ (the redshift at which $\langle D_L \rangle_{\rm sky}$ is calculated). 
We emphasize again that we use the density field on the $z\approx 0$ simulation hypersurface, so this will give an upper limit of the density as smoothed on the observer's past light cone. Further, we approximate the volume element in the average \eref{eq:delta_smooth} as $dV\approx d^3X$ instead of the more general $dV = \sqrt{\gamma}\,  d^3X$ where $\gamma$ is the determinant of the spatial metric of the hypersurface. This approximation is sufficient to gain a rough insight to the local density contrast near the observer. We emphasize that this smoothing method is used \textit{only} for the purposes of visualisation in Figure~\ref{fig:hubble_mean_cut100}.

The inset in Figure~\ref{fig:hubble_mean_cut100} shows a zoom in of the redshift range 0.05--0.3 such that the absolute value of the smoothed density contrast (shown on a log scale) can be seen.
We expect this all-sky average to closely coincide with the monopole (isotropic) contribution to the luminosity distance, since our distribution of ``objects'' fairly samples all points on the sky (to within our \texttt{HEALPix} resolution). This kind of variance is sometimes referred to as ``cosmic variance'', and represents the variation we might expect in distances as we place observers in regions with different local density contrast. 
In Figure~\ref{fig:hubble_mean_cut100}, we do notice some correlation between the smoothed local density contrast and the deviation in $D_L$ at $z\lesssim 0.01$. This correlation is no longer obviously present beyond $z\approx 0.02$ and there is no discernible correlation with variance in $D_L$ and $\delta_{\rm smooth}$ by $z\approx 0.05$.

Aside from cosmic variance, the anisotropic variance of $D_L$ across the observer's sky may bias their cosmological inference if they do not fairly sample all directions. 
To estimate the level of anisotropy across the skies of our 25 observers, we calculate the maximal sky-deviation of the luminosity distance (in a manner similar to MH21),
\bea\label{eq:maxskyvar}
    \Delta D_L \equiv \frac{D_L^{\rm max}-D_L^{\rm min}}{2},
\eea
where $D_L^{\rm max}$ and $D_L^{\rm min}$ are the maximum and minimum values of $D_L$ on an observer's sky, respectively. 
\begin{figure}
    \centering
    \includegraphics[width=0.7\linewidth]{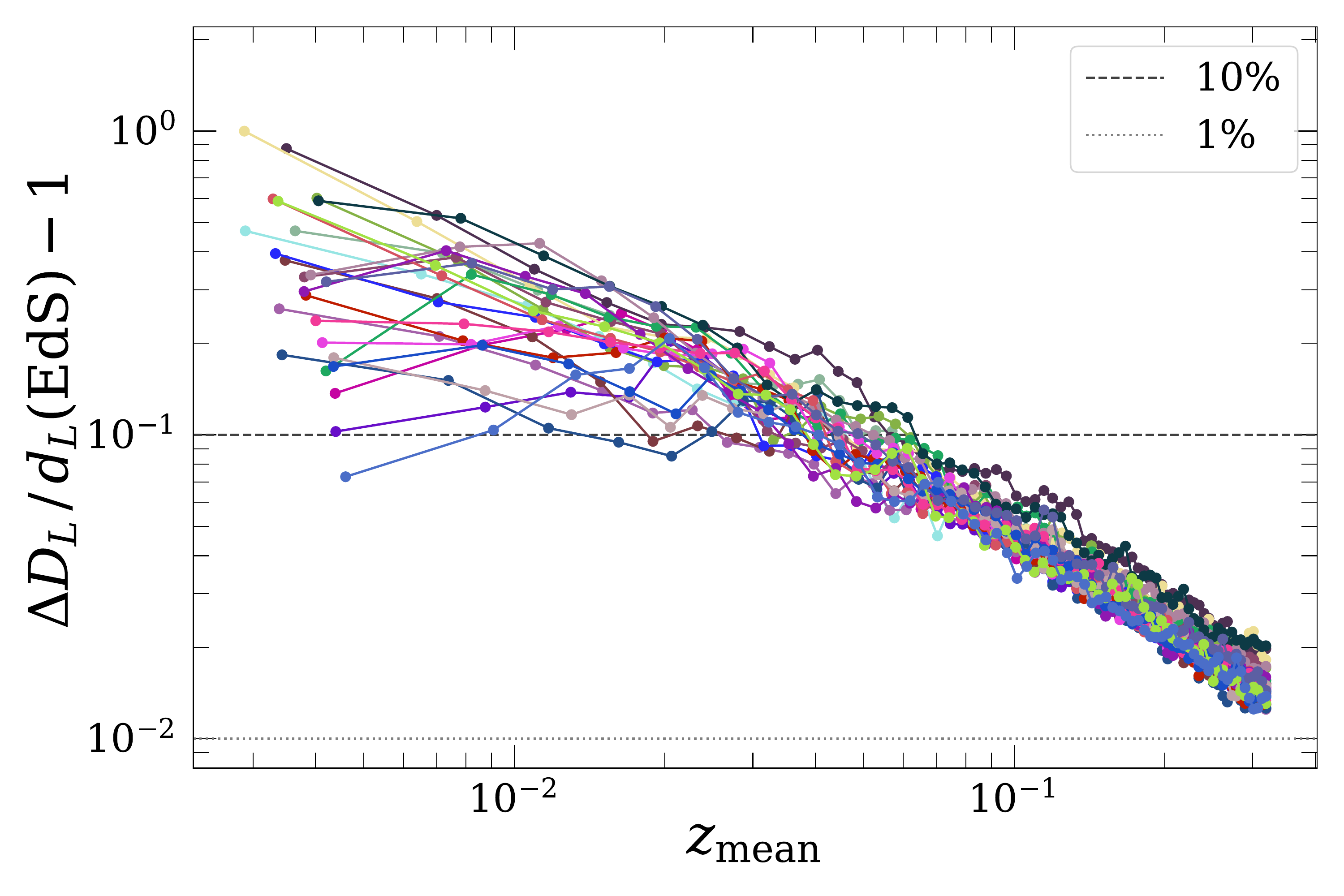}
    \caption{Maximal sky variance as a function of redshift for 25 observers in the simulation shown in Figure~\ref{fig:rho_cut100}. We show $\Delta D_L$ relative to the EdS distance at that redshift, $d_L({\rm EdS})$. 
    Dashed and dotted horizontal lines show a 10\% and 1\% deviation, respectively.}
    \label{fig:maxskyvar_cut100}
\end{figure}
Figure~\ref{fig:maxskyvar_cut100} shows the maximal sky variance \eref{eq:maxskyvar} for all 25 observers in this simulation. We show $\Delta D_L$ relative to the EdS distance as a function of the mean redshift of the slice. $\Delta D_L$ here is computed using the uniform redshift slice of $D_L$, namely, all lines of sight are linearly interpolated to $z_{\rm mean}$ (though $z_{\rm mean}$ varies among observers). 
All observers maintain a consistent maximal sky deviation of $>8\%$ until $z\approx 0.03$, which then remains $>5\%$ until $z\approx 0.06$, and finally the variance is still above $1\%$ out to the maximum redshift studied here, $z\approx 0.3$. At $z\approx 0.3$ the range of variance we see is $1.2$--2\% across all 25 observers. 
In \ref{appx:CMBframe} we compare this result to the same calculation including a boost of the observers to the ``CMB frame'' of the simulation (see also Section~\ref{sec:observers}).

The level of variance we see in Figure~\ref{fig:maxskyvar_cut100} could bias cosmological inferences using data which does not fairly sample the whole sky. 
The precise size of any effect on the resulting cosmological parameters will be dependent on the particular survey geometry or sky coverage in question as well as the redshift ranges considered (see MH21 for an approximate study of the potential impact of this effect). We delay an in-depth investigation of the impact of these effects --- for example on low-redshift supernova data --- to future work.

\section{Caveats}\label{sec:caveats}

The main caveats to our work are as follows:
\begin{enumerate}
    \item Our simulations are matter dominated and contain no cosmological constant. The simulations maintain a matter density $\Omega_m\approx 1$ throughout the simulation on large scales \citep[i.e. there is no significant backreaction on the scale of the simulation domain, see][]{Macpherson:2019}, and the structures that form will therefore grow at a faster rate. Ultimately, this results in overall higher density contrasts in the simulation with respect to an equivalent simulation containing a cosmological constant. In a model universe with typically lower density contrasts, i.e. in \lcdm, we would expect the \textit{amplitude} of anisotropic signatures to be reduced, however, the \textit{qualitative} signatures will be robust to these kinds of changes (see also MH21 for a discussion on this). Any universe with cosmological structure will naturally contain these effects, however, the amplitude is dependent on the particular cosmology in question. 

    \item All simulations we use here sample \textit{only} large-scale structures, due to the limitation of the continuous fluid approximation. We expect the anisotropic signatures to be different with the inclusion of additional small-scale structure. In \citet[][in prep.]{MacHein:2023}, we investigate the impact of small-scale structure on the large-scale angular features in the $D_L(z)$ relation by comparing the two simulations presented in Section~\ref{sec:cut200} and \ref{sec:cut100}. 

    \item We have studied a limited number of observers in this work due to the computational expense of performing all-sky ray tracing. While our observers span a wide range of local density contrasts $\delta_o \approx -0.4$--1 (see Figure~\ref{fig:hubble_mean_cut100}), we have not nearly sampled all possible local environments, nor are these environments especially like ours as observers on Earth. Especially of relevance would be the study of \textit{targeted} observers with similar local environments to what we observe in galaxy surveys, rather than a sample of randomly-placed observers. Our choice for using randomly-placed observers here was motivated on gaining a broad understanding of the variance in these effects throughout the simulation. This method certainly has benefits over studying only one observer, however, the large variance we find among observers should further motivate studies of these effects in constrained simulations of the local Universe \citep{Hess:2013,Hess:2016}. 

\end{enumerate}

\section{Conclusions}\label{sec:conclude}

In this work we have presented an analysis of anisotropy in the luminosity distance at low redshift with ray tracing. Our framework is based on fully nonlinear general relativity, with no assumption of any background cosmology or perturbation theory at any point in the calculation (beyond setting initial data for the simulations themselves).

We assessed the ability of the general cosmography \citep[as truncated at third order in redshift][]{Heinesen:2020b} to capture anisotropies in the luminosity distance as calculated using our new ray-tracing tool. We found the cosmographic distance provides a $<1\%$ accurate prediction out to redshift $z\approx 0.1$ for \textit{strict} smoothing scales of $200 h^{-1}$ Mpc, and to redshift $z\approx 0.034$ for \textit{strict} smoothing scales of $100 h^{-1}$ Mpc. We note that in the latter case, the smoothing scale and the maximum redshift closely coincide with one another in terms of distance scale in an EdS model approximation. In the former case, the maximum redshift represents slightly larger scales than the smoothing scale in the EdS model. We therefore expect the cosmographic prediction to be best at redshift scales closely coinciding with the smoothing scale itself, which (as also stated in Appendix~A of MH21) may pose an issue for using the general cosmography, as truncated at third order, for wide redshift intervals. 

For simulations with small-scale structure beneath $\sim 200\,h^{-1}$ Mpc removed from the initial data --- rather than being strictly excluded for the whole simulation --- we find that the cosmography is accurate to within $\sim 1\%$ to a redshift of $z\approx0.012$. While we cannot explicitly state the size of the ``smoothing scale'' of this simulation at redshift zero, it is clear that the presence of additional small-scale structure spoils the ability of the cosmography (as expanded to third order) to capture $D_L$ to higher redshift. 
This breakdown might be expected \citep{Macpherson:2021}, and therefore some smoothing of the small-scale structure will likely be required to yield an accurate large-scale prediction of anisotropies from the general cosmography as truncated at third order \citep[][in prep.]{MacHein:2023}. 
All simulations studied here are still very far from the real Universe. The minimum scale we sample is $\sim$8--10~$h^{-1}$~Mpc, which is the scale of large galaxy clusters and many structures exist below this scale. 

Therefore, we conclude that achieving a $\sim1\%$ accurate prediction of the luminosity distance (including \textit{all} of its anisotropic contributions) will be difficult for wide redshift ranges. 
As mentioned in Section~\ref{subsec:cg_apps}, such issues can be alleviated by allowing for a decay of an anisotropic signature included in a cosmological fit \citep[such as in, e.g.,][]{Dhawan:2022,Rahman:2022,Rubin:2020,Colin:2019b}. This would ensure that the anisotropies themselves are being constrained by data within the narrow redshift interval in which we expect the series to converge, while higher-redshift objects are contributing to the background cosmological fit. While we cannot smooth the actual structure in the Universe, as we can in simulations, some smoothing of the cosmological data itself may be possible to allow anisotropic constraints \citep[][in prep.]{gevcg:2023}.

We further assessed the level of anisotropy in the luminosity distance in a simulation including structures down to $\sim 100\,h^{-1}$ Mpc in the initial data. %
We found all 25 observers saw a $>8\%$ variance in $D_L$ across their sky to $z\approx 0.03$ %
and $>1\%$ variance at $z\approx 0.3$. 
This kind of variance could be especially important for analysis assuming an isotropic distance law that does not fairly sample the full sky. However, further investigation is necessary to estimate the size of the effect for specific survey geometries and redshift ranges. 
We also found the ``cosmic variance'' of the ray-traced luminosity distance for $z\lesssim 0.01$ is $\pm 10\%$ (relative to the EdS distance) with some correlation with the local density at the observer. Above $z\approx 0.03$, however, all 25 observers have a $<2$\% difference in their all-sky average (isotropic contribution) relative to the EdS model. 
This ``cosmic variance'' is roughly consistent with previous studies on the isotropic variance in the local Hubble parameter using Newtonian N-body simulations \citep{Wu:2017,Odderskov:2016,Wotjak:2014} and NR simulations \citep{Macpherson:2018}.

\section*{Acknowledgements}

HJM would like to thank the anonymous referee whose careful reading and comments improved the quality of this work. 
HJM would also like to thank Julian Adamek, Asta Heinesen, Pierre Fleury, Jim Mertens, Sofie Koksbang, and Eoin \'O Colg\'ain for helpful discussions and comments regarding this manuscript. 
HJM appreciates support received by the Herchel Smith postdoctoral fellowship fund. 
HJM would also like to especially thank Pierre Fleury, Julian Adamek, and Jim Mertens for very useful discussions on light propagation over the years that inspired and formed the beginning of this work. 
This work used the DiRAC@Durham facility managed by the Institute for Computational Cosmology on behalf of the STFC DiRAC HPC Facility (www.dirac.ac.uk). The equipment was funded by BEIS capital funding via STFC capital grants ST/P002293/1, ST/R002371/1 and ST/S002502/1, Durham University and STFC operations grant ST/R000832/1. DiRAC is part of the National e-Infrastructure.
We used GNU parallel in processing data for this work \citep{tange_ole_2018_1146014}.

\appendix

\section{Propagation of the screen vectors}\label{appx:screen_prop}

In Section~\ref{subsec:screen} we introduced the notion of screen space, which is defined by the basis vectors $s^\mu_A$. Not all of the components of $s^\mu_A$ are constrained by \eref{eq:smuconstraints}, with the orientation of the basis remaining arbitrary. A family of observers lying along the geodesic co-moving with $u^\mu$ therefore may define their own basis vectors which satisfy the constraints but are not necessarily aligned with one another. In this work, we will be considering a single basis per observer per geodesic, and we therefore must maintain the same orientation of the basis along the geodesic (although the \textit{initial} orientation remains arbitrary, and our final results are independent of this choice, which we show in \ref{appx:screen_init_test} below). The screen vectors are (by definition) orthogonal to the observer's 4--velocity $u^\mu$, which is not necessarily parallel transported along the geodesic. 
The basis vectors therefore obey a \textit{partial} parallel transport equation \citep[see Section 1.3.3 of][for a proof of this]{Fleury:2015a}, which for convenience can be written as \citep{Sanghai2017}
\bea\label{eq:smu_pptrans}
	\frac{D s^\mu_A}{d\lam} = \frac{k^\mu}{E}  s^\alp_A \frac{D u_\alp}{d\lam},
\eea
where the covariant total derivative $D/d\lam \equiv k^\mu \nabla_\mu$. 
The beams morphology itself is independent of any observer, and so it can be shown that the evolution of the screen has no physical effect on the beam \citep{Sachs:1961}. It can therefore be safely assumed that the 4--velocity $u^\mu$ \textit{is} parallel transported along the geodesic and thus so are the screen vectors. After making this assumption, the screen basis is a ``Sachs basis'' and this can be useful in situations where a clear matter rest-frame is not well-defined \citep[e.g.][]{Lepori:2020}.
In this work, we continue working with a general $u^\mu$ and thus partially parallel transport the screen vectors along the geodesic. 

Similarly to our re-casting of the geodesic equation in Section~\ref{sec:geodesics}, we can re-parametrise the geodesic in terms of coordinate time, $t$. After expanding the covariant total derivative in \eref{eq:smu_pptrans}, 
we can thus write
\bea\label{eq:dsmudt}
	\frac{d s^\mu_A}{dt} = \frac{k^\mu}{k^0 E} s^\alp_A k^\delta\nabla_\delta u_\alp - \frac{1}{k^0}\Gam^\mu_{\beta\gam}k^\beta s^\gam_A.
\eea
From \eref{eq:dsmudt}, we find the evolution equations for the time and space components of the screen vectors to be
\bea\label{eq:ds0dt_step1}
	\fl\frac{ds^0_A}{dt} = \frac{s^0_A}{E} \bigg\{ k^0 \pd_t u_0 + k^i \pd_i u_0 + \frac{1}{\alp}\pd_t(\alp) k^i u_i - E v^i \pd_i\alp + \alp K_{ki} u^k k^i \bigg\} + \frac{s^i_A}{E} \bigg\{ k^0 \pd_t u_i  \nonumber\\ + k^j \pd_j u_i + \frac{1}{\alp} \pd_i(\alp) k^j u_j + E K_{ki} v^k - k^j \Gam^k_{ij} u_k \bigg\},
\eea
and
\bea\label{eq:dsidt_step1}
	\fl\frac{ds^i_A}{dt} = \frac{s^0_A}{E} \bigg\{ k^i \pd_t u_0 + \frac{k^i k^j}{k^0}\pd_j u_0 + \Gam k^i \pd_t\alp  + \frac{k^i k^j}{k^0}\alp u^k K_{kj} - \alp E \gam^{ij} \pd_j\alp + \frac{\alp E}{k^0} K^i_{\ph{i}j} k^j \bigg\} \nonumber\\ + \frac{s^m_A}{E} \bigg\{ k^i \pd_t u_m + \frac{k^i k^j}{k^0} \pd_j u_m + \Gam k^i \pd_m\alp  - \frac{k^i k^j}{k^0} \Gam^k_{jm} u_k + \alp E K^i_{\ph{i}m} \\ - \frac{E}{k^0} \Gam^i_{jm} k^j \bigg\},\nonumber
\eea
respectively. The derivatives of the 4--velocity, $u_\mu$, in the above can be expanded in terms of derivatives of the velocity $v^i=u^i/(\alp u^0)$ as follows
\bea
	\pd_t u_0 &= \frac{u_0\Gam^2}{2} \pd_t (v^i v_i) - \Gam \pd_t \alp, \\
	\pd_l u_0 &= \frac{u_0\Gam^2}{2} \pd_l (v^i v_i) - \Gam \pd_l\alp, \\
	\pd_t u_j &= \Gam \gam_{ij} \pd_t v^i + \frac{\Gam^3}{2} \gam_{ij} v^i \pd_t (v^k v_k) + 2 u_0 v^i K_{ij}, \\
	\pd_l u_j &= \Gam \gam_{ij} \pd_l v^i + \frac{\Gam^3}{2} \gam_{ij}v^i \pd_l (v^k v_k) + \Gam v^i \pd_l \gam_{ij},
\eea
where $\Gam=1/\sqrt{1-v^iv_i}$ is the Lorentz factor, we have used $u_0=-\alp\Gam$, and we have
\bea
	\pd_t (v^i v_i) &= -2\alp K_{ij} v^i v^j + 2 \gam_{ij} v^i \pd_t v^j, \\
	\pd_l (v^i v_i) &= v^i v^j \pd_l \gam_{ij} + 2\gam_{ij} v^i \pd_l v^j.
\eea

For computational convenience, the evolution equations actually solved in \mesc\ are written in terms of some quantities already calculated in other routines, namely 
\bea\label{eq:ds0dt}
	\fl\frac{ds^0_A}{dt} = \frac{s^0_A}{E} \bigg\{ \frac{u_0\Gam^2}{2} \left(k^0 \pd_t v^2 + k^i \pd_i v^2 \right) - \frac{E}{\alp}\pd_t\alp - \pd_i\alp \left(\Gam k^i + E v^i\right) + \alp K_{ij} k^i u^j \bigg\} \nonumber\\ + \frac{s^m_A}{E} \bigg\{ k^0 \Gam \gam_{im} \pd_t v^i + \frac{k^0\Gam^3}{2}\gam_{im} v^i \pd_t v^2 + K_{im}v^i \left(2k^0 u_0 + E\right) \\ + \Gam \gam_{lm} k^i \pd_i v^l
	+ \frac{\Gam^3}{2} \gam_{lm} v^l k^i \pd_i v^2 + \Gam k^i v^l \pd_i \gam_{lm} + \frac{1}{\alp} u_i k^i \pd_m \alp - k^i \Gam^j_{im} u_j \bigg\},\nonumber
\eea
and
\bea\label{eq:dsidt}
	\fl\frac{ds^i_A}{dt} = \frac{s^0_A}{E} \bigg\{ \frac{u_0\Gam^2}{2} \left( k^i \pd_t v^2 + \frac{k^i k^j}{k^0} \pd_j v^2 \right) + \frac{k^i k^j}{k^0} \left( \alp K_{kj}u^k - \Gam \pd_j \alp \right) - \alp E \gam^{ij} \pd_j \alp \nonumber\\ + \frac{\alp E}{k^0} K^i_{\ph{i}j} k^j \bigg\} + \frac{s^m_A}{E} \bigg\{ \Gam k^i \gam_{lm} \left( \pd_t v^l + \frac{k^j}{k^0} \pd_j v^l \right) + \frac{\Gam^3}{2} k^i \gam_{lm} v^l \bigg( \pd_t v^2 \\ + \frac{k^j}{k^0} \pd_j v^2 \bigg) + 2u_0 k^i v^l K_{lm} + \Gam k^i \pd_m \alp + \alp E \gam^{ik}K_{km} + \frac{k^i k^j}{k^0} \Gam v^l \pd_j \gam_{lm} \nonumber\\ - \frac{k^i k^j}{k^0} \Gam^k_{jm} u_k - \frac{E}{k^0} \Gam^i_{jm} k^j \bigg\},\nonumber
\eea
where $\pd_\mu v^2 = \pd_\mu(v^i v_i)$.

\subsection{Observers at rest}\label{appx:screen_obs_rest}

We can show that $s^0_A=0$ for any observer at rest with respect to the coordinates in the case of zero shift. First, expanding the constraint that $s^\mu_A$ must be orthogonal to the observer 4--velocity, i.e. $g_{\mu\nu} s^\mu_A u^\nu = 0$, we have
\bea
    g_{00} s^0_A u^0 + g_{0i}s^0_A u^i + g_{i0} s^i_A u^0 + g_{ij} s^i_A u^j = 0.
\eea
An observer at rest has 4--velocity $u^\mu_{\rm rest}=(1/\alp,{\bf 0})$ in general, namely $u^i=0$ (which further implies $\Gamma=1$). In the above, we then have the remaining non-zero terms
\bea
    g_{00} s^0_A u^0 + g_{i0} s^i_A u^0 = 0,
\eea
and for zero shift we have $g_{i0}=0$, implying $g_{00} s^0_A u^0=0$, and since either $g_{00}$ or $u^0$ cannot be zero, we must have $s^0_A=0$ for the constraint $s^\mu_A u_\mu$ to be satisfied.

\subsection{Initial screen vectors}\label{appx:screen_init}

To calculate the initial values of the screen vectors, we begin with the constraints
\bea\label{eq:initsAconstraints}
	s^\mu_A u_\mu = s^\mu_A e_\mu = g_{\mu\nu} s^\mu_A s^\nu_B = 0,
\eea
which represent five equations for a total of eight unknowns ($s^\mu_A$). We can therefore freely choose three components of $s^\mu_A$ (i.e. the orientation of the basis) and solve the constraints for the remaining five. 
We then must ensure our basis is normalised, i.e. that the following is satisfied: 
\bea
	g_{\mu\nu} s^\mu_A s^\nu_A = g_{\mu\nu} s^\mu_B s^\nu_B = 1.
\eea
We take our solutions to \eref{eq:initsAconstraints} and normalise them to unit length. This results in a set of basis vectors which satisfies all of the required constraints.

\subsection{Test of initial screen vectors}\label{appx:screen_init_test}

The screen vectors are an arbitrary set of basis vectors to describe the screen space, and any set satisfying the constraints (for the same observer in the same environment) should give the same physical result. To test this, we rotate the screen vectors by making a different choice for our free three degrees of freedom and check the physical results are unchanged.

First, we perform a test run for linearly-perturbed EdS spacetime with perturbation amplitude $\phi_0=10^{-5}$ and resolution $N=48$ (see \ref{appx:linperttest} for further details on this test setup). We run the same test twice with all parameters identical except for a different initial set of screen vectors. 
\begin{figure*}[ht!]
	\centering
	\includegraphics[width=0.8\textwidth]{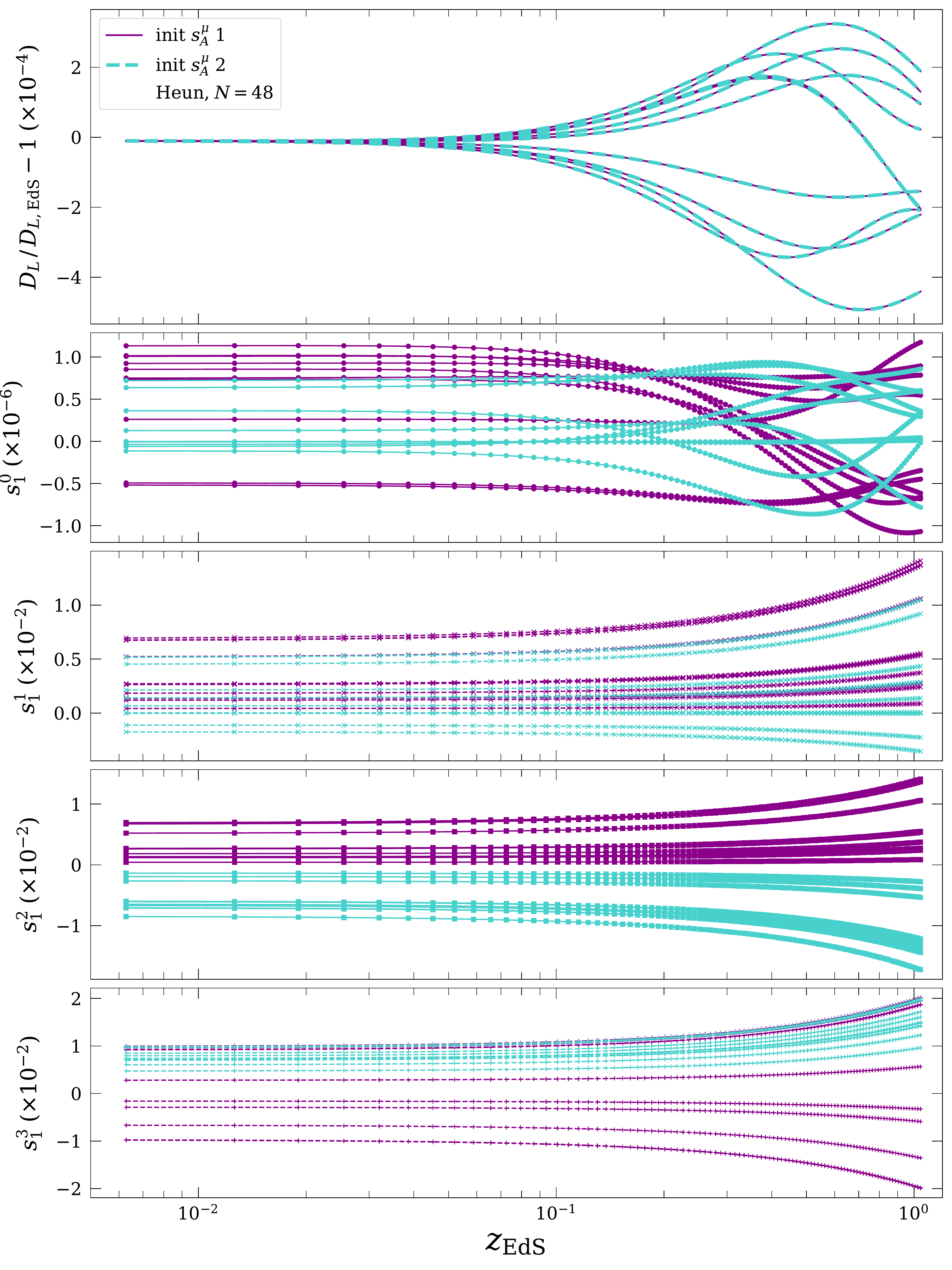}
	\caption{Top panel: luminosity distance as a function of redshift in a linearly-perturbed EdS metric for two test runs with different initial screen vectors $s^\mu_A$ (dashed and solid curves). Bottom four panels: screen vector components for the same two test runs (matching colours). Each of the ten curves show an individual line of sight. The observer and all lines of sight are identical in both tests. The screen vectors change (bottom four panels) but the physical result ($D_L$, top panel) is the same.}
	\label{fig:smurotate_linpert}
\end{figure*}
Figure~\ref{fig:smurotate_linpert} shows the luminosity distance perturbation as a function of redshift for the two tests with different sets of initial screen vectors (solid purple and dashed cyan curves). In the bottom four panels we show each component of the screen vectors (for $A=1$) for the same tests, in the same colours. We can see the screen vector components differ between the tests but the physical result, $D_L$ in the top panel, is consistent. 

We also perform this test with simulation output, using an ET simulation with resolution $N=64$, box size $L=640 h^{-1}$ Mpc and all power beneath 100$h^{-1}$ Mpc removed from the initial data. We perform the ray-tracing to $z\approx 0.1$ for two different sets of initial screen vectors.
\begin{figure*}[ht!]
	\centering
	\includegraphics[width=0.8\textwidth]{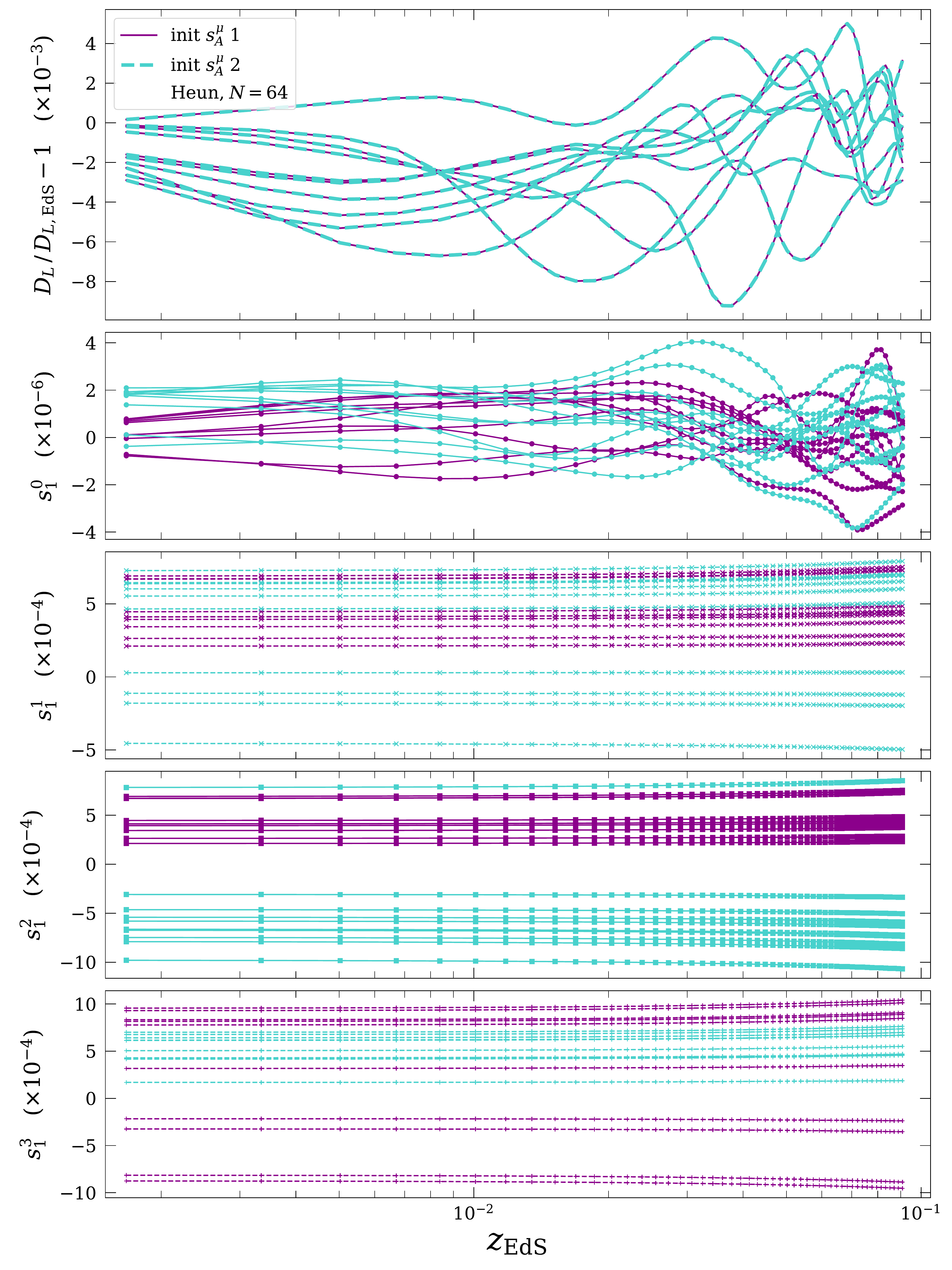}
	\caption{Top panel shows the ray-traced luminosity distance in an ET simulation sampling only large-scale structures for two runs with different initial screen vectors (dashed and solid lines). Each of the ten curves show a randomly drawn line of sight for the same observer. Bottom four panels show the screen vector components for the same pair of runs.}
	\label{fig:smurotate_64pkcut}
\end{figure*}
The top panel of Figure~\ref{fig:smurotate_64pkcut} shows the ray-traced luminosity distance $D_L$, normalised by the EdS distance, as a function of redshift for two different sets of initial screen vectors (dashed cyan and solid purple curves). The bottom four panels show each of the screen vector components for $A=1$. We see the same result that the screen vectors differ between the two runs but the physical result, $D_L$ in the top panel, is consistent. 

We thus conclude from these tests that our choice of initial screen vectors, chosen via the three arbitrary degrees of freedom, has no effect on our physical results.

\section{Boosting data to the hypersurface frame}\label{appx:CMBframe}

In Section~\ref{sec:observers}, we introduced the possibility of approximating the CMB frame --- as used in observational cosmology --- using the simulation hypersurfaces themselves. In this case, our observers are still co-moving with the fluid flow, however, after measuring their distances and redshifts in that co-moving frame, they boost their observations to the ``CMB frame''. Usually, such an observer would infer their velocity with respect to this frame using the dipole they measure in the CMB \citep[as in][]{PlanckDipole:2014a}. For this section, we instead assume the observer's velocity with respect to the fictitious ``CMB frame'' is their peculiar velocity with respect to the Eulerian coordinates, namely, $v^\mu$. If we assume the ``CMB frame'' is well described by an EdS expansion law, 
an observer moving with velocity $v^\mu$ relative to this frame
will see a dipole modulation of their measured luminosity distance of the form \citep[see, e.g.][]{Bonvin:2006b,Hui:2006}
\begin{equation}\label{eq:dLboost}
    d_L^{\rm obs}(z) = d_{L,{\rm EdS}}(z) + v^\mu e_\mu \frac{(1+z)}{\mathcal{H}(z)},
\end{equation}
where $e^\mu$ is the direction of observation on the sky and $\mathcal{H}(z)$ is the conformal Hubble parameter.
Peculiar motion of the observer with respect to an FLRW frame induces a dipole in \textit{both} the measured redshift, $z$, and the luminosity distance, $d_L$. However, both of these effects have been accounted for in deriving the above expression \citep[see][for more details]{Hui:2006}, and thus $d_L(z)$ and $d_{L,{\rm EdS}}(z)$ are calculated at the same (observed) redshift $z$. 

To apply a boost to each observer's data in our simulations, we calculate $d_L^{\rm obs}$ above using the peculiar velocity, $v^i$, of the fluid at their location on the grid. We then subtract the dipole component of $d_L^{\rm obs}$ dipole from the ray-traced luminosity distance $D_L$ to ``correct'' for the motion of the observer. We confirm this process removes the majority of the dipole contribution to the luminosity distance, however, some dipolar modulation still remains. 

We would also like to emphasize that this process is not exactly the same as that done with real observational data. In our observations, we measure the redshift but do not usually have direct access to the luminosity distance. Peculiar velocity corrections are thus usually applied to our redshift measurements. However, here we are working with simulated $D_L(z)$ which has been interpolated along each line of sight, in order to arrive at a surface of constant $z$. The intention of this type of data is to directly probe the $D_L(z)$ relation with respect to the FLRW theory prediction. 
For this type of data, applying the boost to the luminosity distance, using the method described above, allows us to estimate the impact of the observer's peculiar velocity on the anisotropy measured in $D_L$. 
In future studies, we aim to work with synthetic observational catalogues instead of shells of constant $z$, wherein we will consider corrections which may be more in line with observational data analysis methods. 

We might expect %
this correction to reduce the amount of anisotropy for our observers.
\begin{figure}
    \centering
    \begin{subfigure}{.5\textwidth}
        \centering
        \includegraphics[width=\linewidth]{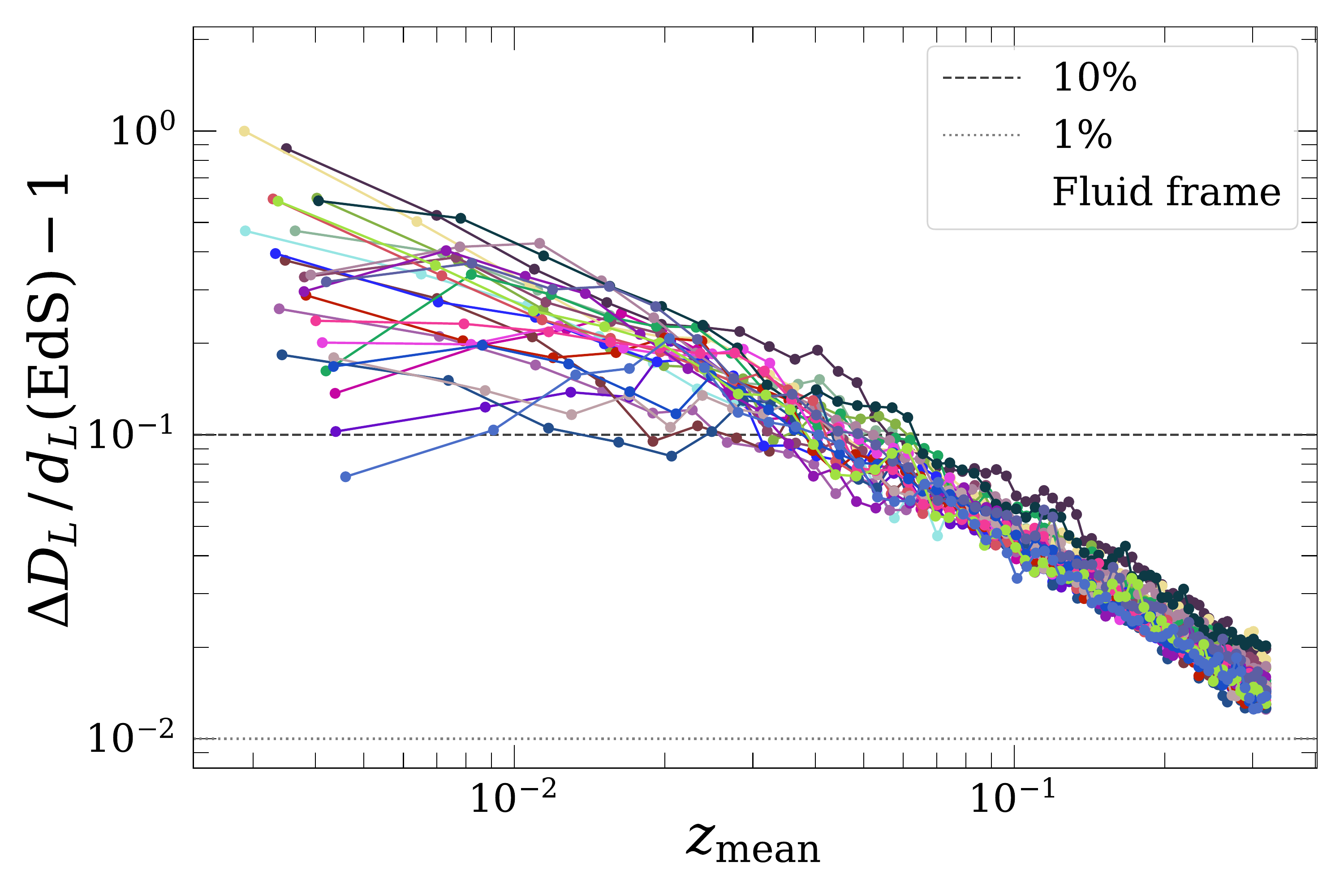}
    \end{subfigure}%
    \begin{subfigure}{.5\textwidth}
         \centering
        \includegraphics[width=\linewidth]{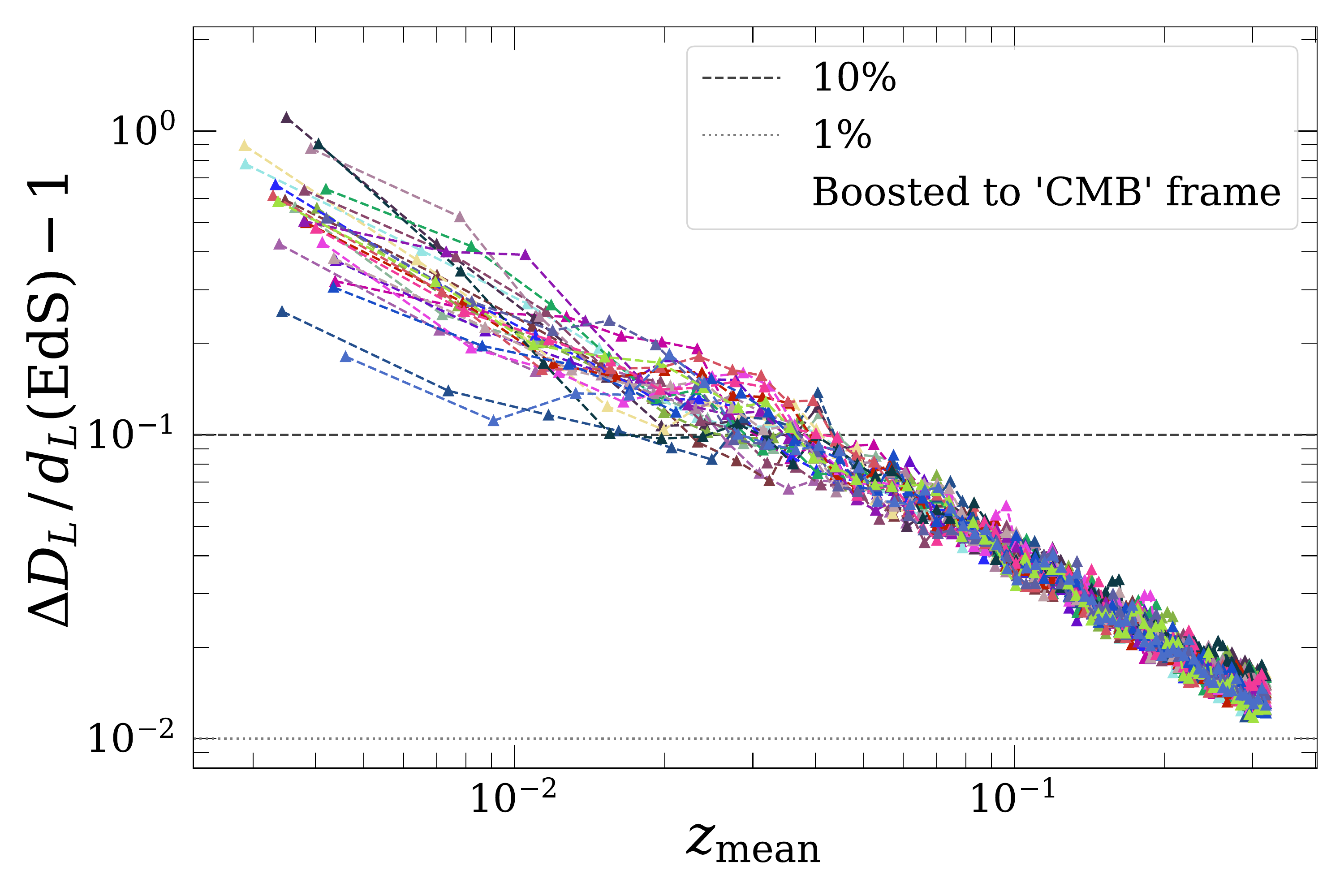}
    \end{subfigure}
    \caption{Maximal sky variance as a function of redshift for 25 observers in the simulation shown in Figure~\ref{fig:rho_cut100}. We show $\Delta D_L$ relative to the EdS distance at that redshift, $d_L({\rm EdS})$. 
    Left panel shows the variance for observations in the frame co-moving with the fluid flow and the right panel shows the same observations boosted to the ``CMB frame'' using each observers coordinate peculiar velocity. 
    Dashed and dotted horizontal lines in both panels show 10\% and 1\% deviation, respectively.}
    \label{fig:maxskyvar_cut100_wboost}
\end{figure}
In Figure~\ref{fig:maxskyvar_cut100_wboost} we compare the results for the maximal sky variance in $D_L$ calculated in the fluid frame (as shown in Figure~\ref{fig:maxskyvar_cut100}, reproduced in the left panel) to the boosted observations in the ``CMB frame'' (right panel). 
We see a reduction in the spread of variance across observers, however, no significant reduction in the maximal sky variance is seen. 
We note again that this ``CMB frame'' is not necessarily coincident with the way in which we define the CMB frame in observational cosmology. The correct way to define such a frame would be to infer the velocity with respect to the CMB that each observer would see. However, as mentioned in the main text, this would require ray tracing to very high redshift in order to obtain a map to mimic the CMB radiation. 

We note that the peculiar velocites of the ``sources'' (with respect to the simulation hypersurface) considered here may be able to account for the anisotropies we see in the right panel of Figure~\ref{fig:maxskyvar_cut100_wboost}. However, since we find that the simulation slice remains very close to the background EdS model we choose for the initial slice \citep[when averaging over large scales, see][]{Macpherson:2019} this is perhaps not surprising.
Usually, in cosmological data analysis we might also apply some corrections for the peculiar velocity of low-redshift sources with respect to the CMB frame \citep[see][]{Peterson:2021}. Investigating the effect of such corrections on our simulation data falls beyond the scope of this work. 
Connecting the observational CMB frame to ray-traced simulation data is an important avenue to pursue if we wish to make predictions for future observations.

\section{Spatial interpolation}\label{appx:interp}

\texttt{Mescaline} reads in HDF5 data at a distinct set of time slices as output by the ET simulation. The code itself is thus tracking the propagation of light beams through this simulated space-time, at the specific coordinate times defined by the constant-$t$ surfaces. %
The simulation is also discretised in space and in general the light beam will not be positioned on a grid cell at each time step; it will land somewhere in between. At every step, we therefore must use spatial interpolation to calculate the simulation quantities (e.g. metric tensor, curvature) appearing on the right-hand-side of the evolution equations at the position of the light beam. 
For this purpose, we use the B-spline Fortran library\footnote{\url{http://jacobwilliams.github.io/bspline-fortran}}, which provides one to six dimensional B-spline interpolation on a regular grid. In \mesc, the order of the spatial interpolation is left as a free variable. However, for all purposes in this paper we use fifth-order spline interpolation. We have found that a higher-order interpolation offers very little improvement in the accuracy of the results.
Calculating all source terms at the fifth-order stencil is an expensive part of the \mesc\ ray tracer, however, reducing to fourth order interpolation only yields a $\sim 4\%$ increase in speed.

\section{Transformation of vectors from local Fermi frame}\label{appx:fermi}

Quantities output from the raytracer, such as redshift and luminosity distance, are calculated as seen by observers moving with 4--velocity $u^\mu$. In this work, we take this to coincide with the 4--velocity of the fluid flow. However, initial direction vectors we specify ($e^\mu$) are defined in our coordinate system, which is not comoving with the fluid. We will generally be given the directions of objects {in the observer's frame}, e.g., the right ascension and declination from a given survey. These coordinates are represented by the Cartesian direction vector $m^{\hat i}$. We must transform these to the direction vector in our coordinate system, $m^i$, which is then used to calculate $e^\mu$. When plotting sky-distributions of redshift or distance we use $m^{\hat i}$ to ensure angular power spectra, direction of multipoles, etc., are correctly defined in the observer's frame. 

In this section, we describe the transformation from a vector defined in the observer's local Fermi frame\footnote{The author would like to thank Jim Mertens for correspondence aiding with the calculations in this section.} --- also known as the local inertial frame, or proper reference frame of an observer who is not necessarily geodesic \citep[see pg. 327 of ][]{MTW} %
--- to the simulation coordinate system.
We consider the observer's metric to be {conformally} Minkowski, with local ``scale factor'' equal to $\gam^{1/3}$ --- since the observer's ``scale factor'' may not necessarily follow the convention $a(z=0)=1$ within the simulation. 
We transform the local metric at the observer, $\tilde{\eta}^{\alp\beta}=\gam^{-1/3}\eta^{\alp\beta}$, to the metric tensor in our coordinate system using the following basis (where ${\hat \alp}$ represents the observer's coordinate $x^{\hat \alp}$)
\bea
	g^{\mu\nu} &= e^\mu_{{\hat \alp}} e^\nu_{\hat \beta} g^{{\hat \alp} {\hat \beta}}, \\
		&= e^\mu_{{\hat \alp}} e^\nu_{\hat \beta} \gam^{-1/3} \eta^{{\hat \alp} {\hat \beta}},
\eea
such that
\bea
		\gam^{1/3} g^{\mu\nu} &= e^\mu_{{\hat 0}} e^\nu_{\hat 0} \eta^{{\hat 0} {\hat 0}} + e^\mu_{{\hat i}} e^\nu_{\hat j} \eta^{{\hat i} {\hat j}}, \\
		&= - u^\mu u^\nu + e^\mu_{{\hat i}} e^\nu_{\hat j} \delta^{{\hat i} {\hat j}},
\eea
where in the last line we have used $e^\mu_{\hat 0}=u^\mu$, which is the observer's time coordinate in our coordinate system, i.e. the 4--velocity of the observer. %

The transformation of $m^{\hat i}$ is given by
\bea
	m^i = e^i_{\hat j} m^{\hat j},
\eea
where we do not need to transform the time component since $e^0$ is set via the normalisation condition of $e^\mu$ in \mesc. We therefore only need to solve six equations, namely
\bea
	\tilde{g}^{ij} \equiv \gam^{1/3}  g^{ij} = - u^i u^j + e^i_{k} e^j_{l} \delta^{kl},
\eea
(where we have now removed the hats for simplification) for the nine components of $e^i_j$. We therefore have three degrees of freedom remaining, and without loss of generality we can set some components to zero:
\begin{equation}
	e^i_j =
	\pmatrix{%
	e^x_x & 0 & 0 \cr %
	e^x_y & e^y_y & 0 \cr %
	e^x_z & e^y_z & e^z_z}.
\end{equation}
From this choice, we find the remaining components are
\bea
	e^z_z &= \sqrt{\tilde{g}^{zz} + (u^z)^2}, \\
	e^y_z &= \frac{\tilde{g}^{yz} + u^y u^z}{e^z_z}, \\
	e^y_y &= \sqrt{\tilde{g}^{yy} + (u^y)^2 - (e^y_z)^2}, \\
	e^x_z &= \frac{\tilde{g}^{xz} + u^x u^z}{e^z_z}, \\
	e^x_y &= \frac{\tilde{g}^{xy} + u^x u^y - e^x_z e^y_z}{e^y_y}, \\
	e^x_x &= \sqrt{\tilde{g}^{xx} + (u^x)^2 - (e^x_y)^2 - (e^x_z)^2}.
\eea
This transformation is implemented 
after determining the directions $m^{\hat j}$ but
before calculating the direction vector $e^\mu$, which is defined in the simulation coordinate system.

\section{Tests of the \mesc\ ray tracer}\label{appx:tests}

We must test the accuracy and precision of the \mesc\ ray tracer on known analytic solutions. Tests such as these are necessary in order for us to trust results from simulation output, where there is not necessarily a known analytic solution. 
Here we present the results of ray tracing in two test metrics. In \ref{appx:EdStest} we test within a pure EdS space-time: a flat FLRW metric tensor with a matter-dominated source (no dark energy). In \ref{appx:linperttest}, we test within a linearly-perturbed EdS space-time.

For all tests we use an RK2 (Heun) integration with Courant factor dt/dx$=0.1$ and a cubic domain with $L=4\,h^{-1}$~Gpc and three resolutions $N=12,24$, and 48.
In all tests, we feed an analytic form of the metric into \mesc\ before passing this into the regular (and unchanged) ray-tracing routines. In the following sections, we will present the analytic metric for both cases and the analytic solutions for the quantities we evolve, namely $k^\mu$, $s^\mu_A$, and $D_A$ (and $z$ in the case of pure EdS). We always compare three resolutions to the analytic solutions to ensure the error is converging at the expected rate with a decrease in time step size.

\subsection{Definition of error and convergence}

We define the relative error with respect to the relevant analytic solutions, e.g., for the luminosity distance in an EdS space-time, using
\bea\label{eq:errordef}
    {\rm err}(d_L) \equiv \frac{d_{L,{\rm num}}}{d_{L,{\rm EdS}}} - 1,
\eea
where $d_{L,{\rm num}}$ is the numerical luminosity distance and $d_{L,{\rm EdS}}$ is the analytic luminosity distance. 

We calculate the rate of convergence by evaluating the error at all three resolutions with grid spacing $\Delta x_1$, $\Delta x_2$, and $\Delta x_3$. For an integration method of order $p$, the error will be $\mathcal{O}(\Delta x^p)$, and the convergence rate of the error in \eref{eq:errordef} is given by
\bea\label{eq:convergence}
    \mathcal{C} = \frac{{\rm err}_{\Delta x_1} - {\rm err}_{\Delta x_2}}{{\rm err}_{\Delta x_2} - {\rm err}_{\Delta x_3}},
\eea
and the expected convergence rate is
\bea\label{eq:convergence-expected}
    \mathcal{C}_{\rm exp} = \frac{\Delta x_1^p - \Delta x_2^p}{\Delta x_2^p - \Delta x_3^p}\,.
\eea
For the RK2 method we have $p=2$ and therefore $\mathcal{C}_{\rm exp}=4$ for the case of $\Delta x_1 = 2\Delta x_2 = 4\Delta x_3$.

\subsection{EdS spacetime}\label{appx:EdStest}

For our first test we will use the EdS space-time, which has metric tensor
\bea\label{eq:EdSmetric}
	ds^2 = -a^2 d\eta^2 + a^2\delta_{ij} dx^i dx^j,
\eea
where $\eta$ is the conformal time, and $a=a_{\rm init} (\eta/\eta_{\rm init})^2$ is the scale factor, and $a_{\rm init}$ and $\eta_{\rm init}$ are the initial values of the scale factor and conformal time, respectively.

\subsubsection{Analytic solutions.}

We adopt the convention $a_{\rm init}=1$ for the tests, implying $a(z=0) = 1+z_{\rm init}$. The scale factor and redshift are then simply related by
\bea\label{eq:EdSz}
	z(\eta) = \frac{a_{\rm init}}{a(\eta)} - 1.
\eea
The luminosity distance in an EdS spacetime, with cosmological density parameters $\Omega_m=1,\Omega_\Lambda=\Omega_k=0$, is
\bea\label{eq:EdSdL}
	d_{L,{\rm EdS}}(z) = \frac{2}{\mH_0} \left(1+z - \sqrt{1+z}\right),
\eea
where $\mH(z) \equiv a'/a$ is the conformal Hubble parameter, where a dash is a derivative with respect to conformal time, and $\mH_0 \equiv \mH(z=0)$. 
Substituting the EdS metric into the evolution equations \eref{eqs:kmuevol} for $k^0$ and $k^i$ we find
\bea
	\frac{dk^0}{dt} &= - 2k^0 \mathcal{H}, \quad\quad
	\frac{dk^i}{dt} &= -2 k^i \mathcal{H},
\eea
which gives 
\bea\label{eq:kmuEdS}
    k^0 &= \frac{k^0_{\rm ini}}{a^2}, \quad\quad
    k^i &= \frac{k^i_{\rm ini}}{a^2}
\eea
for some arbitrary initial data, $k^0_{\rm ini}$ and $k^i_{\rm ini}$. Similarly, for the propagation of the screen vectors \eref{eq:ds0dt} and \eref{eq:dsidt} we find
\bea
	\frac{ds^i_A}{dt} &= - s^i_A \mathcal{H},
\eea
and therefore 
\bea\label{eq:smuEdS}
    s^i_A = \frac{s^i_{A,{\rm ini}}}{a},
\eea
for some arbitrary initial $s^i_{A,{\rm ini}}$. Since our observers are co-moving with the EdS matter content, they have 4--velocity components $u^i=0$ and thus $s^0_A=0$ and ${ds^0_A}/{dt} = 0$ (see \ref{appx:screen_obs_rest}).

\begin{figure*}[ht!]
	\centering
	\includegraphics[width=0.8\textwidth]{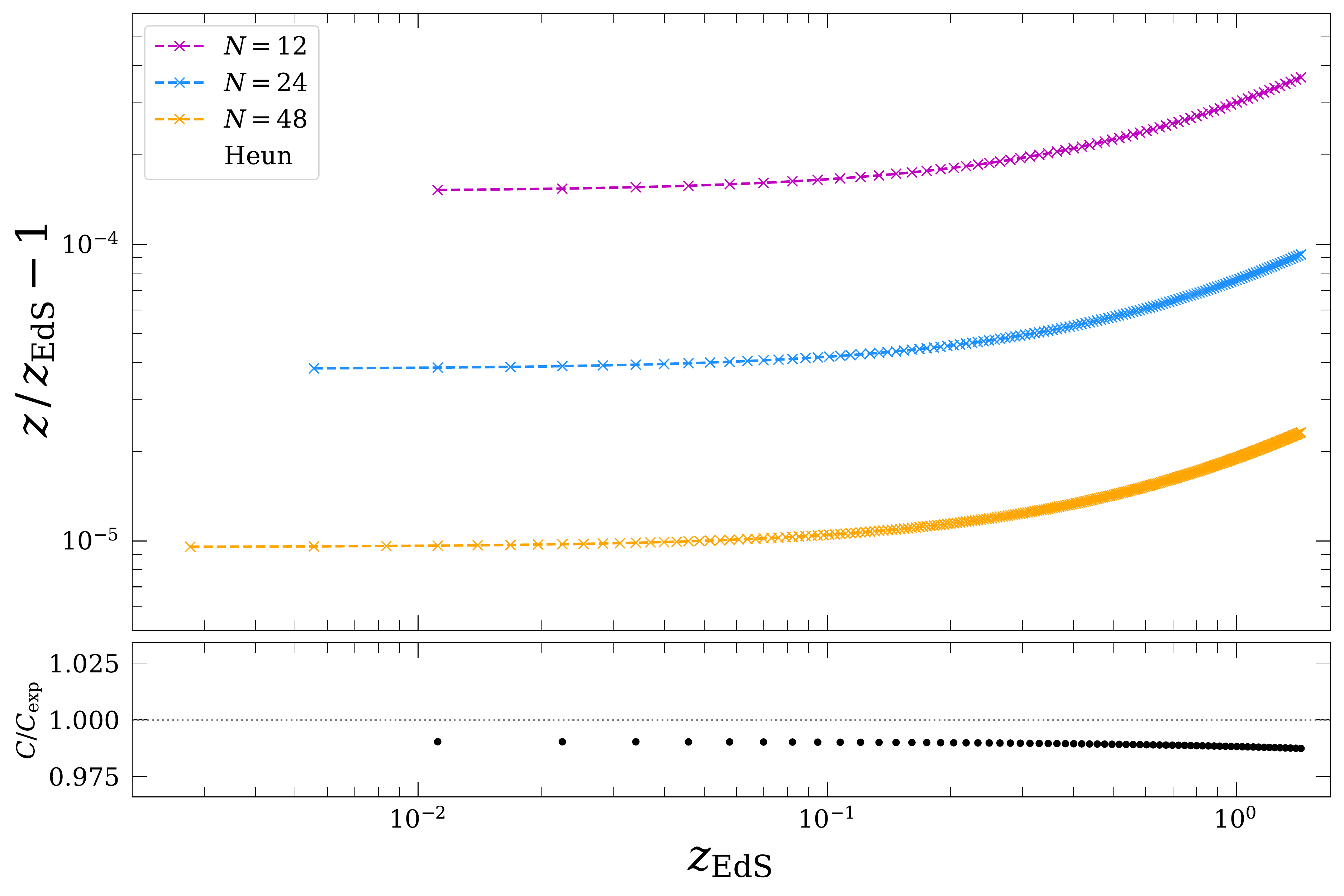}
	\caption{\texttt{Mescaline} redshift, $z$, relative to the EdS redshift, $z_{\rm EdS}$, as a function of $z_{\rm EdS}$. We show the relative error for three resolutions $N=12,24$, and 48 in the top panel, and the convergence rate, $C$, relative to the expected ($C_{\rm exp}=4$ for RK2) in the bottom panel.}
	\label{fig:zvsEdS}
\end{figure*}
\begin{figure*}[ht!]
	\centering
	\includegraphics[width=0.8\textwidth]{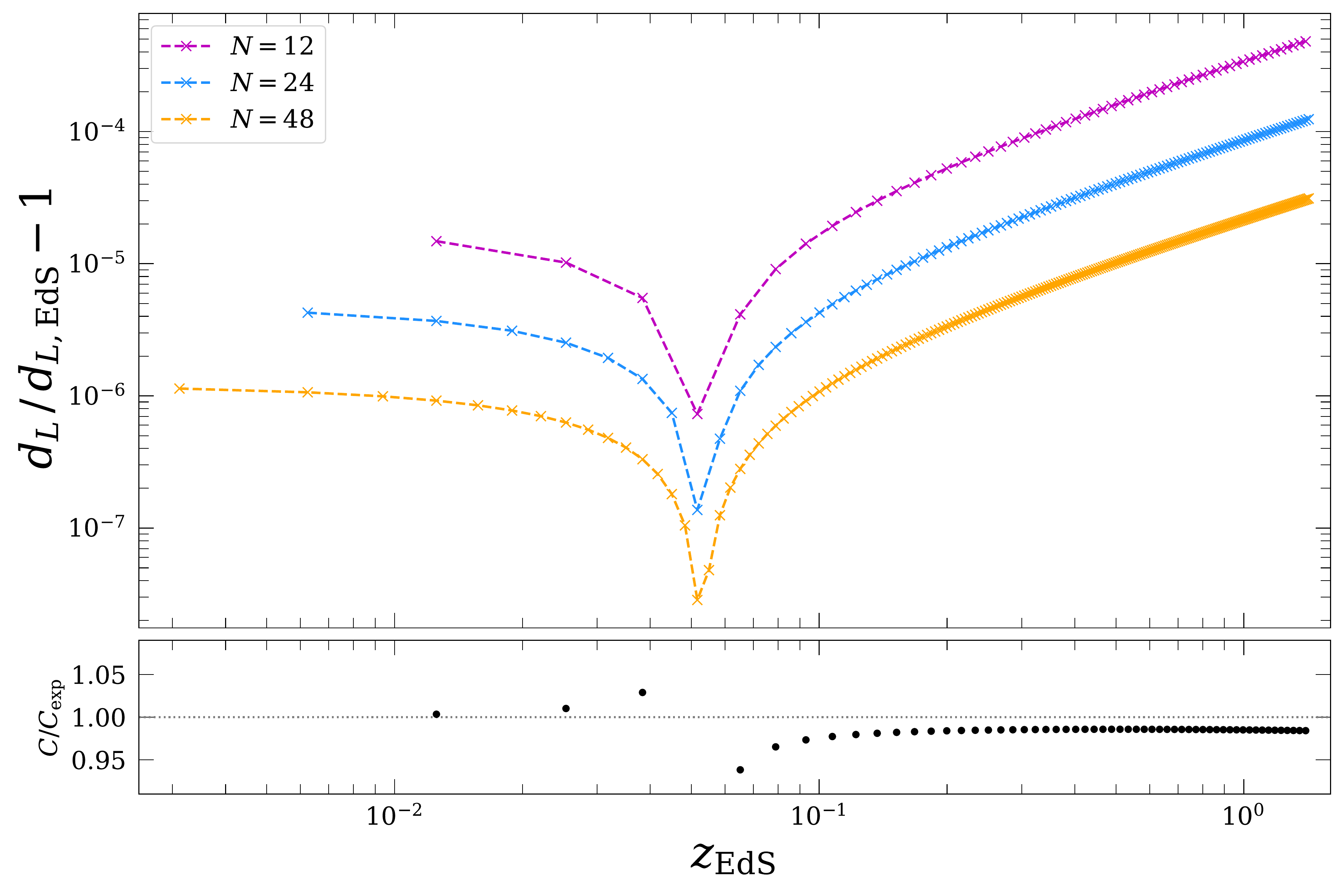}
	\caption{Ray-traced luminosity distance, $d_L$, in an analytic EdS metric relative to the EdS analytic solution, $d_{L,{\rm EdS}}$, as a function of $z_{\rm EdS}$. We show the relative error for three resolutions $N=12,24$, and 48 in the top panel, and the convergence rate, $C$, relative to the expected value for RK2 in the bottom panel.}
	\label{fig:dLvsEdS}
\end{figure*}
\subsubsection{\texttt{Mescaline} vs. analytic solutions.}

We set up the pure EdS metric \eref{eq:EdSmetric} and send this into \mesc, using the same routines which calculate $z$ and $D_A$ for a general metric evolved using NR.
We compare the redshift $z$, luminosity distance $d_L$, photon 4--momentum $k^\mu$, and screen vectors $s^\mu_A$ to the analytic solutions given in the previous section. The initial data for both $k^\mu$ and $s^\mu_A$ ($k^\mu_{\rm ini}$ and $s^\mu_{\rm ini}$) is dependent on the direction of observation and therefore is arbitrary. We also check the null condition and the constraints on both $e^\mu$ and $s^\mu_A$ are satisfied to within round-off error after generating initial data.
We follow one line of sight only, since all directions are equivalent in the EdS space-time, for each resolution until $z\approx 1.3$. 

Figure~\ref{fig:zvsEdS} shows the error in the redshift as calculated with \mesc\ relative to the EdS solution \eref{eq:EdSz} calculated using the conformal time. We show three resolutions (different colours) in the top panel, and the respective convergence rate \eref{eq:convergence} in the bottom panel relative to the expected $C_{\rm exp}=4$ for RK2. 
For our highest (yet still coarse) resolution of $N=48$ the error remains below $0.01\%$, and the convergence rate is as expected.

Figure~\ref{fig:dLvsEdS} shows the error in the luminosity distance as calculated with \mesc\ for an analytic EdS metric relative to the EdS solution. The top panel shows the relative error for each resolution, and the bottom panel shows the convergence rate for the three curves, again relative to the expected rate for RK2. Error in the luminosity distance lies below $\sim 0.005\%$ for $N=48$ at redshift $z_{\rm EdS}\sim 1$, and the error converges as expected.

\begin{figure*}[ht!]
	\centering
	\includegraphics[width=\textwidth]{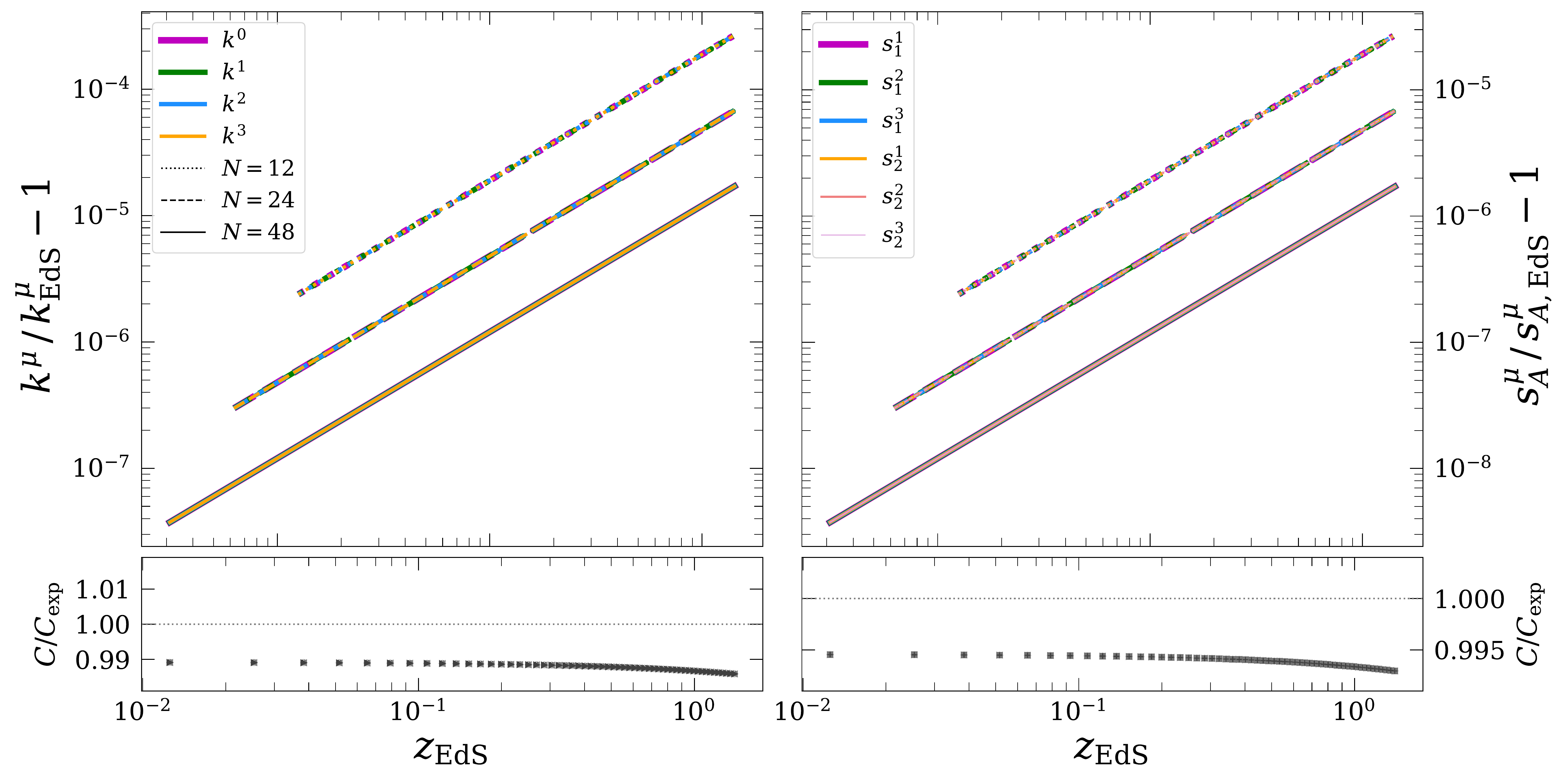}
	\caption{Photon 4--momentum, $k^\mu$, and screen vectors, $s^\mu_A$, as calculated with \mesc\ relative to their analytic EdS evolution. The top left panel shows the relative error in each component of $k^\mu$ (different colours) for three resolutions (different line styles), and the bottom left panel shows the convergence rate for all $k^\mu$ components, relative to the expected rate for RK2. The top right panel shows the relative error in $s^i_A$ for three resolutions, with the bottom right panel showing the respective convergence rates.}
	\label{fig:ksvsEdS}
\end{figure*}
Figure~\ref{fig:ksvsEdS} shows the error in the photon 4--momentum, $k^\mu$, and the screen vectors, $s^\mu_A$, with respect to their EdS analytic solutions \eref{eq:kmuEdS} and \eref{eq:smuEdS}, respectively, as a function of EdS redshift. 
The top left panel of Figure~\ref{fig:ksvsEdS} shows the relative error in each component of $k^\mu$ (different colours) for three resolutions (different line styles), and the bottom left panel shows the convergence rate for each set of three resolutions as different point styles (which all closely coincide).
The top and bottom right panels shows relative error and convergence for the components $s^i_A$. We note again that $s^0_A=0$ for EdS, which we have checked is the case in these tests.
The error converges at the expected rate in all cases shown here.

\subsection{Linearly-perturbed EdS spacetime}\label{appx:linperttest}

We now test the \mesc\ ray tracer in the presence of inhomogeneities. For this we will use a linearly-perturbed EdS space-time, which (for scalar-only perturbations) has metric tensor
\bea\label{eq:linpertmetric}
    ds^2 = -a^2(1+2\phi)d\eta^2 + a^2(1-2\phi)\delta_{ij} dx^i dx^j,
\eea
where $\phi=\phi(x^i)$ is a small perturbation with amplitude $\phi_0\ll1$. Since our observers are co-moving with the fluid flow, they also must be perturbed with respect to the background. 
For zero shift, the components of the fluid 4--velocity are $u^i = \Gamma v^i$ and $u^0=\Gamma/\alp$, where $\Gamma=1/\sqrt{1-v^i v_i}$ is the Lorentz factor and $v^i$ is the peculiar velocity of the fluid with respect to the Eulerian observer.
In the limit of linear perturbations, we therefore have $u^i=v^i$ and $u^0=1/\alp$. 
The observer 4--velocity is thus $u^\mu=(1/\alp,v^i)$, where the velocity perturbation is given in \citet{Macpherson:2019}. 

In the next section, we will derive the geodesic equation and the Jacobi evolution equations for the linearly-perturbed EdS metric. %
We will then numerically solve this system using an RK4 integrator to give a semi-analytic solution to compare with the \mesc\ results. We use RK4 such that the analytic solution is closer to the true solution and thus we can still check if our \mesc\ solutions (advanced with RK2) are converging at the expected rate. 

\subsubsection{Semi-analytic solutions.}

The geodesic equation in the perturbed space-time with metric \eref{eq:linpertmetric} is
\bea
	\frac{dk^0}{dt} &= -2\mH k^0 - \frac{2k^j\pd_j \phi}{1+2\phi}, \label{eq:dk0dt_linpert}\\
	\frac{dk^i}{dt} &= -2\mH k^i - \frac{k^0\pd^i\phi}{1-2\phi} - \frac{k^k}{k^0}\Gam^i_{kj} k^j, \label{eq:dkidt_linpert} %
\eea
which is correct to all orders in $\phi$. The Christoffel symbols are
\bea\label{eq:gamijk_linpert}
    \Gamma^i_{jk} = \frac{1}{(1-2\phi)}\left[\delta^{il}\pd_l(\phi) \delta_{jk} -\pd_j (\phi) \delta^i_k - \pd_k (\phi) \delta^i_j \right],
\eea
which is also correct to all orders. The position of the photon is calculated via \eref{eq:dxidt}, namely, $dx^i/dt=k^i/k^0$ with $k^i$ and $k^0$ given as solutions to \eref{eq:dk0dt_linpert} and \eref{eq:dkidt_linpert}, respectively.

For the perturbed EdS metric \eref{eq:linpertmetric}, the evolution equations for the components of the screen vectors $s^\mu_A$, to linear order in $\phi$ and all of its derivatives, become
\bea
	\fl\frac{ds^0_A}{dt} = 
	\frac{s^0_A}{E} \left( -\mH k^0 \alp - ak^i\pd_i\phi \right)
	+ \frac{s^m_A}{E} \bigg[ a'\delta_{im} v^i \big( 2a k^0 - E\big) 
	+ k^0 a^2 \delta_{im}\pd_t v^i \nonumber\\ + a^2\delta_{lm}k^i \pd_i v^l \bigg],
\eea
and
\bea
	\fl\frac{ds^i_A}{dt} = \frac{s^0_A}{E} \bigg[-a\frac{k^i k^j}{k^0} (a' v^k \delta_{kj} + \pd_j\phi) - E \pd^i\phi - E\mH \frac{k^i}{k^0} \bigg] + \frac{s^m_A}{E} \bigg[ a^2 k^i \delta_{lm}\bigg( \frac{k^j}{k^0} \pd_j v^l  + \pd_t v^l\bigg) \nonumber\\ + 2aa' k^i v^l \delta_{lm} + a k^i\pd_m\phi - E\mH\delta^i_{\ph{i}m} - \frac{E}{k^0} \Gam^i_{jm} k^j \bigg].
\eea
The Jacobi matrix equation is, also to linear order in $\phi$ and all of its derivatives, 
\bea\label{eq:dtPAB_linperturb}
	\frac{d}{dt} \mathcal{P}^A_{\ph{A}B} &= \left(2\mH + \frac{2}{k^0}k^i\pd_i\phi\right) \mathcal{P}^A_{\ph{A}B} + \frac{1}{(k^0)^2} \mathcal{R}^A_{\ph{A}C} \mathcal{D}^C_{\ph{C}B},
\eea
where we have used the null condition to substitute
\bea
	\frac{\delta_{ij}k^ik^j}{(k^0)^2} = \frac{\alp^2}{a^2(1-2\phi)} = \frac{1+2\phi}{1-2\phi}.
\eea

We also must build the optical tidal matrix $\mathcal{R}^A_{\ph{A}C}$ for the linearly-perturbed space-time to advance \eref{eq:dtPAB_linperturb}. The components of the Riemann tensor we need, correct to first order in $\phi$ and all of its derivatives, are
\numparts\bea
	R_{0i0j} &= \left[ (a')^2 - aa''\right] (1-2\phi) \delta_{ij} + a^2 \pd_i \pd_j \phi, \nonumber\\
	R_{0xxy} &= - aa' \pd_y \phi,\nonumber\\
	R_{0xxz} &= -aa' \pd_z\phi, \nonumber\\
	R_{0yxy} &= aa' \pd_x\phi, \nonumber\\
	R_{0yyz} &= -aa' \pd_z\phi, \nonumber\\
	R_{0zxz} &= aa' \pd_x\phi, \nonumber\\
	R_{0zyz} &= aa' \pd_y\phi, \nonumber\\
	R_{xyxz} &= a^2 \pd_y \pd_z \phi,\nonumber\\
	R_{xyxy} &= (a')^2 (1-6\phi) + a^2\left(\pd_y^2\phi + \pd_x^2 \phi\right), \nonumber\\
	R_{xyyz} &= -a^2 \pd_x \pd_z \phi, \nonumber\\
	R_{xzyz} &= a^2 \pd_x \pd_y \phi, \nonumber\\
	R_{xzxz} &= (a')^2 (1-6\phi) + a^2\left(\pd_z^2\phi + \pd_x^2 \phi\right), \nonumber\\
	R_{yzyz} &= (a')^2 (1-6\phi) + a^2\left(\pd_z^2\phi + \pd_y^2 \phi\right),\nonumber
\eea\endnumparts
with $R_{0xyz}=R_{0yxz}=R_{0zxy}=0$. The components of the Ricci tensor are
\bea
    R_{00} &= -3\mH' + \nabla^2\phi, \\
    R_{0i} &= 2\mH \pd_i \phi, \\
    R_{ij} &= \left[\left( \mH^2 + \frac{a''}{a}\right)\left(1-4\phi\right) + \nabla^2\phi\right]\delta_{ij}.
\eea
In deriving all of the equations in this section, we have made use of the following relations:
\bea
	\pd_i \alp = \frac{a^2 \pd_i \phi}{\alp}, \quad
	\pd_t \alp = \frac{a'\alp}{a},\quad
	v^j \pd_l \gam_{ij} = -2a^2 v^j \pd_l (\phi) \delta_{ij}.\nonumber
\eea

\subsubsection{\texttt{Mescaline} vs. linearly-perturbed EdS.}

We set up the linearly-perturbed EdS metric \eref{eq:linpertmetric} in \mesc\ and send it into the ray tracer. %
We use a functional form of $\phi(x^i)$ of
\bea
    \phi(x^i) = \phi_0 \sum_i {\rm sin}\left(\frac{2\pi x^i}{L}\right),
\eea
where $L$ is the box length of the test domain and corresponds to the wavelength of the perturbation. 
For the analytic solutions, we use the same metric and advance $k^\mu, x^\mu, s^\mu_A$, and $\mathcal{D}^A_{\ph{A}B}$ according to the evolution equations given in the previous section. We store the semi-analytic solutions for the angular diameter distance, photon 4--momentum, and screen vectors at the same time steps as the ray-tracer output. 
We use three resolutions $N=24,48$, and 96, a perturbation amplitude of $\phi_0=10^{-8}$ to ensure second-order terms can be safely neglected, and a 4~$h^{-1}$~Gpc box size so that we can shoot rays to $z\approx 1$ without sampling the structure more than once. 
We place an observer at the centre of the domain, and shoot ten randomly drawn lines of sight. We calculate the relative difference between the fully nonlinear \mesc\ results (e.g. $D_A$) and the semi-analytic solutions (e.g. $D_{A, {\rm (semi-ana)}}$) for each line of sight, and average the resulting error over all lines of sight. %

\begin{figure*}[ht!]
	\centering
	\includegraphics[width=0.8\textwidth]{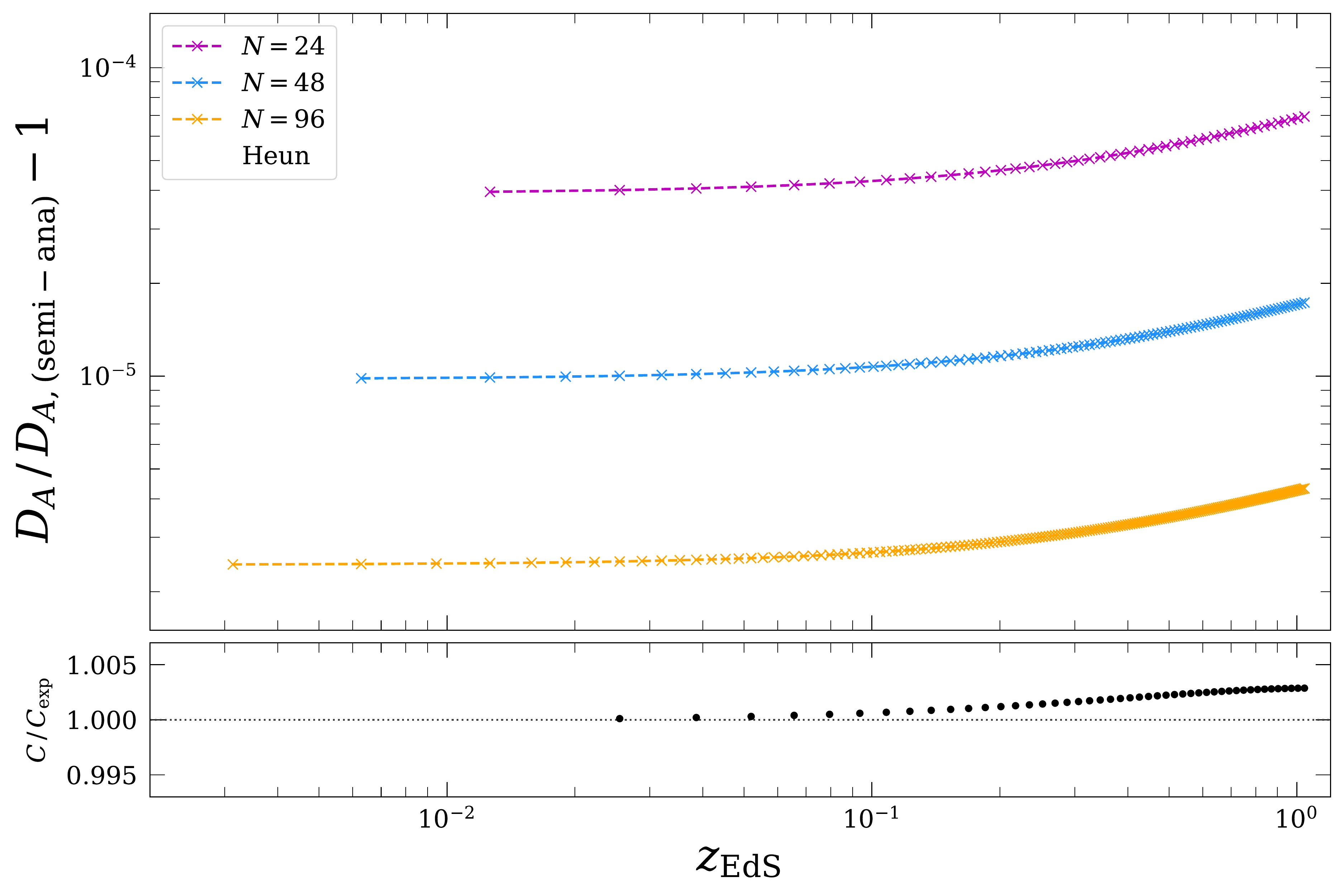}
	\caption{Top panel: Relative difference between ray-traced $D_A$ and the semi-analytic solution for a linearly-perturbed EdS spacetime. We show the average error over ten lines of sight for three resolutions $N=24,48$, and 96 as a function of the background redshift, $z_{\rm EdS}$. Bottom panel: convergence rate $C$, relative to the expected rate $C_{\rm exp}=4$ for RK2, for the three curves in the top panel.}
	\label{fig:DA_linpert}
\end{figure*}
The top panel of Figure~\ref{fig:DA_linpert} shows the ray-traced angular diameter distance relative to the semi-analytic distance for the same linearly-perturbed EdS metric. Different colours show resolutions $N=24,48$, and 96 and the bottom panel shows the convergence rate $C$ for the three curves relative to the expected for RK2 of $C_{\rm exp}=4$. 
The error is always below $\sim 0.003\%$ for resolution $N=96$, and converges at the expected rate.

\begin{figure*}[ht!]
	\centering
	\includegraphics[width=\textwidth]{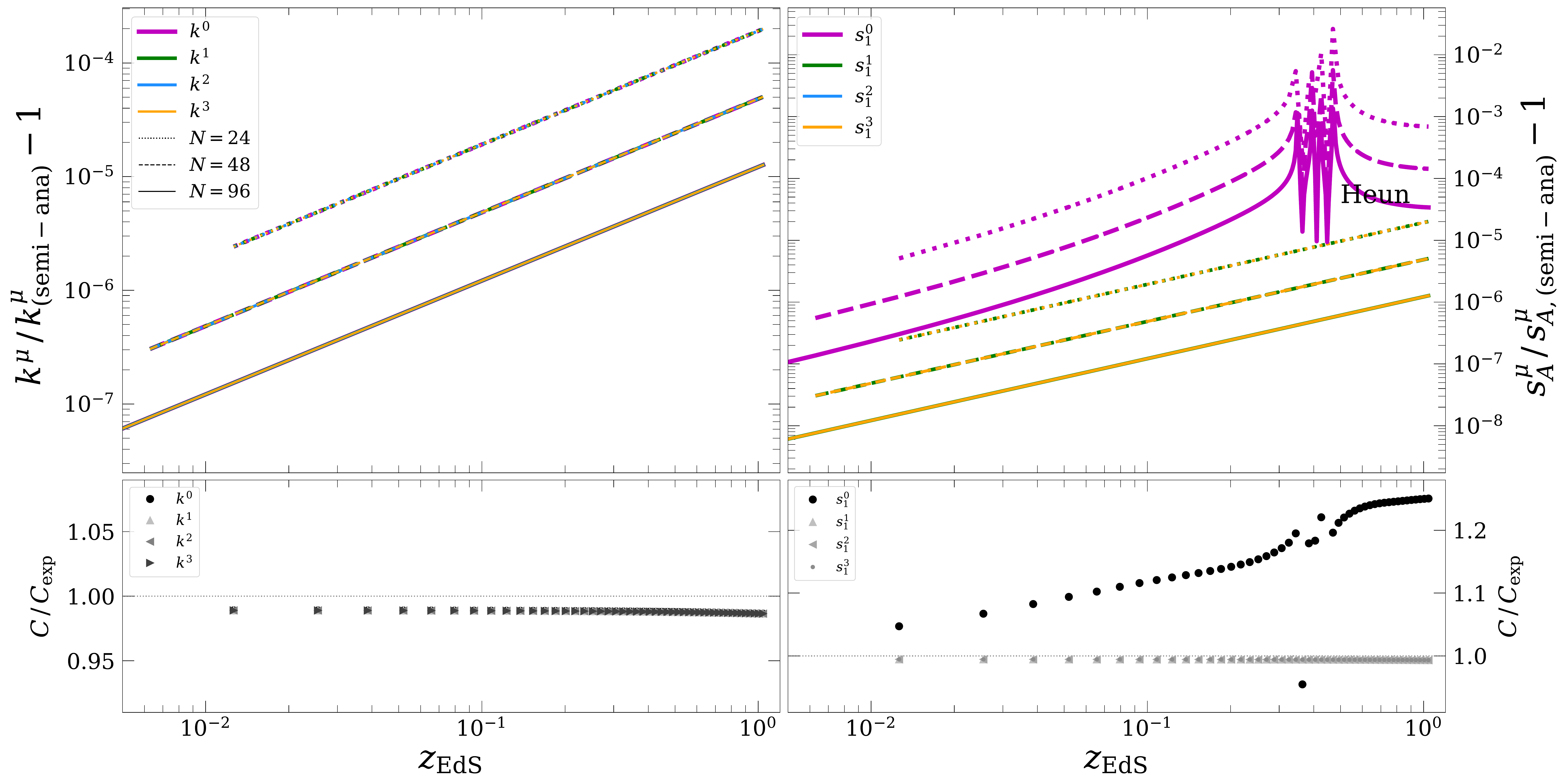}
	\caption{Top panels: Relative difference between $k^\mu$ (left) and $s^\mu_1$ (right) calculated with \mesc\ and the semi-analytic solutions for a linearly-perturbed EdS space-time. We show the average difference over ten lines of sight for three resolutions $N=24,48$, and 96 as a function of the background redshift. Bottom panels: convergence rate $C$, relative to the expected $C_{\rm exp}=4$, for the curves in the respective top panels.}
	\label{fig:kmu_smu_linpert}
\end{figure*}
Figure~\ref{fig:kmu_smu_linpert} shows the relative difference between the photon 4--momentum components $k^\mu$ (top left) and screen vector components $s^\mu_1$ (top right) and the semi-analytic solutions for the linearly-perturbed EdS metric.
The bottom panels show the convergence rate $C$ relative to the expected $C_{\rm exp}=4$ for RK2 for the respective top panels. 
All $k^\mu$ and $s^i_1$ components have very small errors and converge at the expected rate. 
We note that we only show $s^\mu_1$ (and not $s^\mu_2$) because both cases $A=1,2$ are evolved using the same source terms. %

The error in $s^0_1$ is a few orders of magnitude larger than the spatial components. For the background we have $s^0_A=0$, and so this component is perturbative only and thus has a smaller magnitude than the spatial components, all of which have background contributions. This further implies that a larger fraction of the error is made up of round-off error, and so we might expect the relative error to be larger than the other components. Additionally, the noise we see in the error in $s^0_1$ can be attributed to a sign change in the error in some of the lines of sight that make up the average shown here.

\subsection{Errors in analysis of simulation data}\label{appx:sim_conv}

We are also interested in the level of error in our ray-tracing calculations in a situation for which we do not have an analytic solution. In such a case, we can estimate the error in our calculation via a Richardson extrapolation. This is based on the expectation that our calculations should approach the ``true'' solution as we increase our numerical resolution towards inifinty at a rate which is determined by the order of accuracy of the implemented scheme. 

We use three NR simulations with resolutions $N=64,128$, and 256 which are initialised with a power spectrum of large-scale only perturbations within a $L=1.28\,h^{-1}$ Gpc domain. Specifically, all power beneath $\sim 200 \,h^{-1}$ Mpc is removed from the initial data, and thus the simulations remain \textit{close to} the linear regime for the extent of the matter-dominated simulations from the initial redshift of $z=1000$ until $z=0$. The initial data is identical between resolutions, and thus we expect minimal difference between the simulations at redshift zero for the large scales we sample. These simulations are thus suitable for a Richardson extrapolation (for quantities calculated at individual positions in the domain), for which we require derivatives to be the same between resolutions. 

For all simulations, we place 10 observers randomly and propagate an isotropic sky of $12\times N_{\rm side}^2$ lines of sight for $N_{\rm side}=16$ until $z\approx 0.1$ using an RK2 integrator. All lines of sight and observer positions are identical for all simulations. We thus expect $D_L$ and $z$ calculated along each line of sight, and for each observer, to converge at a rate $\propto N^{-2}$. For each observer, each iteration, and each line of sight, we fit a curve of the form $D_L(N) = D_{L,{\rm inf}} + X / N^2$, where $D_{L,{\rm inf}}$ is the $N\rightarrow\infty$ luminosity distance (or the ``true'' solution) and $X$ is a constant. We determine both $D_{L,{\rm inf}}$ and $X$ using the \texttt{curve\_fit} function as a part of the SciPy\footnote{\url{https://scipy.org}} Python package, using a similar relation for the redshift $z$. The relative error in $D_L$ (similarly for $z$) is then
\begin{equation}
    {\rm err}(D_L) \equiv \frac{D_{L,N=256}}{D_{L,{\rm inf}}} - 1,
\end{equation}
namely, the relative difference between our highest-resolution calculation and the ``true'' solution at $N\rightarrow\infty$. We average the error over each observer's sky at each iteration to track the level of error for all observers at all redshifts, namely $\langle {\rm err}(D_L)\rangle_\Omega(z_{\rm mean})$ for each observer. 

\begin{figure*}[ht]
    \centering
    \includegraphics[width=0.7\textwidth]{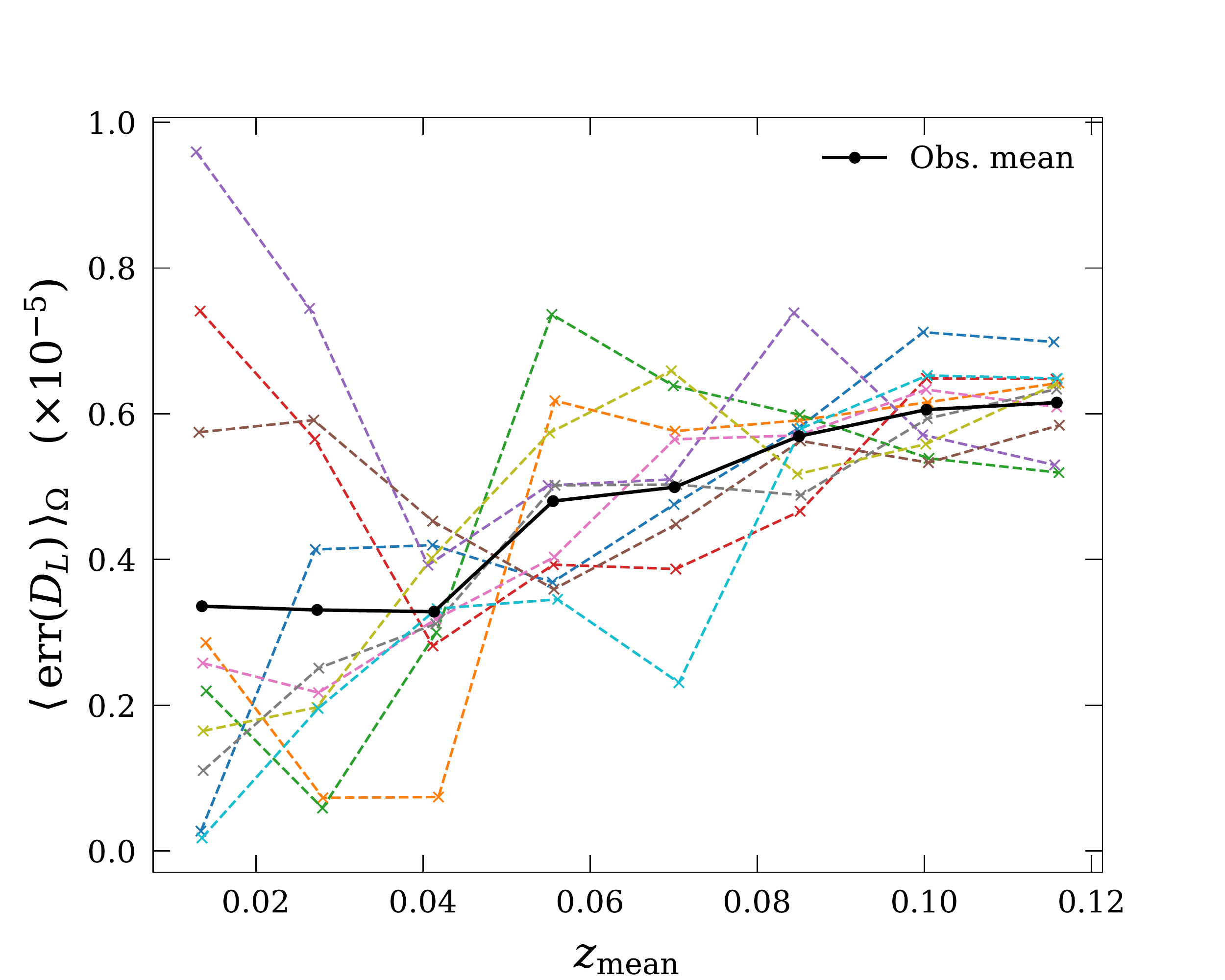}
    \caption{Sky-average of the relative error in $D_L$ for 10 observers (coloured dashed curves) as a function of the mean redshift of the slice. The thick black curve shows the average over all observers.}
    \label{fig:DLerr_Rich}
\end{figure*}

\begin{figure*}[ht]
    \centering
    \includegraphics[width=0.7\textwidth]{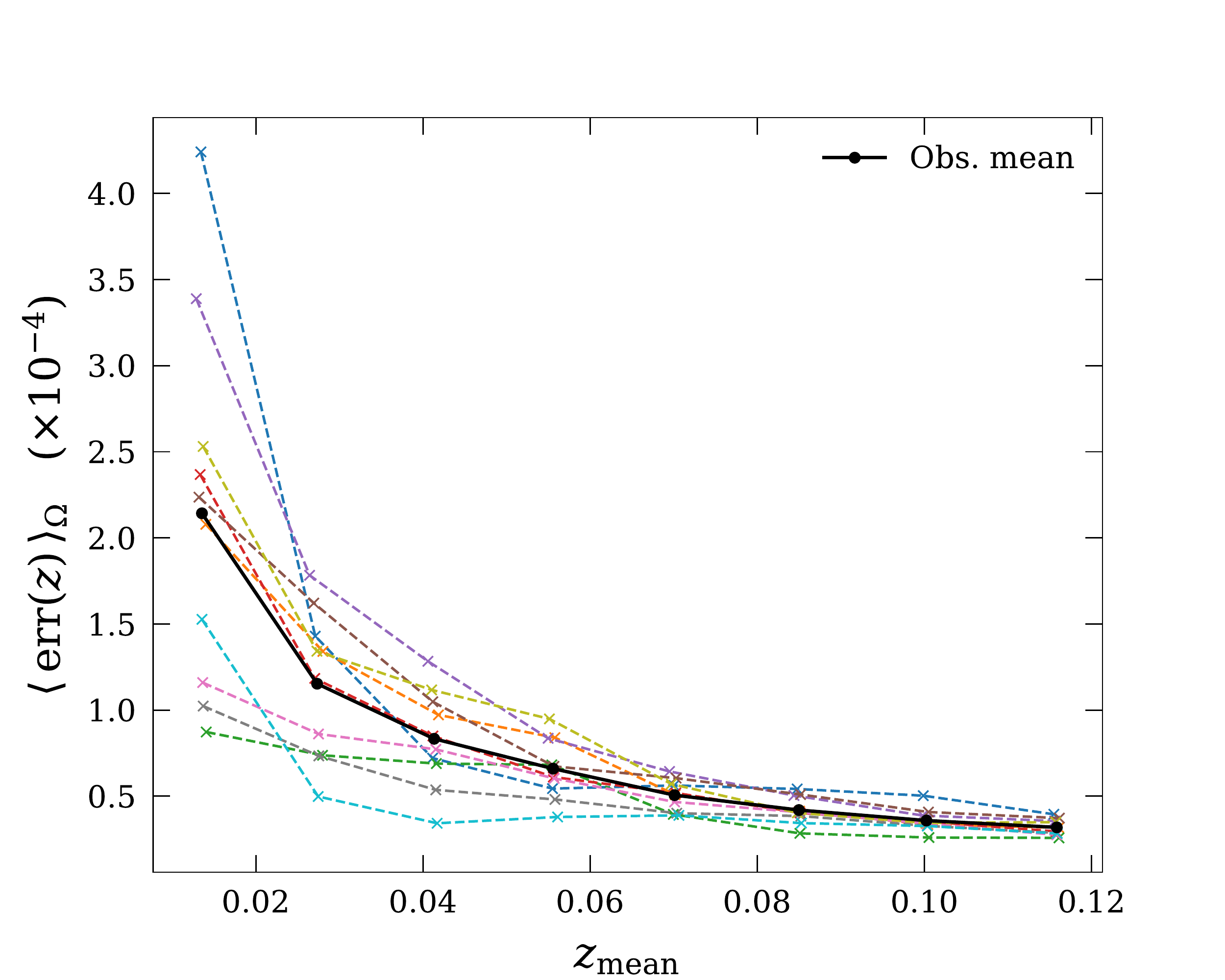}
    \caption{Sky-average of the relative error in the redshift $z$ for 10 observers (coloured dashed curves) as a function of the mean redshift of the slice. The thick black curve shows the average over all observers.}
    \label{fig:zerr_Rich}
\end{figure*}
Figures~\ref{fig:DLerr_Rich} and \ref{fig:zerr_Rich} show the sky-averaged relative error in $D_L$ and $z$, respectively, as a function of the mean redshift on that observer's slice, $z_{\rm mean}$. The relative error in $D_L$ is always of order $10^{-5}$ and the relative error in $z$ is always of order $10^{-4}$.

\subsection{Violation of the null constraint}\label{appx:null_test}

In \mesc, the propagation of the components of the photon 4--momentum $k^\mu$, as well as the setting of initial data, is done independently of the null condition requirement for photons, namely $k^\mu k_\mu = 0$. We can thus use this condition as a numerical test of the accuracy of the numerical evolution of $k^\mu$. 

In this appendix, we assess the violation of the null condition in the \mesc\ ray tracer. First, we test this using the analytic linear-perturbed metric tensor from \ref{appx:linperttest}, and second we test this using a set of NR simulations similar to the one used in \ref{appx:screen_init_test}.

\subsubsection{Linearly-perturbed metric. }

We use the test functionality of \mesc\ to pass the linearly-perturbed FLRW metric tensor \eref{eq:linpertmetric} into the ray tracing routines. For each case, we place a single observer in the model universe and propagate 10 randomly-drawn lines of sight for a set time interval. We repeat this process for three resolutions $N=12,24$, and 48 with the same spacing in time (resulting in different numbers of iterations). We also use four different cases of perturbation size, controlled by the amplitude of the metric perturbation $\phi_0$. We choose $\phi_0=10^{-2}, 10^{-4}, 10^{-6}$, and $10^{-8}$ for these tests. 

\begin{figure*}[ht]
    \centering
    \includegraphics[width=0.9\textwidth]{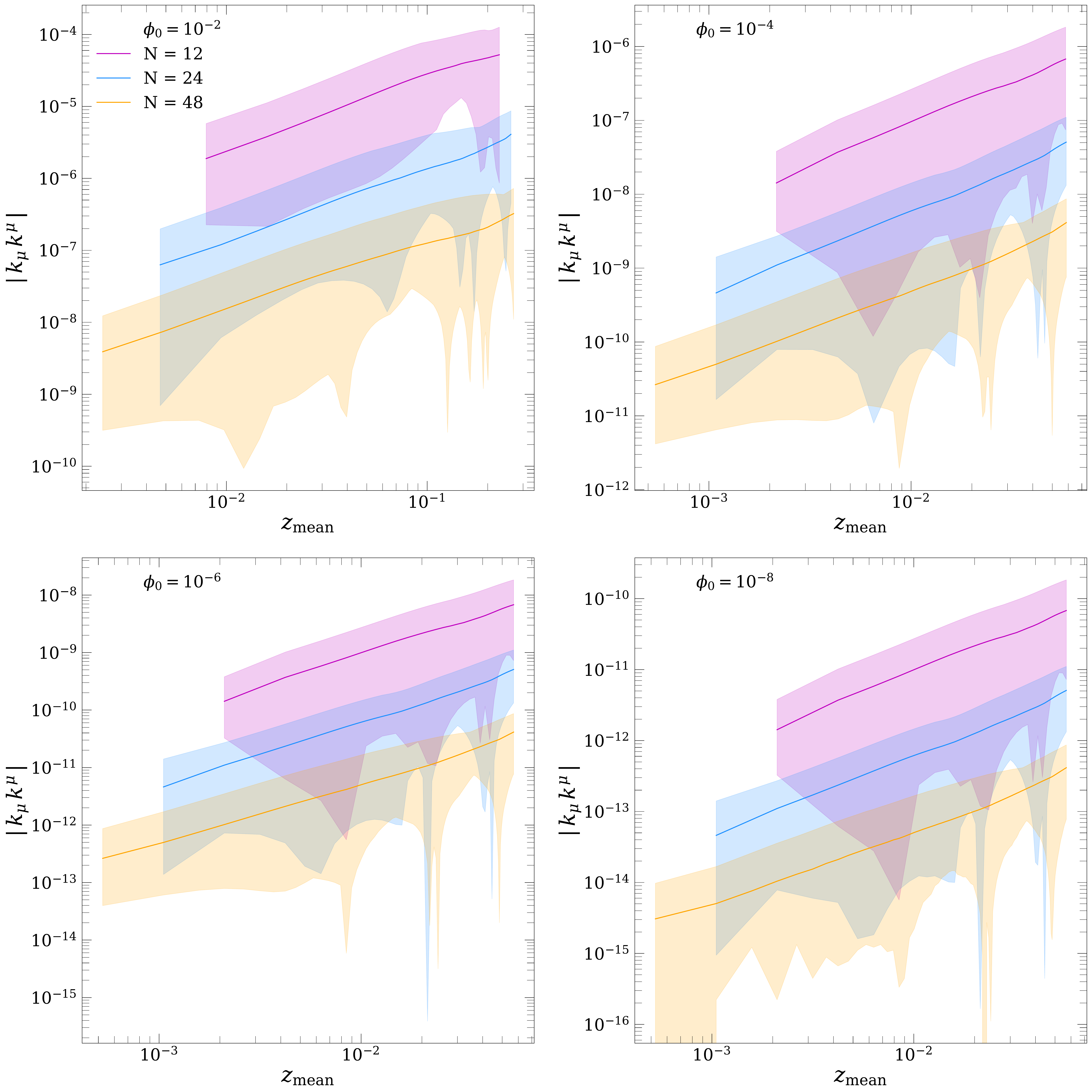}
    \caption{Absolute value of the violation of the null condition as a function of redshift (mean across 10 lines of sight) for a linearly-perturbed FLRW test metric tensor. Each panel shows a different amplitude of perturbation, all with three resolutions $N=12,24$, and 48 (magenta, blue, and orange curves, respectively). Solid curves show the mean over 10 lines of sight for one observer, and the shaded regions show the range of minimum violation to maximum violation.}
    \label{fig:null_linpert}
\end{figure*}
Figure~\ref{fig:null_linpert} shows the absolute value of the null condition violation, $|k^\mu k_\mu|$, as a function of redshift for four cases of $\phi_0$ (panels). Different colours represent different resolutions, as indicated in the legend, and solid curves show the mean over all 10 lines of sight for this observer and the shaded region represents the range from minimum violation to maximum violation across all 10 lines of sight. All panels show the same observer, and the redshift on the x-axis is the redshift averaged over all lines of sight. 
We note that for $\phi_0=10^{-2}$, the final redshift is $z\approx 0.27$ whereas all other perturbation sizes give a final redshift of $z\approx 0.06$ for the same simulation time, due to the larger gravitational field the photon must travel through. 

The violation in the null condition reduces as we increase resolution, as would be expected of a numerical error. The ratio $|k^\mu k_\mu|/\phi_0$ is approximately constant at $\approx 10^{-6}$--$10^{-5}$ for $N=48$ across all cases. Since the violation reduces with resolution, we conclude that it is dominated by numerical error. 

\subsubsection{Simulation test.}

We also test the violation of the null condition in a set of NR simulations. In this case, we randomly place 10 observers within the simulation $z\approx 0$ hypersurfaces, and propagate 50 randomly-drawn lines of sight for each observer until $z\approx 0.06$--0.09 using an RK2 integrator, tracking the violation in $k^\mu k_\mu$ along the way. 
We use two simulations, with resolution $N=64$ and 128, which have identical initial data and box size $640\,h^{-1}$ Mpc. The initial perturbations are drawn from a CLASS power spectrum and have all modes below $100\,h^{-1}$ Mpc removed to ensure modes are sampled with sufficiently many grid points. 
This set of simulations are similar to those used in \ref{appx:sim_conv} except they sample smaller-scale structures initially. Therefore, the physical structure in these simulations at $z\approx 0$ differs between resolutions even though the initial data is identical.
We maintain the same spacing in iteration for the runs at different resolution, to ensure we are studying convergence of both spatial and time derivatives. 

\begin{figure*}[ht]
    \centering
    \includegraphics[width=0.9\textwidth]{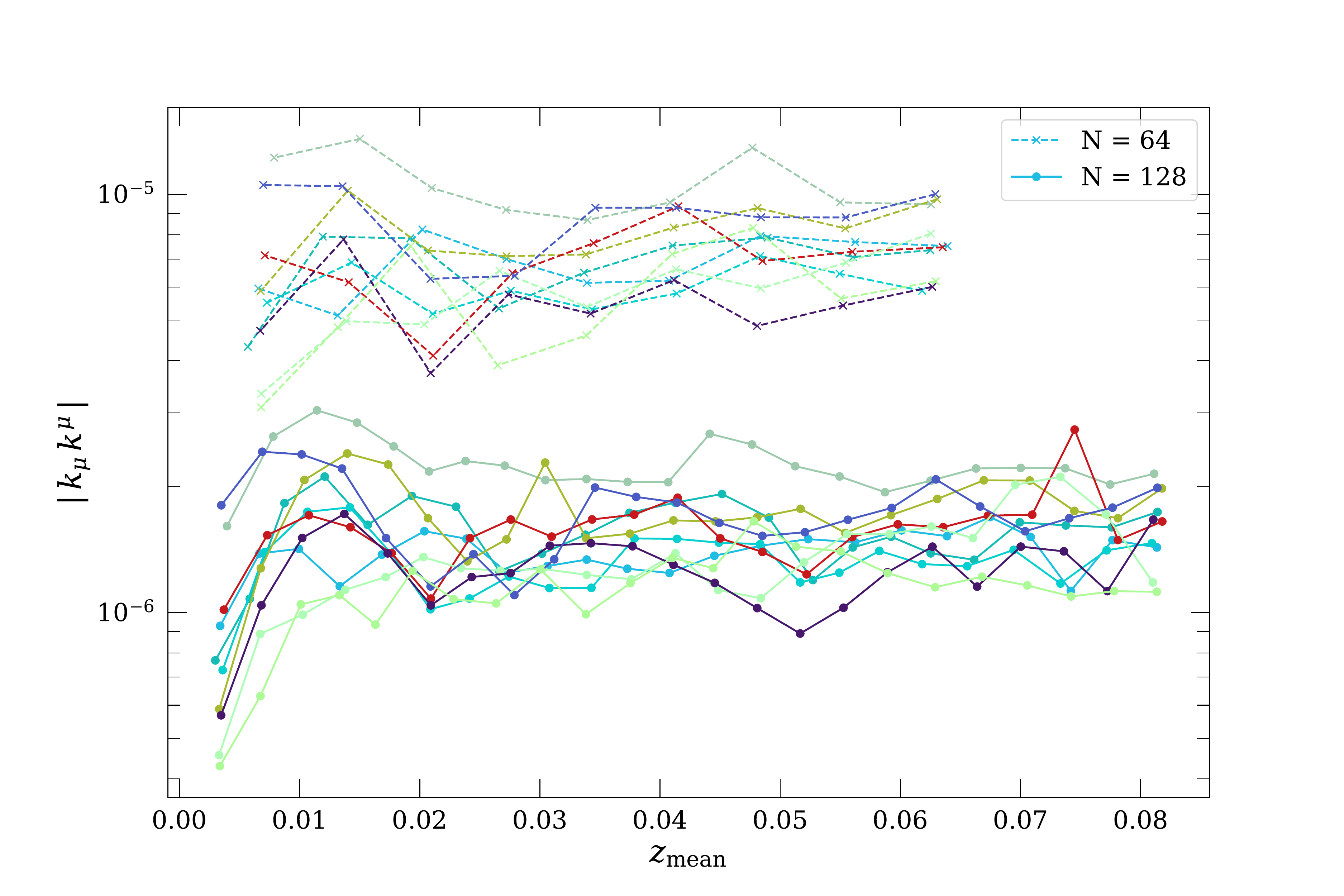}
    \caption{Absolute value of the violation of the null condition as a function of redshift for 10 observers placed in two NR simulations. Solid and dashed curves show the violation for 10 observers as averaged over 50 lines of sight for resolutions $N=128$ and 64, respectively. We show the violation as a function of each observer's mean redshift for that time slice, $z_{\rm mean}$. %
    }
    \label{fig:null_sim}
\end{figure*}
Figure~\ref{fig:null_sim} shows the absolute value of the violation in the null condition for 10 observers averaged across 50 lines of sight each for resolutions $N=128$ (solid curves) and $N=64$ (dashed curves). 
The violation in the null condition remains $<10^{-5}$ for all cases studied here, and reduces by about a factor of four with increasing resolution $N=64\rightarrow 128$, as is expected based on a second-order accurate scheme.

\newcommand{\newblock}{}
\bibliographystyle{apalike}
\bibliography{refs.bib}

\end{document}